\pdfoutput=1
\documentclass[a4paper,11pt]{article}

\usepackage{jheppub}
\usepackage[T1]{fontenc} 
\usepackage[numbers,sort&compress]{natbib}
\usepackage{multirow}
\usepackage{subfigure}
\def\subfigtopskip{4pt}
\def\subfigcapskip{-10pt}
\def\subfigcapmargin{0pt}
\usepackage{lineno}
\usepackage{rotating}
\usepackage{booktabs}
\usepackage{epstopdf}

\def\pt{\ensuremath{p_{\mathrm{T}}}} 
\def\pT{\ensuremath{p_{\mathrm{T}}}} 
\def\ET{\ensuremath{E_{\mathrm{T}}}} 
\def\TeV{\ifmmode {\mathrm{\ Te\kern -0.1em V}}\else
                   \textrm{Te\kern -0.1em V}\fi}%
\def\GeV{\ifmmode {\mathrm{\ Ge\kern -0.1em V}}\else
                   \textrm{Ge\kern -0.1em V}\fi}%
\def\MeV{\ifmmode {\mathrm{\ Me\kern -0.1em V}}\else
                   \textrm{Me\kern -0.1em V}\fi}%
\def\keV{\ifmmode {\mathrm{\ ke\kern -0.1em V}}\else
                   \textrm{ke\kern -0.1em V}\fi}%
\def\eV{\ifmmode  {\mathrm{\ e\kern -0.1em V}}\else
                   \textrm{e\kern -0.1em V}\fi}%

\def\tbull{ \textbullet}
\def\MET{\ensuremath{E_\mathrm{T}^\mathrm{miss}}}
\def\mT{\ensuremath{m_{\mathrm{T}}}}

\def\Ztautau{\ensuremath{Z\to\tau\tau}}
\def\Zmumu{\ensuremath{Z\to\mu^+\mu^-}}

\def\bbbar{\ensuremath{b\bar{b}}~}
\def\ttbar{\ensuremath{t\bar{t}}~}

\def\MMC{\ensuremath{m_{\tau\tau}^{\mathrm{MMC}}}}

\def\thad{\ensuremath{\tau_{\mathrm{ had}}}}

\def\tlep{\ensuremath{\tau_{\mathrm{ lep}}}}
\def\t{\ensuremath{\tau}}

\def\thadhad{\ensuremath{\thad\thad}}
\def\tlhad{\ensuremath{\tlep\thad}}
\def\tll{\ensuremath{\tlep\tlep}}

\def\Wjets{ \ensuremath{W{\rm+jets}} }
\def\Zjets{ \ensuremath{Z{\rm+jets}} }
\def\Zmumu{ \ensuremath{Z(\to\mu\mu){\rm+jets}} }

\def\Zll{ \ensuremath{Z \to \ell \ell} }

\def\atk{anti-$k_t$} 

\newcommand{\hgg}{\mbox{$H \to  \gamma \gamma$}}
\newcommand{\hbb}{\mbox{$H \to  b \bar{b}$}}

\newcommand{\hzzs}{\mbox{$H \to  Z Z^*$}}

\newcommand{\hwws}{\mbox{$H \to  W W^{*}$}}

\newcommand{\htautau}{\mbox{$H \to  \tau \tau$}}

\newcommand{\fbs}{\mbox{$\rm{fb}^{-1}$}}

\def\mathswitchr#1{\relax\ifmmode{\mathrm{#1}}\else$\mathrm{#1}$\fi}

\def\mathswitch#1{\relax\ifmmode#1\else$#1$\fi}

\usepackage{amsmath}
\usepackage{preprintcover}  
\PreprintCoverPaperTitle{Evidence for the Higgs-boson Yukawa coupling to tau leptons with the ATLAS detector}  
\PreprintIdNumber{CERN-PH-EP-2014-262}  
\PreprintCoverAbstract{
Results of a search for $\htautau$ decays are presented, based on the full set
of proton--proton collision data recorded by the ATLAS experiment at the LHC
during 2011 and 2012. The data correspond to integrated luminosities of
4.5~\fbs\ and 20.3~\fbs\ at centre-of-mass energies of $\sqrt{s}$ = 7~\TeV\ and
$\sqrt{s}$ = 8~\TeV\ respectively. All combinations of leptonic
($\tau \to \ell \nu   \bar  \nu$ with $\ell = e, \mu$) and hadronic ($\tau \to \rm{hadrons}~ \nu$) tau decays
are considered. An excess of events over the expected background from other
Standard Model processes is found with an observed (expected) significance of
4.5 (3.4) standard deviations.  This excess provides evidence for the direct
coupling of the recently discovered Higgs boson to fermions.
The measured signal strength, normalised to the Standard Model
expectation, of $\mu = 1.43 \ ^{+0.43}_{-0.37}$
is consistent with the predicted Yukawa coupling strength in the Standard
Model.}  

\PreprintJournalName{Journal of High Energy Physics}  

\title{\boldmath Evidence for the Higgs-boson Yukawa coupling to tau leptons with the ATLAS detector}

\author[]{The ATLAS Collaboration}

\abstract{
Results of a search for $\htautau$ decays are presented, based on the full set
of proton--proton collision data recorded by the ATLAS experiment at the LHC
during 2011 and 2012. The data correspond to integrated luminosities of
4.5~\fbs\ and 20.3~\fbs\ at centre-of-mass energies of $\sqrt{s}$ = 7~\TeV\ and
$\sqrt{s}$ = 8~\TeV\ respectively. All combinations of leptonic
($\tau \to \ell \nu   \bar  \nu$ with $\ell = e, \mu$) and hadronic ($\tau \to
\rm{hadrons}~ \nu$) tau decays
are considered. An excess of events over the expected background from other
Standard Model processes is found with an observed (expected) significance of
4.5 (3.4) standard deviations.  This excess provides evidence for the direct
coupling of the recently discovered Higgs boson to fermions. 
The measured signal strength, normalised to the Standard Model
expectation, of $\mu = 1.43 \ ^{+0.43}_{-0.37}$
is consistent with the predicted Yukawa coupling strength in the Standard
Model.}

\makeatletter
\g@addto@macro\bfseries{\boldmath}
\makeatother

\begin{document}

\maketitle
\flushbottom

\section{Introduction}
\label{sec:intro}

The investigation of the origin of electroweak symmetry breaking and, related
to this, the experimental confirmation of the Brout--Englert--Higgs
mechanism~\cite{Englert:1964et,Higgs:1964ia,Higgs:1964pj,Guralnik:1964eu,Higgs:1966ev,Kibble:1967sv}
is one of the prime goals of the physics programme at the Large Hadron Collider
(LHC)~\cite{1748-0221-3-08-S08001}.  With the discovery of a Higgs boson with a
mass of approximately 125~\GeV\ by the ATLAS~\cite{paper2012ichep} and
CMS~\cite{CMSpaper} collaborations, an important milestone has been reached.
More precise measurements of the properties of the discovered
particle~\cite{Aad:2013wqa,Chatrchyan:2013lba} as well as tests of the
spin--parity quantum numbers~\cite{Aad:2013xqa,Chatrchyan:2012jja,Khachatryan:2014kca} 
continue to be consistent with the predictions for the Standard Model (SM) Higgs boson.

These measurements rely predominantly on studies of the bosonic decay modes,
\hgg, \hzzs\ and \hwws.  To establish the mass generation mechanism for fermions as
implemented in the SM, it is of prime importance to demonstrate the direct
coupling of the Higgs boson to fermions and the proportionality of its
strength to
mass~\cite{Weinberg:1967tq}.  The most promising candidate decay modes are the
decays into tau leptons, \htautau, and bottom quarks ($b$-quarks), \hbb.  Due to the high background, the
search for decays to $\bbbar$ is restricted to Higgs bosons produced
in modes which have a more distinct signature but a lower cross-section, 
such as $H$ production with an associated vector boson.
The smaller rate of these processes in the presence of still large background
makes their detection challenging. More favourable signal-to-background
conditions are expected for \htautau\ decays.  
Recently, the CMS Collaboration published evidence for \htautau\ decays at a significance in terms of standard deviations  of $3.2\sigma$~\cite{Chatrchyan:2014nva}, and an excess corresponding to a significance of $2.1 \sigma$ in the search for $H \to b \bar{b}$ decays~\cite{Chatrchyan:2013zna}. The combination of channels provides evidence for fermionic couplings with a significance of $3.8 \sigma$~\cite{Chatrchyan:2014vua}.  The yield of events in the search for $H \to b \bar{b}$ decays observed by the ATLAS Collaboration has a signal significance of $1.4 \sigma$~\cite{ATLAS_Hbb}.  The Tevatron experiments have observed an excess corresponding to $2.8 \sigma$ in the  $H \to b \bar{b}$ search~\cite{Aaltonen:2012qt}. 

In this paper, the results of a search for \htautau\ decays are presented,
based on the full proton--proton dataset collected by the ATLAS experiment
during the 2011 and 2012 data-taking periods, corresponding to integrated
luminosities of  4.5~\fbs\ at a centre-of-mass energy of $\sqrt{s}=7\TeV$ and
20.3~\fbs\ at $\sqrt{s}=8\TeV$.
These results supersede the earlier upper limits on the cross section times the branching ratio obtained with the 7 TeV data~\cite{Aad:2012mea}.
All combinations of leptonic ($\tau \to \ell \nu   \bar  \nu$ with $\ell = e,
\mu$) and hadronic ($\tau \to \rm{hadrons} \ \nu$) tau  decays
are considered.\footnote{Throughout this paper
the inclusion of charge-conjugate decay modes is implied.} The corresponding
three analysis channels are denoted by $\tlep \tlep$,  $\tlep \thad$, and
$\thad \thad$ in the following.  The search is designed to be sensitive to the
major production processes of a SM Higgs boson, i.e.~production via gluon
fusion (ggF)~\cite{Alioli:2008tz}, vector-boson fusion (VBF)~\cite{Nason:2009ai},
 and associated production ($VH$) with $V=W$ or
$Z$. These production processes lead to different final-state signatures, which
are exploited by defining an event categorisation. 
Two dedicated categories are considered to achieve both a good
signal-to-background ratio and good resolution for the reconstructed 
$\tau \tau$ invariant mass.
The VBF category, enriched in events produced via
vector-boson fusion, is defined by the presence of two jets with a large
separation in pseudorapidity.\footnote{The ATLAS experiment uses a right-handed
coordinate system with its origin at the nominal interaction point (IP) in
the centre of the detector and the $z$-axis along the beam direction. The
$x$-axis points from the IP to the centre of the LHC ring, and the $y$-axis
points upward. Cylindrical coordinates $(r,\phi)$ are used in the
transverse $(x,y)$ plane, $\phi$ being the azimuthal angle around the beam
direction. The pseudorapidity is defined in terms of the polar angle
$\theta$ as $\eta=-\ln\tan(\theta/2)$. The distance $\Delta{}R$ in the
$\eta$--$\phi$ space is defined as $\Delta{}R=\sqrt{({\Delta\eta})^2 +
({\Delta\phi})^2}$.} The boosted category contains events where the reconstructed Higgs boson candidate has a large transverse momentum. It is
dominated by events produced via gluon fusion with additional jets from
gluon radiation. In view of the signal-to-background conditions, and in
order to exploit correlations between final-state observables, a
multivariate analysis technique, based on boosted decision trees
(BDTs)~\cite{bdt, bdt:gradboost, bdt:adaboost}, is used to extract the
final results. As a cross-check, a separate analysis where cuts on
kinematic variables are applied is carried out.

\section{The ATLAS detector and object reconstruction}
\label{sec:atlas}

The ATLAS detector~\cite{DetectorPaper:2008} is a multi-purpose detector with a
cylindrical geometry.  It comprises an inner detector (ID) surrounded by a thin
superconducting solenoid, a calorimeter system and an extensive muon
spectrometer  in a toroidal magnetic field. The ID tracking system
consists of a silicon pixel detector, a silicon microstrip detector (SCT), and
a transition radiation tracker (TRT).  It provides precise position and
momentum measurements for charged particles and allows efficient identification
of jets containing $b$-hadrons ($b$-jets) in the pseudorapidity range $|\eta|<2.5$. The ID
is immersed in a $2~\mathrm{T}$ axial magnetic field and is surrounded by high-granularity
 lead/liquid-argon (LAr) sampling electromagnetic calorimeters which
cover the pseudorapidity range $ |\eta|< 3.2$. A steel/scintillator tile
calorimeter provides hadronic energy measurements in the central pseudorapidity
range ($ |\eta|< 1.7$). In the forward regions ($1.5 < |\eta| < 4.9$), the
system is complemented by two end-cap calorimeters using LAr as active material
and copper or tungsten as absorbers. The muon spectrometer (MS) surrounds the
calorimeters and consists of three large superconducting eight-coil toroids, a
system of tracking chambers, and detectors for triggering.  The deflection of
muons is measured  within $|\eta|< 2.7$ by three layers of precision drift
tubes, and cathode strip chambers in the innermost layer for $|\eta| > 2.0$.
The trigger chambers consist of resistive plate chambers in the barrel ($|\eta|
< 1.05$) and thin-gap chambers in the end-cap regions ($1.05<|\eta|<2.4$).

A three-level trigger system~\cite{TriggerPapaer:2012} is used to select events.
A hardware-based Level-1 trigger uses a subset of detector information to reduce the event
rate to a value to $75$~kHz or less. The rate of accepted events is then reduced to about
$400$~Hz by two software-based trigger levels, Level-2 and the Event Filter.

The reconstruction of the basic physics objects used in this
analysis is described in the following. The primary vertex referenced below
is chosen as the proton--proton vertex candidate with the highest sum of the squared transverse
momenta of all associated tracks.

Electron candidates are reconstructed from energy clusters in the
electromagnetic calorimeters matched to a track in the ID. They are required to
have a transverse energy, $\ET = E\sin\theta$, greater than $15\GeV$, to be within the
pseudorapidity range $|\eta|<2.47$, and to satisfy the medium shower shape and
track selection criteria defined in ref.~\cite{Aad:2014fxa}.  Candidates found
in the transition region between the end-cap and barrel calorimeters ($1.37<|\eta|<1.52$) are not considered.
Typical reconstruction and identification efficiencies for electrons satisfying
these selection criteria range between 80\% and 90\% depending on $\ET$ and $\eta$.

Muon candidates are reconstructed using an
algorithm~\cite{Aad:2014rra} 
that combines information from the ID and the MS.  The distance between the
$z$-position of the point of closest approach of the muon inner-detector track to
the beam-line and the $z$-coordinate of the primary vertex is required to be less
than 1 cm. 
 This requirement reduces the contamination due to cosmic-ray muons
and beam-induced backgrounds. Muon quality criteria such as inner detector hit
requirements are applied to achieve a precise measurement of the muon
momentum and reduce the misidentification rate.  Muons are required to have a
momentum in the transverse plane $\pT >10\GeV$ and to be within $|\eta|<$~2.5.
Typical efficiencies for muons satisfying these selection criteria are above
$95\%$~\cite{Aad:2014rra}.

Jets are reconstructed using the \atk\ jet clustering algorithm~\cite{Antikt1,
Antikt2} with a radius parameter $R=0.4$, taking topological energy
clusters~\cite{ATL-LARG-PUB-2008-002} in the calorimeters as inputs.  Jet
energies are corrected for the contribution of multiple interactions using a
technique based on jet area~\cite{Cacciari:2007fd} and are calibrated using $\pT$-
and $\eta$-dependent correction factors determined from data and
simulation~\cite{Aad:2011he,Aad:2014bia,Aad:2012vm}.
Jets are required to be reconstructed
in the range $|\eta|<4.5$ and to have $\pT > 30\GeV$.
To reduce the contamination of jets by additional interactions in the same or
neighbouring bunch crossings (pile-up), tracks originating from the primary vertex 
must contribute a large fraction of the $\pT$ when summing the scalar $\pT$ of all tracks
in the jet.  This jet vertex fraction (JVF) is required to be at least 75\% (50\%) for jets
with  $| \eta | < 2.4$ in the $7\TeV$ ($8\TeV$) dataset.
Moreover, for the 8~TeV dataset, the JVF selection
is applied only to jets with $\pT < 50\GeV$.  Jets with no associated tracks
are retained.

In the pseudorapidity range $\left|\eta\right|<2.5$, $b$-jets are selected
using a tagging algorithm~\cite{ATLAS-CONF-2014-046}.  The $b$-jet tagging
algorithm  has an efficiency of 60--70\% for $b$-jets in simulated $\ttbar$
events. 
The corresponding light-quark jet
misidentification probability is 0.1--1\%, depending on the jet's $\pt$ and
$\eta$.

Hadronically decaying  tau  leptons are reconstructed starting from clusters of
energy in the electromagnetic and hadronic calorimeters. The \thad\
\footnote{In the following, the \thad\ symbol always refers to the visible
decay products of the \t\ hadronic decay.} reconstruction is seeded by the \atk\
jet finding algorithm with a radius parameter $R = 0.4$. 
Jet-specific cleaning selection such as the JVF requirement is not needed for the  tau candidate seeds, due to stricter vertex requirements.
Tracks with $\pt >1\GeV$ within a cone of
radius 0.2 around the cluster barycentre are matched to the
\thad\ candidate, and the \thad\ charge is determined from the sum of the
charges of its associated tracks.
The rejection of jets is provided in a separate identification step using
discriminating variables based on tracks with $\pt > 1\GeV$ and the energy deposited in calorimeter cells found in the core region ($\Delta{}R < 0.2$) and in the region $0.2 <
\Delta{}R < 0.4$ around the \thad\ candidate's direction. Such discriminating
variables are combined in a boosted decision tree and three working points,
labelled tight, medium and loose~\cite{ATLAS_Tauperf},
are defined, corresponding to different \thad\ identification efficiency
values.

In this analysis, \thad\ candidates with $p_\mathrm{T}>20\GeV$ and
$|\eta|<2.47$ are used. The \thad\ candidates are required to have charge
$\pm{}1$, and must be $1$- or $3$-track (prong) candidates. In addition, a
 sample without the charge and track multiplicity requirements is retained for
background modelling in the $\thadhad$ channel, as described in section~\ref{subsec:qcdbkg}.
The identification efficiency for \thad\ candidates satisfying the medium
 criteria is of the order of 55--60\%.  Dedicated
criteria~\cite{ATLAS_Tauperf} to separate \thad\ candidates from
misidentified electrons are also applied, with a selection efficiency for true
\thad\ decays of $95\%$.  The probability to misidentify a jet with
$p_\mathrm{T}>20\GeV$ as a \thad\ candidate is typically 1--2\%.

Following their reconstruction, candidate leptons, hadronically decaying taus
and jets may point to the same energy deposits in the calorimeters (within
$\Delta{}R<0.2$). Such overlaps are resolved by selecting objects in the following order of
priority (from highest to lowest): muons, electrons, \thad, and jet candidates.  For all channels, the
leptons that are considered in overlap removal with \thad\ candidates need
to only satisfy looser criteria than those defined above, to reduce
misidentified \thad\ candidates from leptons. The \pt\ threshold of muons
considered in overlap removal is also lowered to $4\GeV$.

The missing transverse momentum (with magnitude \MET)
is reconstructed using the energy deposits in calorimeter cells calibrated
according to the reconstructed physics objects ($e$, $\gamma$, \thad, jets and
$\mu$) with which they are
associated~\cite{springerlink:10.1140/epjc/s10052-011-1844-6}.
The transverse momenta of reconstructed muons are included in the \MET\
calculation, with the energy deposited by these muons in the calorimeters taken
into account.  The energy from calorimeter cells not associated with any physics 
objects is scaled by a soft-term vertex fraction and also included in the
\MET\ calculation. This fraction is the ratio of the summed scalar \pt\ of
tracks from the primary vertex not matched with objects to the summed scalar \pt\ of
all tracks in the event also not matched to objects. This method
allows to achieve a more accurate 
reconstruction of the \MET\ in high pile-up
conditions~\cite{ATLAS-CONF-2013-082}.

\section{Data and simulated samples}
\label{sec:samples}

After data quality requirements, the integrated luminosities of the samples
used are 4.5~\fbs\ at  \mbox{$\sqrt{s}=7\TeV$} and 20.3~\fbs\ at
\mbox{$\sqrt{s}=8\TeV$}.

Samples of signal and background events were simulated using various Monte
Carlo (MC) generators, as summarised in table~\ref{tab:MCGenerator}. The
generators used for the simulation of the hard-scattering process and the model
used for the simulation of the parton shower, of the hadronisation and of the
underlying-event activity are listed. In addition, the cross section values to
which the simulation is normalised and the perturbative order in QCD of the
respective calculations are provided.

The signal contributions considered include the three main processes for Higgs
boson production at the LHC: ggF, VBF, and associated $VH$
production processes. The contributions from the associated $t \bar{t} H$
production process are found to be small and are neglected.  The ggF and
 VBF production processes are simulated with
\textsc{Powheg}~\cite{Nason:2004rx,Frixione:2007vw,Alioli:2010xd,Bagnaschi:2011tu}
interfaced to \textsc{Pythia8}~\cite{Sjostrand:2007gs}.  In the
\textsc{Powheg} event generator, the \textsc{CT10}~\cite{PDF-CT10}
parameterisation of the parton distribution functions (PDFs) is used.  The overall
normalisation of the  ggF process is taken from a calculation at
next-to-next-to-leading-order (NNLO)~\cite{Djouadi:1991tka, Dawson:1990zj,
Spira:1995rr, Harlander:2002wh, Anastasiou:2002yz, Ravindran:2003um} in QCD,
including soft-gluon resummation up to next-to-next-to-leading
logarithm terms (NNLL)~\cite{Catani2003}.  Next-to-leading order (NLO) electroweak
(EW) corrections are also included~\cite{Aglietti:2004nj,Actis:2008ug}.  Production by
VBF  is normalised to a cross section calculated with full NLO
QCD and EW corrections~\cite{Ciccolini:2007jr, Ciccolini:2007ec, Arnold:2008rz}
with an approximate NNLO QCD correction applied~\cite{Bolzoni:2010xr}. The
associated $VH$
production process is simulated with \textsc{Pythia8}. The
\textsc{Cteq6L1}~\cite{Pumplin:2002vw} parameterisation of PDFs is used for the
\textsc{Pythia8} event generator.  The predictions for $VH$ production are
normalised to cross sections calculated at NNLO in QCD~\cite{Brein:2003wg},
with NLO EW radiative corrections~\cite{Ciccolini:2003jy} applied.

Additional corrections to the shape of the generated \pt\ distribution of Higgs
bosons produced via ggF are applied to match the distribution from a
calculation at NNLO including the NNLL corrections provided by the
\textsc{HRes2.1}~\cite{hres_2} program. In this calculation, the effects of
finite masses of the top and bottom quarks~\cite{hres_1,hres_2} are included
and dynamical renormalisation and factorisation scales,
$\mu_\mathrm{R},\mu_\mathrm{F} = \sqrt{m_{H}^2 + \pt^2}$, are used.  A
reweighting is performed separately for events with no more than one
jet at particle level and for events with two or more jets. In the latter case,
the Higgs boson \pt\ spectrum is reweighted to match the \textsc{MinLo HJJ}
predictions~\cite{minlo_hjj}.
The reweighting is derived such that the inclusive Higgs boson \pt\ spectrum
and the \pt\ spectrum of events with at least two jets match the
\textsc{HRes2.1} and \textsc{MinLo HJJ} predictions respectively, and that the
jet multiplicities are in agreement with (N)NLO calculations from
\textsc{JetVHeto}~\cite{JetVHeto1,JetVHeto2,JetVHeto3}.

The NLO EW corrections for VBF production depend on the $\pT$ of the
Higgs boson, varying from a few percent at low $\pT$ to $\sim 20\%$ at
$\pT$~=~300~\GeV~\cite{Dittmaier:12013084}.  The $\pT$ spectrum of the
VBF-produced Higgs boson is therefore reweighted, based on the difference between the
\textsc{Powheg}+\textsc{Pythia} calculation and the
\textsc{Hawk}~\cite{Ciccolini:2007jr,Ciccolini:2007ec} calculation which
includes these corrections.

The main and largely irreducible $Z/\gamma^{*} \to \tau \tau$ background is
modelled using $Z/\gamma^{*}\to\mu \mu$ events from data,\footnote{These
processes are hereafter for simplicity denoted by $Z \to \tau \tau$ and
$Z\to\mu \mu$ respectively, even though the whole continuum above and below
the $Z$ peak is considered. } where the muon tracks and associated energy
depositions in the calorimeters are replaced by the corresponding simulated
signatures of the final-state particles of the  tau  decay.  In this approach,
essential features such as the modelling of the kinematics of the produced
boson, the modelling of the hadronic activity of the event (jets and underlying
event) as well as contributions from pile-up are taken from data. Thereby the
dependence on the simulation is minimised and only the $ \tau$ decays and the
detector response to the  tau-lepton decay products are based on simulation.
By requiring two isolated, high-energy muons with opposite charge and a dimuon
invariant mass $m_{\mu \mu} >$ 40~\GeV, $Z  \to \mu \mu$ events can be selected
from the data with high efficiency and purity.  To replace the muons
in the selected events, all tracks associated with the muons are removed and
calorimeter cell energies associated with the muons are corrected by subtracting
the corresponding energy depositions in a single simulated $Z \to \mu \mu $
event with the same kinematics. Finally, both the track information and the
calorimeter cell energies from a simulated $Z \to \tau \tau$ decay are added to
the data event.
The decays of the tau leptons are simulated by \textsc{Tauola}~\cite{Was:2000st}.
  The  tau lepton kinematics 
are matched to the kinematics of the muons they are replacing,  including
 polarisation and spin correlations~\cite{Czyczula:2012ny}, and the mass difference between the muons
 and the tau leptons is accounted for.
 This hybrid sample is referred to as
 embedded data in the following.

Other background processes are simulated using different generators, each
interfaced to \textsc{Pythia}~\cite{Pythia,Sjostrand:2007gs} or
\textsc{Herwig}~\cite{HERWIG1, HERWIG2} to provide the parton shower, hadronisation and
the modelling of the underlying event, as indicated in
table~\ref{tab:MCGenerator}.  For the \textsc{Herwig} samples, the decays of
tau leptons are also simulated using \textsc{Tauola}~\cite{Was:2000st}.
Photon radiation from charged
leptons for all samples is provided by \textsc{Photos}~\cite{Davidson:2010ew}.  The samples for $W/Z$+jets production are generated
with \textsc{Alpgen}~\cite{alpgen}, employing the \textsc{MLM} matching
scheme~\cite{MLM} between the hard process (calculated with LO matrix elements
for up to five partons) and the parton shower.  For $WW$ production, the
loop-induced $gg\to{}WW$ process is also generated, using the
\textsc{gg2WW}~\cite{Binoth:2006mf} program.  In the
\textsc{AcerMC}~\cite{Kersevan:2004yg}, \textsc{Alpgen}, and \textsc{Herwig}
event generators, the \textsc{Cteq6L1} parameterisation of the PDFs is used,
while the \textsc{CT10} parameterisation is used for the generation of events
with \textsc{gg2WW}.  The normalisation of these background contributions is
either estimated from control regions using data, as described in
section~\ref{sec:background}, or the cross sections quoted in
table~\ref{tab:MCGenerator} are used.

For all samples, a full simulation of the ATLAS detector
response~\cite{Aad:2010ah} using the \textsc{Geant4}
program~\cite{Agostinelli:2002hh} was performed. In addition, events from
minimum-bias interactions were simulated using the
\textsc{AU2}~\cite{atlasmctunes} parameter tuning of \textsc{Pythia8}. 
The \textsc{AU2} tune includes the set of optimized parameters for 
the parton shower, hadronisation, and multiple parton interactions.
They are overlaid
on the simulated signal and background events according to the luminosity
profile of the recorded data. The contributions from these pile-up interactions
are simulated both within the same bunch crossing as the hard-scattering
process and in neighbouring bunch crossings. Finally, the resulting simulated
events are processed through the same reconstruction programs as the data.

\begin{table}
  \begin{center}
  \small
    \begin{tabular}{|l|l|lrl|}
      \hline
      \multirow{2}{*}{Signal  ($m_H=125$~\GeV)}   & \multirow{2}{*}{MC generator}  & \multicolumn{3}{l|}{$\sigma \times {\cal B}$ [pb]} \\
                                 &                                  &  \multicolumn{3}{l|}{$\sqrt{s}=8~\TeV$}  \\
      \hline \hline
      ggF, $H\to\tau\tau$   & \textsc{Powheg}~\cite{Nason:2004rx,Frixione:2007vw,Alioli:2010xd,Bagnaschi:2011tu}  & 1.22   & NNLO+NNLL & \cite{Djouadi:1991tka, Dawson:1990zj, Spira:1995rr, Harlander:2002wh, Anastasiou:2002yz, Ravindran:2003um,YellowReportIII}\\
         & \quad + \textsc{Pythia8}~\cite{Sjostrand:2007gs} &   &  & \\
       VBF, $H\to\tau\tau$     &  \textsc{Powheg + Pythia8}                                             & 0.100  & (N)NLO    & \cite{Ciccolini:2007jr, Ciccolini:2007ec, Arnold:2008rz,YellowReportIII}\\
      $WH$, $H\to\tau\tau$    &  \textsc{Pythia8}                                                      & 0.0445 & NNLO & \cite{Brein:2003wg,YellowReportIII}\\
      $ZH$, $H\to\tau\tau$    &  \textsc{Pythia8}                                                      & 0.0262 & NNLO & \cite{Brein:2003wg,YellowReportIII}\\
      \hline \hline
      \multirow{2}{*}{Background} & \multirow{2}{*}{MC generator} & \multicolumn{3}{l|}{$\sigma \times {\cal B}$ [pb]}\\
                 &              &  \multicolumn{3}{l|}{$\sqrt{s}=8~\TeV$}            \\
      \hline \hline
      $W (\to \ell \nu$), ($\ell = e, \mu, \tau)$                  & \textsc{Alpgen}~\cite{alpgen}+\textsc{Pythia8}      & 36800  & NNLO & \cite{PhysRevLett.103.082001,PhysRevLett.98.222002}\\
      $Z/\gamma^{*}(\to \ell\ell)$,& \multirow{2}{*}{\textsc{Alpgen+Pythia8}}   & \multirow{2}{*}{3910} & \multirow{2}{*}{NNLO} & \multirow{2}{*}{\cite{PhysRevLett.103.082001,PhysRevLett.98.222002}}\\
      60~\GeV$<m_{\ell\ell}<2$~\TeV   & & & &\\
      $Z/\gamma^{*}(\to \ell\ell)$,   & \multirow{2}{*}{\textsc{Alpgen+Herwig}~\cite{herwig}}       & \multirow{2}{*}{13000}  & \multirow{2}{*}{NNLO} & \multirow{2}{*}{\cite{PhysRevLett.103.082001,PhysRevLett.98.222002}}\\
      10~\GeV$<m_{\ell\ell}<60$~\GeV  & & & &\\
      VBF $Z/\gamma^{*}$($\to\ell\ell$)                           & \textsc{Sherpa}~\cite{sherpa}              &  1.1  & LO & \cite{sherpa}\\
      $t\bar{t}$                         & \textsc{Powheg + Pythia8}                                          & 253$^{\dagger}$  & NNLO+NNLL & \cite{ttbar:xsec1,ttbar:xsec2,ttbar:xsec3,ttbar:xsec4,ttbar:xsec5,ttbar:xsec6}\\
      Single top : $Wt$                  & \textsc{Powheg + Pythia8}                                          & 22$^{\dagger}$ & NNLO & \cite{singletop:Wt}\\
      Single top : $s$-channel           & \textsc{Powheg + Pythia8}                                   & 5.6$^{\dagger}$ & NNLO & \cite{singletop:s-ch}\\
      Single top : $t$-channel           & AcerMC~\cite{Kersevan:2004yg}+\textsc{Pythia6}~\cite{Pythia}             & 87.8$^{\dagger}$ & NNLO & \cite{singletop:t-ch}\\
      $q\bar{q} \to WW$                  & \textsc{Alpgen+Herwig}                                           & 54$^{\dagger}$ & NLO & \cite{MCFMVV}\\
      $gg \to WW $                       & \textsc{gg2WW}~\cite{Binoth:2006mf}+\textsc{Herwig}                       & 1.4$^{\dagger}$ & NLO & \cite{Binoth:2006mf}\\
      $WZ,ZZ$                            & \textsc{Herwig}                                                  & 30$^{\dagger}$ & NLO & \cite{MCFMVV}\\
      $H \to WW$                         & same as for $H\to\tau\tau$ signal                                & 4.7$^{\dagger}$  &  & \\
      \hline
    \end{tabular}
\caption{
Monte Carlo generators used to model the signal and the background processes at
$\sqrt{s}=8\TeV$. The cross sections times branching fractions ($\sigma \times
{\cal B}$) used for the normalisation of some processes (many of these are
subsequently normalised to data) are included in the last column together with
the perturbative order of the QCD calculation.  For the signal processes the
$H\to\tau\tau$ SM branching ratio is included, and for the $W$ and $Z/\gamma^{*}$
background processes the branching ratios for leptonic decays ($\ell = e, \mu,
\tau$) of the bosons are included.  For all other background processes,
inclusive cross sections are quoted (marked with a $\dagger$).}
\label{tab:MCGenerator}
  \end{center}
\end{table}

\section{Event selection and categorisation}
\label{sec:selection}

\subsection{Event selection}

Single lepton, dilepton and di-\thad\ triggers were used to select the
events for the analysis.
A summary of the triggers used by each channel at the two centre-of-mass
energies is reported in table~\ref{tab:triggers}. Due to the increasing
luminosity and the different pile-up conditions, the online \pT\ thresholds
increased during data-taking in 2011 and again for 2012, and more stringent
identification requirements were applied for the data-taking in 2012.
The \pT\ requirements on the objects in the analysis are usually 2~\GeV\ higher
than the trigger requirements, to ensure that the trigger is fully efficient.

\begin{table}
\begin{center}
\footnotesize
\begin{tabular}{|l|c|cccccc|}
\hline
\multicolumn{8}{|c|}{$\sqrt{s}=7~\TeV$}\\
\hline \hline
\multirow{4}{*}{Trigger} & Trigger & \multicolumn{6}{c|}{\multirow{2}{*}{Analysis level thresholds [\GeV]}} \\
& level & & & & & &\\ 
& thresholds, & \multicolumn{2}{c}{\multirow{2}{*}{\tll}} & \multicolumn{2}{c}{\multirow{2}{*}{\tlhad}} & \multicolumn{2}{c|}{\multirow{2}{*}{\thadhad}} \\
& \pT~[\GeV] & & & & & &\\ 
\hline\hline
\multirow{2}{*}{Single electron}  & \multirow{2}{*}{20$-$22}
& \multirow{2}{*}{$e\mu$:} & $\pt^{e} > 22-24$
& \multirow{2}{*}{$e\tau$:} & $\pt^{e} > 25$ & \multicolumn{2}{c|}{\multirow{2}{*}{--}} \\
& & & $\pt^{\mu} > 10$ & &  $\pt^{\tau} > 20$ & & \\ \hline
\multirow{4}{*}{Single muon}  & \multirow{4}{*}{18}
& \multirow{2}{*}{$\mu\mu$:} & $\pt^{\mu_1} > 20$
& \multirow{4}{*}{$\mu\tau$:} & & \multicolumn{2}{c|}{\multirow{4}{*}{--}} \\
& & & $\pt^{\mu_2} > 10$ & &  $\pt^{\mu} > 22$  & & \\ 
& & \multirow{2}{*}{$e\mu$:}   & $\pt^{\mu} > 20$ & &$\pt^{\tau} > 20$ & & \\ 
& & & $\pt^{e} > 15$ & & & & \\ \hline
\multirow{2}{*}{Di-electron}  & \multirow{2}{*}{12/12}
& \multirow{2}{*}{$ee$:} & $\pt^{e_1} > 15$
& \multicolumn{2}{c}{\multirow{2}{*}{--}} & \multicolumn{2}{c|}{\multirow{2}{*}{--}} \\
& & & $\pt^{e_2} > 15$ & & & & \\ \hline
\multirow{2}{*}{Di-\thad}  & \multirow{2}{*}{29/20}
& \multicolumn{2}{c}{\multirow{2}{*}{--}} & \multicolumn{2}{c}{\multirow{2}{*}{--}} & \multirow{2}{*}{$\tau\tau$:} & $\pt^{\tau_1} > 35$\\
& & & & & & & $\pt^{\tau_2} > 25$ \\ \hline

\hline\hline
\multicolumn{8}{|c|}{$\sqrt{s}=8~\TeV$}\\
\hline \hline
\multirow{4}{*}{Trigger} & Trigger & \multicolumn{6}{c|}{\multirow{2}{*}{Analysis level thresholds [\GeV]}} \\
& level & & & & & &\\ 
& thresholds, & \multicolumn{2}{c}{\multirow{2}{*}{\tll}} & \multicolumn{2}{c}{\multirow{2}{*}{\tlhad}} & \multicolumn{2}{c|}{\multirow{2}{*}{\thadhad}} \\
& \pT~[\GeV] & & & & & &\\ 
\hline\hline
\multirow{4}{*}{Single electron}  & \multirow{4}{*}{24}
& \multirow{2}{*}{$e\mu$:} & $\pt^{e} > 26$
& \multirow{4}{*}{$e\tau$:} & & \multicolumn{2}{c|}{\multirow{4}{*}{--}} \\
& & & $\pt^{\mu} > 10$ & &  $\pt^{e} > 26$ & & \\
& & \multirow{2}{*}{$ee$:} & $\pt^{e_1} > 26$ & & $\pt^{\tau} > 20$ & & \\
& & & $\pt^{e_2} > 15$ & & & & \\ \hline
\multirow{2}{*}{Single muon}  & \multirow{2}{*}{24}
& \multicolumn{2}{c}{\multirow{2}{*}{--}} & \multirow{2}{*}{$\mu\tau$:} & $\pt^{\mu} > 26$ & \multicolumn{2}{c|}{\multirow{2}{*}{--}} \\
& & & & &  $\pt^{\tau} > 20$ & & \\ \hline
\multirow{2}{*}{Di-electron}  & \multirow{2}{*}{12/12}
& \multirow{2}{*}{$ee$:} & $\pt^{e_1} > 15$
& \multicolumn{2}{c}{\multirow{2}{*}{--}} & \multicolumn{2}{c|}{\multirow{2}{*}{--}} \\
& & & $\pt^{e_2} > 15$ & & & & \\ \hline
\multirow{2}{*}{Di-muon}  & \multirow{2}{*}{18/8}
& \multirow{2}{*}{$\mu\mu$:} & $\pt^{\mu_1} > 20$
& \multicolumn{2}{c}{\multirow{2}{*}{--}} & \multicolumn{2}{c|}{\multirow{2}{*}{--}} \\
& & & $\pt^{\mu_2} > 10$ & & & & \\ \hline
\multirow{2}{*}{Electron+muon}  & \multirow{2}{*}{12/8}
& \multirow{2}{*}{$e\mu$:} & $\pt^{e} > 15$
& \multicolumn{2}{c}{\multirow{2}{*}{--}} & \multicolumn{2}{c|}{\multirow{2}{*}{--}} \\
& & & $\pt^{\mu} > 10$ & & & & \\ \hline
\multirow{2}{*}{Di-\thad}  & \multirow{2}{*}{29/20}
& \multicolumn{2}{c}{\multirow{2}{*}{--}} & \multicolumn{2}{c}{\multirow{2}{*}{--}} & \multirow{2}{*}{$\tau\tau$:} & $\pt^{\tau_1} > 35$\\
& & & & & & & $\pt^{\tau_2} > 25$ \\ \hline
\end{tabular}
\caption{Summary of the triggers used to select events for the different analysis channels at the two centre-of-mass energies. The transverse momentum thresholds applied at trigger level and in the analysis are listed. When more than one trigger is used, a logical OR is taken and the trigger efficiencies are calculated accordingly.}
\label{tab:triggers}
\end{center}
\end{table}

In addition to applying criteria to ensure that the detector was functioning
properly, requirements to increase the purity and quality of the data sample
are applied by rejecting non-collision events such as cosmic rays and beam-halo
events. At least one reconstructed vertex is required with at least
four associated tracks with $\pt>400\MeV$ and a position consistent with the beam spot.
It has been verified that, after object selection cuts, contributions from other
 primary vertices are negligible.

With respect to the object identification requirements described in
section~\ref{sec:atlas}, tighter criteria are applied to address the different
background contributions and compositions in the different analysis channels.
Higher \pt\ thresholds are applied to electrons, muons, and \thad\ candidates
according to the trigger conditions satisfied by the event, as listed in
table~\ref{tab:triggers}.  For the channels involving leptonic  tau decays,
$\tll$ and $\tlhad$, additional isolation criteria for electrons and muons,
based on tracking and calorimeter information, are used to suppress the
background from misidentified jets or from semileptonic decays of charm and
bottom hadrons. The calorimeter isolation variable $I (\ET,\Delta R)$ is
defined as the sum of the total transverse energy in the calorimeter in a 
cone of size $\Delta{}R$ around the electron cluster or the muon track, divided
by the $\ET$ of the electron cluster or the $\pT$ of the muon respectively.
The track-based isolation $I (\pt,\Delta R)$ is defined as the sum of the
transverse momenta of tracks within a cone of $\Delta{}R$ around the electron
or muon track, divided by the $\ET$ of the electron cluster or the
muon $\pT$ 
respectively. The isolation requirements applied are slightly different for the
two centre-of-mass energies and are listed in table~\ref{tab:isol}.

\begin{table}
\begin{center}
\begin{tabular}{|lc|c|c|}
\hline
& & \tll & \tlhad \\
\hline\hline
\multirow{4}{*}{Electrons}
& \multirow{2}{*}{7~\TeV}
    & $I(p_\mathrm{T},0.4)<0.06$ & $I(p_\mathrm{T},0.4)<0.06$\\
& & $I(E_\mathrm{T},0.2)<0.08$ & $I(E_\mathrm{T},0.2)<0.06$ \\ \cline{3-4}
& \multirow{2}{*}{8~\TeV}
    & $I(p_\mathrm{T},0.4)<0.17$ & $I(p_\mathrm{T},0.4)<0.06$ \\
& & $I(E_\mathrm{T},0.2)<0.09$ & $I(E_\mathrm{T},0.2)<0.06$ \\
\hline\hline
\multirow{4}{*}{Muons}
& \multirow{2}{*}{7~\TeV}
    & $I(p_\mathrm{T},0.4)<0.06$ & $I(p_\mathrm{T},0.4)<0.06$ \\
& & $I(E_\mathrm{T},0.2)<0.04$ & $I(E_\mathrm{T},0.2)<0.06$ \\ \cline{3-4}
& \multirow{2}{*}{8~\TeV}
    & $I(p_\mathrm{T},0.4)<0.18$ & $I(p_\mathrm{T},0.4)<0.06$ \\
& & $I(E_\mathrm{T},0.2)<0.09$ & $I(E_\mathrm{T},0.2)<0.06$ \\
\hline
\end{tabular}
\caption{Summary of isolation requirements applied for the selection of isolated electrons and muons at the two centre-of-mass
energies. The isolation variables are defined in the text.}
\label{tab:isol}
\end{center}
\end{table}

In the $\thadhad$ channel, isolated taus are selected by requiring that there are no tracks with
$\pT > 0.5\GeV$ in an isolation region of $0.2 < \Delta{}R < 0.6$ around the
tau direction. This requirement leads to a 12\% (4\%) efficiency loss
for 
hadronic taus, while 30\% (10\%) of contamination from jets is rejected in
$8~(7)\TeV$ data.

After the basic lepton selection, further channel-dependent cuts are applied, as
detailed in the following.  The full event selection is summarised in
table~\ref{tab:selection}.

\paragraph{The $\boldsymbol{\tll}$ channel:}
exactly two isolated leptons with opposite-sign (OS) electric charges, passing
the $\pt$ threshold listed in table~\ref{tab:triggers}, are required. Events
containing a \thad\ candidate are vetoed. For the $\thad$ candidates considered,
the criteria used to reject electrons misidentified as \thad\ candidates are
tightened to a working-point of 85\% signal
efficiency~\cite{ATLAS_Tauperf}.

In addition to the irreducible $Z \to \tau \tau$ background, sizeable
background contributions from $Z \to \ell \ell$ and from  \ttbar\ production
are expected in this channel.  Background contributions from $Z$ decays, but
also from low mass resonances (charmonium and bottomonium), are rejected by requirements
on the invariant mass $m^\mathrm{vis}_{\tau\tau}$ of the visible  tau  decay
products, on the angle $\Delta\phi_{\ell \ell}$ between the two leptons in the
transverse plane and on  $\MET$.  To
reject the large $Z \to \ell \ell$ contribution in events with same-flavour
(SF) leptons ($ee, \mu \mu$), more stringent cuts on the visible mass and on
\MET\ are applied for these events than for events with different-flavour (DF)
leptons ($e \mu$).  For SF final states, an additional variable named {\em high-\pt\ objects \MET}
 ($E_\mathrm{T}^\mathrm{miss,HPTO}$) is also used to
reject background from $Z/\gamma^{*}$ production. It is calculated from the
high-\pt\ objects in the event, i.e.\ from the two leptons and from jets with
$\pt>25\GeV$. Due to the presence of real neutrinos, the two \MET\
variables are strongly correlated for signal events but only loosely
correlated for background from $Z \to ee$ and $Z \to \mu \mu$ decays.

To further suppress background contributions from misidentified
leptons\footnote{Misidentified leptons and \thad\ candidates are also referred to
as ``fake'' in this paper.} a minimum value of the
scalar sum of the transverse momenta of the two leptons is required.
Contributions from $\ttbar$ events are further reduced by rejecting events with
a $b$-jet with $\pt > 25$~\GeV.

Within the collinear approximation \cite{collinear}, i.e.\ assuming that the
tau  directions are given by the directions of the visible  tau decay products
and that the momenta of the neutrinos constitute the missing transverse
momentum, the  tau  momenta can be reconstructed.  For  tau  decays, the
fractions of the  tau  momenta carried by the visible decay
products,\footnote{The variable $p_{\mathrm{vis}}$ is defined as the total momentum of the
visible decay products of the  tau  lepton while $p_{\mathrm{mis}}$ is defined as
the momentum of the neutrino system reconstructed using the collinear approximation.}
\mbox{$ x_{\tau, i} = p_{\mathrm{vis}, i} / (p_{\mathrm{vis}, i} +
p_{\mathrm{mis}, i})$}, with $i=1,2$, are expected to lie in the interval $0 < x_{\tau, i}
< 1$, and hence corresponding requirements are applied to further reject
non-tau background contributions.

Finally, to avoid overlap between this analysis and the search for  $\hwws \to
\ell \nu \ell \nu$ decays, the $\tau\tau$ mass in the collinear approximation is
required to satisfy the condition $m_{\tau\tau}^{\mathrm{coll}}  > m_Z - 25\GeV$.

\paragraph{The $\boldsymbol{\tlhad}$ channel:}
exactly one isolated lepton and one \thad\ candidate with OS charges, passing
the \pt\ thresholds listed in table~\ref{tab:triggers}, are required. The
criteria used to reject electrons misidentified as \thad\ are also tightened in
this channel to a working-point of 85\% signal
efficiency~\cite{ATLAS_Tauperf}.

The production of $\Wjets$ and of top quarks constitute the dominant reducible
background in this channel. To substantially reduce the $\Wjets$ contribution,
a cut on the transverse mass\footnote{
$m_{\mathrm{T}}=\sqrt{2p_{\mathrm{T}}^{\ell} \
E_{\mathrm{T}}^{\mathrm{miss}} \cdot (1-\cos{\Delta{\phi}})}$, where
$\Delta{\phi}$ is the azimuthal separation between the directions of the
lepton and the missing transverse momentum.} constructed from the lepton
and the missing transverse momentum is applied and events with $m_{\mathrm{T}}>70~\GeV$ are rejected.
Contributions from $\ttbar$ events are reduced by rejecting events with a
$b$-jet with $\pt > 30$~\GeV.

\paragraph{The $\boldsymbol{\thadhad}$ channel:}
one isolated medium \thad\ candidate  and one isolated tight \thad\ candidate with OS charges are required.
Events with electron or muon candidates are rejected.  For all data, the missing transverse momentum 
must satisfy \MET\ $>$ 20~\GeV\ and its direction must either be between the two
visible \thad\ candidates in $\phi$ or within $\Delta\phi < \pi/4$ of the
nearest \thad\ candidate.  To further reduce the background from
multijet production, additional cuts on the $ \Delta R$ and pseudorapidity
separation $ \Delta \eta$ between the two \thad\ candidates are applied.

With these selections, there is no overlap between the individual channels.

\begin{table}
\begin{center}
\small
\begin{tabular}{|c|l|}
\hline
Channel & Preselection cuts \\
\hline\hline
\multirow{10}{*}{\tll}
    & Exactly two isolated opposite-sign leptons \\
& Events with \thad~candidates are rejected \\
& $30~\GeV < m^\mathrm{vis}_{\tau\tau} < 100$ $(75)\GeV$ for DF (SF) events \\
& $\Delta\phi_{\ell \ell} < 2.5$ \\
& $\MET > 20 \ (40)\GeV$ for DF (SF) events\\
& $E_\mathrm{T}^\mathrm{miss,HPTO}$ > 40~\GeV\ for SF events \\
& $\pt^{\ell_1}+\pt^{\ell_2} > 35\GeV$ \\
& Events with a $b$-tagged jet with \pt\ $>$ 25~\GeV\ are rejected \\
& $0.1<x_{\tau_1},x_{\tau_2}<1$ \\
& $m_{\tau\tau}^\mathrm{coll} > m_Z - 25\GeV$ \\
\hline
\multirow{3}{*}{\tlhad}
& Exactly one isolated lepton and one medium \thad~candidate with opposite charges \\
& $m_{\mathrm{T}}< 70\GeV$ \\
& Events with a $b$-tagged jet with \pt\ $>$ 30~\GeV\ are rejected \\
\hline
\multirow{8}{*}{\thadhad}
& One isolated medium and one isolated tight opposite-sign $\thad$-candidate \\
& Events with leptons are vetoed \\
& $\MET>20\GeV$ \\
& \MET\ points between the two visible taus in $\phi$, or min$[\Delta \phi (\tau, \MET)] < \pi /4$\\
& $0.8<\Delta{}R(\tau_{\rm had_1},\tau_{\rm had_2}) < 2.4$ \\
& $\Delta{}\eta(\tau_{\rm had_1},\tau_{\rm had_2}) < 1.5$ \\
\hline\hline
Channel &  VBF category selection cuts \\ \hline\hline
\multirow{2}{*}{\tll}
    & At least two jets with $\pt^{j_1}$ $>$ 40~\GeV\ and $\pt^{j_2}$ $>$ 30~\GeV \\
& $\Delta\eta(j_1,j_2) > 2.2 $\\
\hline
\multirow{3}{*}{\tlhad}
    & At least two jets  with $\pt^{j_1}$ $>$ 50~\GeV\ and $\pt^{j_2}$ $>$ 30~\GeV \\
& $\Delta\eta(j_1,j_2) > 3.0 $\\
 & $m^\mathrm{vis}_{\tau\tau} > 40\GeV$\\
 \hline
\multirow{3}{*}{\thadhad}
& At least two jets  with $\pt^{j_1}$ $>$ 50~\GeV\ and $\pt^{j_2}$ $>$ 30~\GeV \\
& $\pt^{j_2}$ $>$ 35~\GeV\ for jets with $|\eta| > 2.4$\\
& $\Delta\eta(j_1,j_2) > 2.0 $\\

\hline\hline
Channel & Boosted category selection cuts \\ \hline\hline
\tll & At least one jet with \pt $>$ 40~\GeV\ \\\hline
\multirow{2}{*}{All}
& Failing the VBF selection \\
& $\pt^H > 100\GeV$\\
\hline
\end{tabular}
\caption{
Summary of the event selection for the three analysis channels. The requirements used
in both the preselection and for the definition of the analysis categories are
given. The labels (1) and (2) refer to the leading (highest $\pT$) and
subleading final-state objects (leptons, \thad, jets).  The variables are
defined in the text.}
\label{tab:selection}
\end{center}
\end{table}

\subsection{Analysis categories}

To exploit signal-sensitive event topologies, two analysis categories
are defined in an exclusive way.
\begin{itemize}
\item The VBF category targets events with a Higgs boson produced via
    vector boson fusion and is characterised by the presence of two high-\pT\
    jets with a large pseudorapidity separation (see table
    \ref{tab:selection}). The $\Delta\eta(j_1,j_2)$ requirement is applied
    to the two highest-\pt\ jets in the event.
    In the \tlhad\ channel, there is an additional requirement that
    $m^\mathrm{vis}_{\tau\tau}>40\GeV$, to eliminate low-mass $Z/\gamma^{*}$
    events.
    Although this category is dominated by VBF events, it also includes
    smaller contributions from ggF and $VH$ production.

\item The boosted category targets events with a boosted Higgs boson
    produced via ggF. Higgs boson candidates are therefore required to have 
    large transverse momentum, $p_\mathrm{T}^H>100\GeV$. The $p_\mathrm{T}^H$
    is reconstructed using the vector sum of the missing transverse momentum and the transverse momentum
    of the visible tau decay products.  In the \tll\ channel, at least one jet
    with $\pT > 40\GeV$ is required. The jet requirement  selects a region of 
    the phase space where the $\MET$ of same-flavour events is well modelled by simulation. 
    In order to define an orthogonal
    category, events passing the VBF category selection are not considered.
    This category also includes small contributions from VBF and {\em VH}
    production.
\end{itemize}

While these categories are conceptually identical across the three channels,
differences in the dominant background contributions
require different selection criteria.
For both categories, the requirement on jets is inclusive and additional jets,
apart from those passing the category requirements, are allowed.

For the \thadhad\ channel, the so-called {\em rest category} is used as a
control region. In this category, events passing the preselection requirements
but not passing the VBF or boosted category selections are considered. This
category is used to constrain the $Z \to \tau \tau$ and multijet background
contributions. The signal contamination in this category is negligible.

\subsection{Higgs boson candidate mass reconstruction}
\label{sec:mmc}

The di-tau invariant mass (\MMC)  is reconstructed using the missing mass
calculator (MMC)~\cite{mmc}. This requires solving an underconstrained system
of equations for six to eight unknowns, depending on the number of neutrinos in
the $\tau\tau$ final state. These unknowns include the $x$-, $y$-, and
$z$-components of the momentum carried by the neutrinos for each of
the two tau  leptons in the event, and the invariant mass of the two neutrinos
from any leptonic  tau  decays. The calculation uses the constraints from the
measured $x$- and $y$-components of the missing transverse momentum, and the visible masses of both tau
candidates.  A scan is performed over the two components of the missing transverse momentum vector
and the yet undetermined variables. Each scan point is weighted by its
probability according to the \MET\ resolution and the tau decay topologies. The
estimator for the $\tau\tau$ mass is defined as the most probable value of the
scan points.

The MMC algorithm provides a solution for $\sim$99\%  of the $H \to \tau \tau$
and $Z \to \tau \tau$ events. This is a distinct advantage compared to the mass
calculation using the collinear approximation where the failure rate is higher
due to the implicit collinearity assumptions.  The small loss rate of about 1\%
for signal events is due to large fluctuations of the $\MET$ measurement or
other scan variables.

\begin{figure}[htbp]
\centering
\hspace*{\fill}
\subfigure[]{\includegraphics[width=6.5cm]{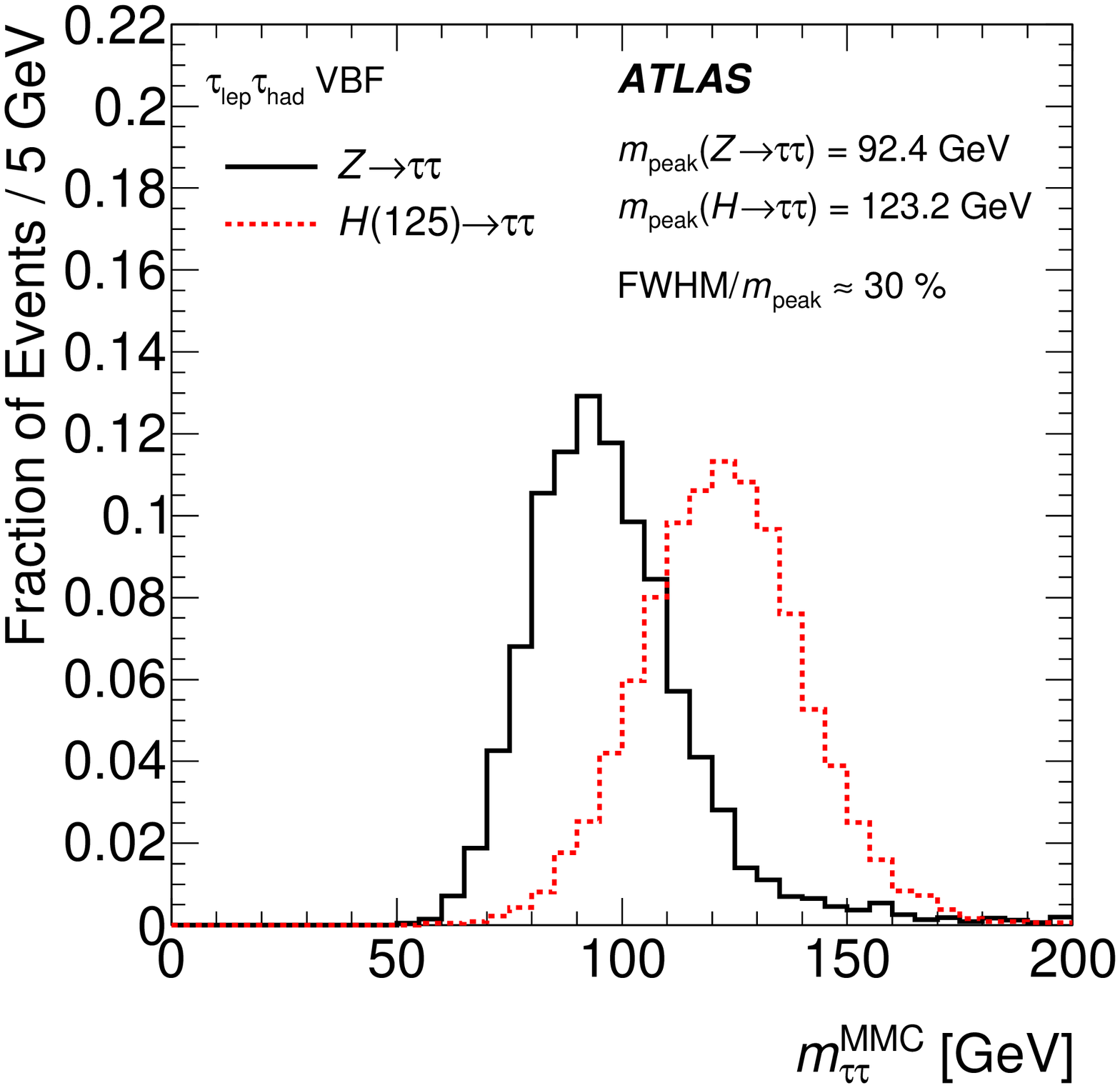}}\hfill
\subfigure[]{\includegraphics[width=6.5cm]{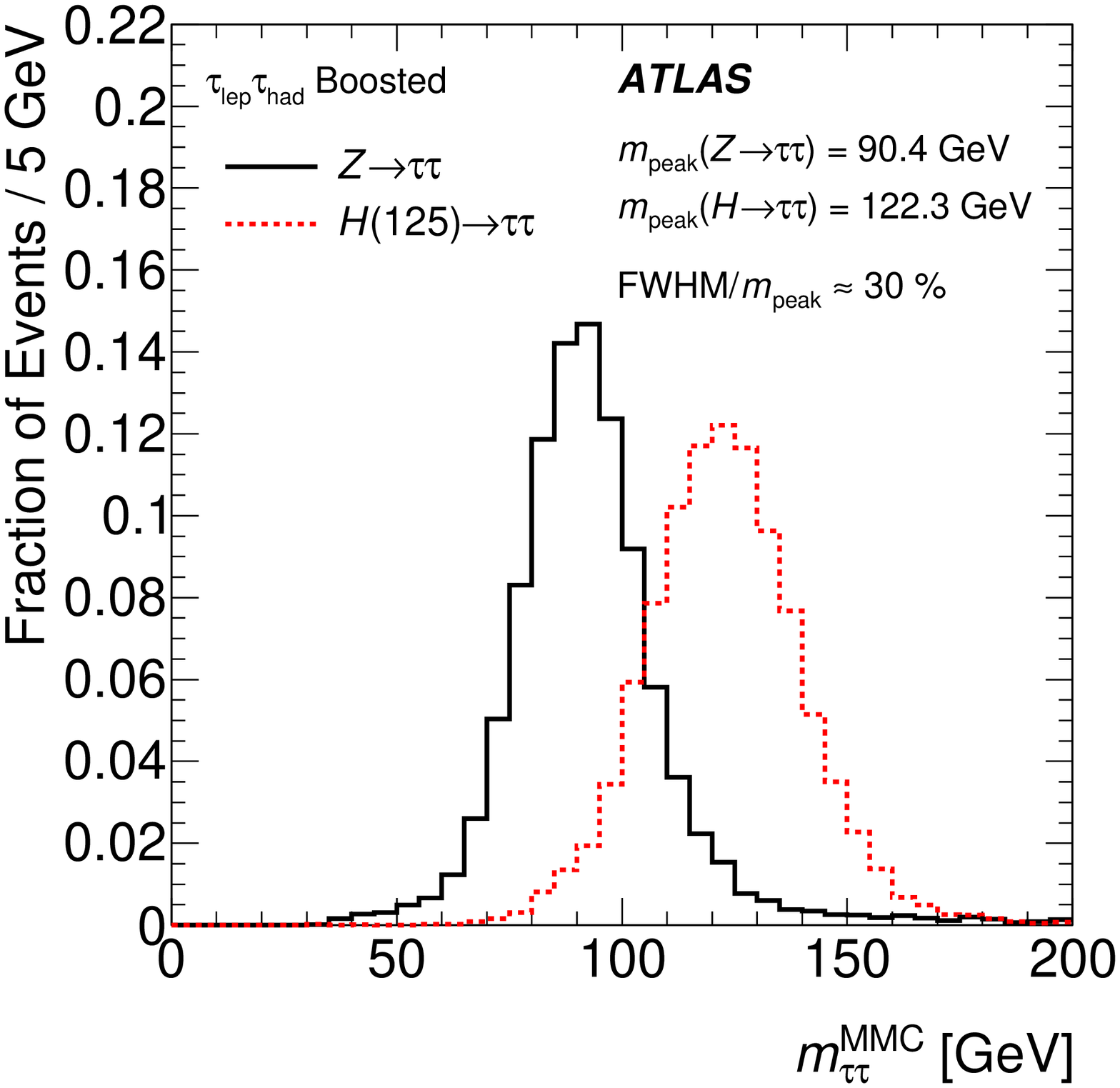}}
\hspace*{\fill}

\caption{
The reconstructed invariant $\tau\tau$ mass, \MMC\, for $H \to \tau \tau$
($m_H$~=~125~\GeV) and $Z \to \tau \tau$ events in MC simulation and
embedding 
respectively, for events passing (a) the VBF category selection and (b)
the boosted category selection in the $\tlhad$ channel. 
}
\label{fig:MMC-resolution}
\end{figure}

Figure~\ref{fig:MMC-resolution} shows reconstructed \MMC\ mass distributions
 for $H \to \tau \tau$ and $Z \to \tau \tau$ events in the  $\tlhad$  
VBF and  boosted categories.  The mass resolution, defined
as the ratio between the full width at half maximum (FWHM) and the peak value
of the mass distribution ($m_{\mathrm{peak}}$), is found to be $\approx{} 30\%$
for all categories and channels.

\section{Boosted decision trees}
\label{sec:bdts}

Boosted decision trees are used in each category to extract the Higgs boson
signal from the large number of background events.  Decision trees~\cite{bdt}
recursively partition the parameter space into multiple regions where signal or
background purities are enhanced. Boosting is a method which improves the
performance and stability of decision trees and involves the combination of
many trees into a single final discriminant~\cite{bdt:gradboost, bdt:adaboost}.
After boosting, the final score undergoes a transformation to map the scores on
the interval $-1$ to $+1$. The most signal-like events have scores near 1 while
the most background-like events have scores near $-1$.

Separate BDTs are trained for each analysis category and channel with signal
and background samples, described in section \ref{sec:background}, at
$\sqrt{s}=8\TeV$. They are then applied to the analysis of the data at both
centre-of-mass energies. The separate training naturally exploits differences
in event kinematics between different Higgs boson production modes. It also
allows different discriminating variables to be used to address the different
background compositions in each channel. 
A large set of potential variables was investigated, in each channel 
separately, and only those variables
which led to an improved discrimination performance of the BDT were kept.
For the training in the VBF
category, only a VBF Higgs production signal sample is used, while 
training in the boosted
category uses ggF, VBF, and $VH$ signal samples. The
Higgs boson mass is chosen to be $m_H=125\GeV$ for all signal samples.
The BDT input variables used at both centre-of-mass energies are listed in
table~\ref{tab:bdtvars}.  Most of these variables have straightforward
definitions, and the more complex ones are defined in the following.

\begin{itemize}

\item $\Delta{}R(\tau_1,\tau_2)$: the distance $\Delta{}R$ between
    the two leptons, between the lepton and $\thad$, or between the two $\thad$
    candidates, depending on the decay mode.

\item $p_\mathrm{T}^{\mathrm{Total}}$: magnitude of the vector sum of the transverse momenta of the visible tau decay products, the two leading jets, and
    \MET.

\item Sum $\pt$: scalar sum of the \pt\ of the visible components of the  tau
    decay products and of the jets.

\item $\MET\phi$ centrality: a variable that quantifies the relative
  angular position of the missing transverse momentum with respect to
  the visible tau decay products in the
    transverse plane. The transverse plane is transformed such that the
    direction of the  tau  decay products are orthogonal, and that the smaller
    $\phi$ angle between the  tau  decay products defines the positive quadrant
    of the transformed plane. The $\MET\phi$ centrality is defined as the sum of
    the $x$- and $y$-components of the \MET\ unit vector in this transformed
    plane.

\item Sphericity: a variable that describes the isotropy of the energy flow in
    the event~\cite{Hanson:1975fe}.  It is based on the quadratic momentum
    tensor
    \begin{equation}
    S^{\alpha\beta}= \frac{\sum_ip^\alpha_ip^\beta_i}{\sum_i|\vec{p_i}^2|}.
    \end{equation}
    In this equation, $\alpha$ and $\beta$ are the indices of the tensor. The
    summation is performed over the momenta of the selected leptons and jets in
    the event. The sphericity of the event ($S$) is then defined in terms of
    the two smallest eigenvalues of this tensor, $\lambda_2$ and $\lambda_3$,
    \begin{equation}
    S= \frac{3}{2}(\lambda_2+\lambda_3).
    \end{equation}

\item min$(\Delta \eta_{\ell_1\ell_2,\mathrm{jets}})$: the minimum $\Delta \eta$ between the dilepton system and either of the two jets.

\item Object $\eta$ centrality: a variable that quantifies the $\eta$ position
    of an object (an isolated lepton, a \thad\ candidate or a jet) with respect
    to the two leading jets in the event. It is defined as
    \begin{equation}
    C_{\eta_1, \eta_2}(\eta) = \exp \left[ \frac{-4}{(\eta_1 - \eta_2)^2}\left(\eta - \frac{\eta_1 + \eta_2}{2}\right)^2\right] ,
    \end{equation}
    where $\eta$, $\eta_1$ and $\eta_2$ are the pseudorapidities of the object and
    the two leading jets respectively.  This variable has a value of $1$ when the
    object is halfway in $\eta$ between the two jets, $1/$e when the object is
    aligned with one of the jets, and $<1/$e when the object is not between the jets in $\eta$.
    In the  $\tll$ channel the $\eta$ centrality of a third jet in the event,
    $C_{\eta_1, \eta_2}(\eta_{ j_3})$, and the product of the $\eta$ centralities of the
    two leptons are used as BDT input variables, while in the $\tlhad$ channel the
    $\eta$ centrality of the lepton, $C_{\eta_1, \eta_2}(\eta_{\ell})$, is used, and in
    the $\thadhad$ channel the $\eta$ centrality of each $\tau$, $C_{\eta_1,
    \eta_2}(\eta_{\tau_1})$ and $C_{\eta_1, \eta_2}(\eta_{\tau_2})$, is used.
    Events with only two jets are assigned a dummy value of $-0.5$ for $C_{\eta_1,
    \eta_2}(\eta_{ j_3})$.
\end{itemize}

\begin{figure}[htbp]
  \centering
  \hspace*{\fill}
  \subfigure[]{ \includegraphics[width=6.0cm]{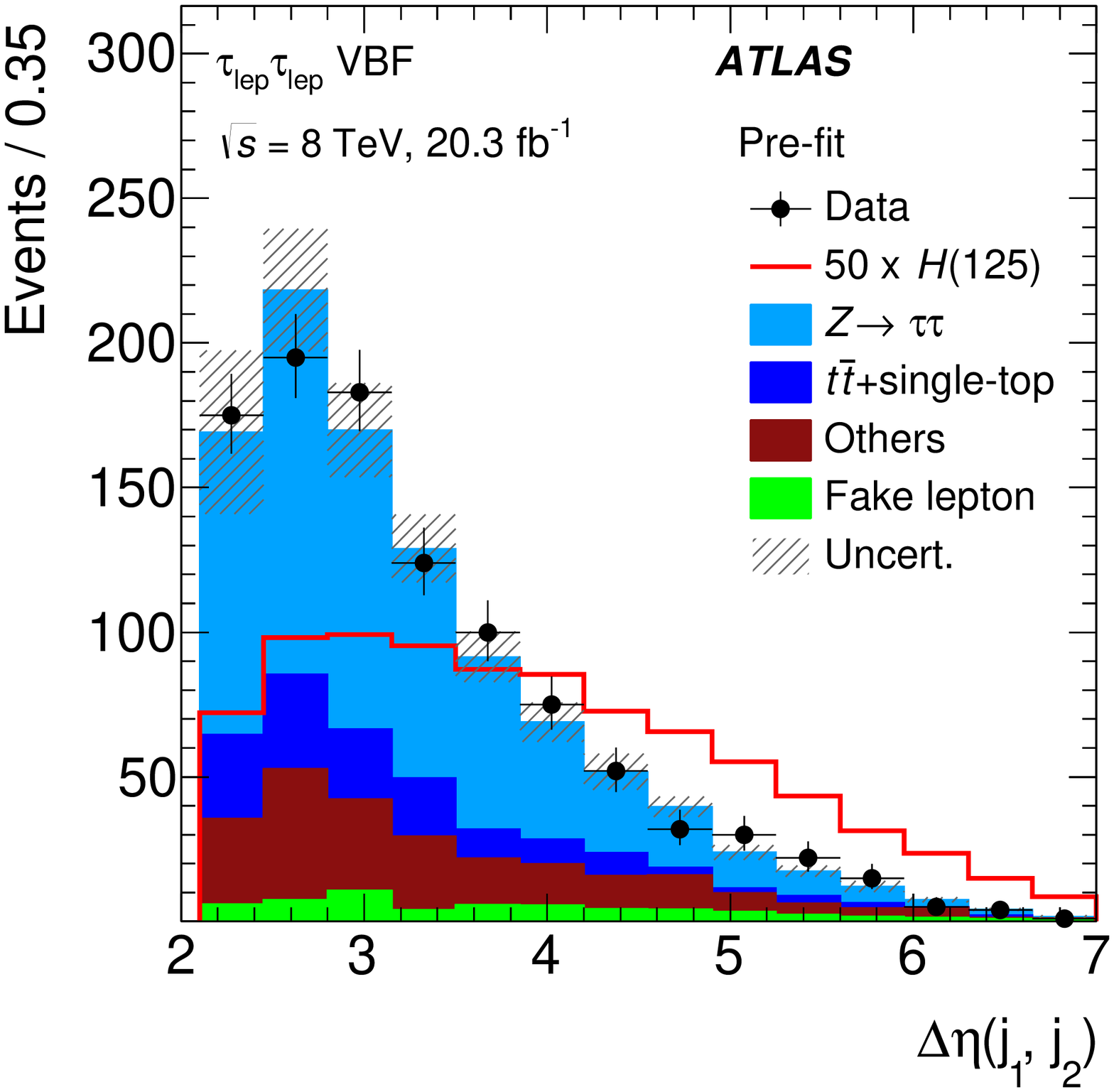}}\hfill
  \subfigure[]{ \includegraphics[width=6.0cm]{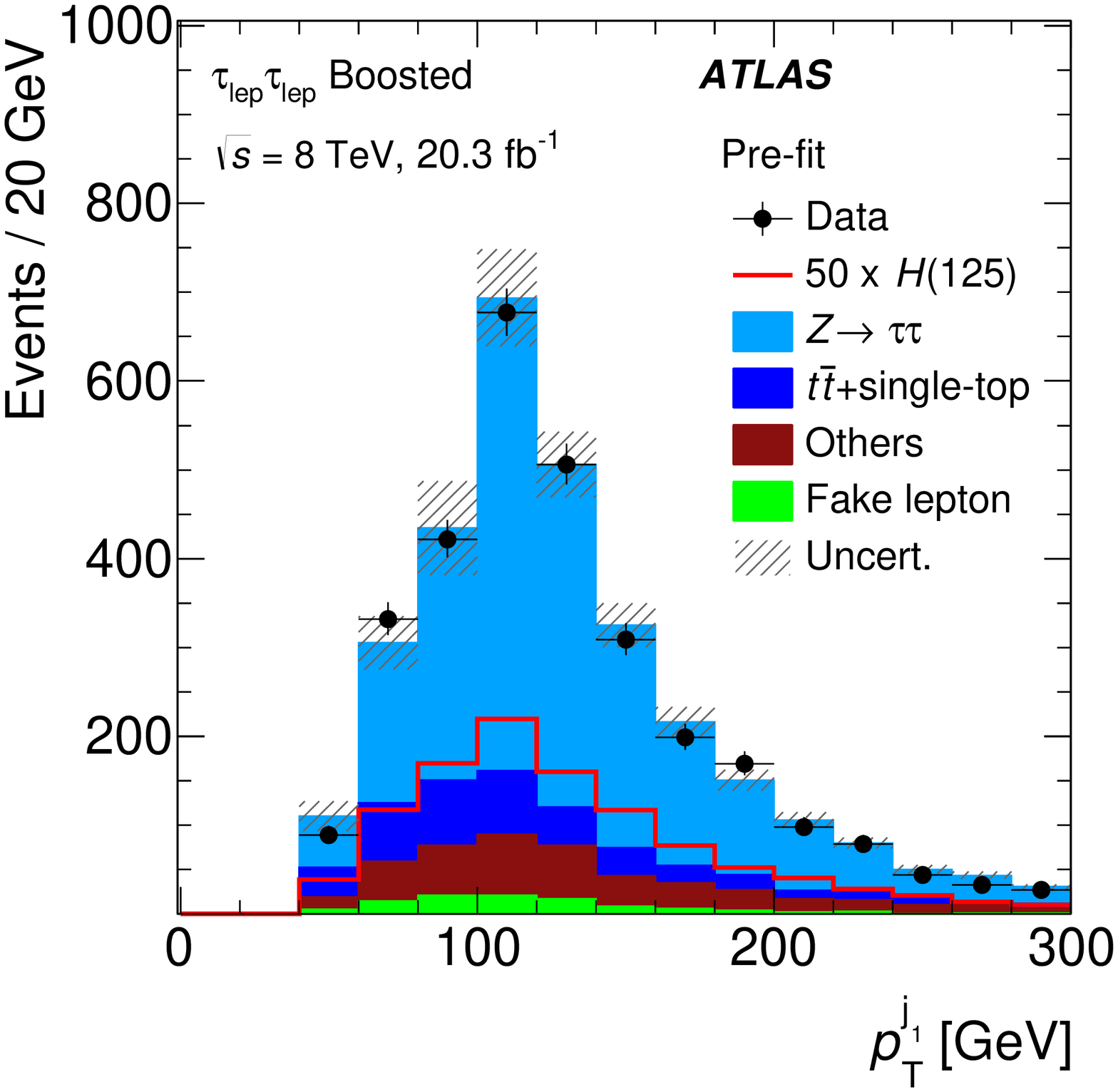}}
  \hspace*{\fill}

  \hspace*{\fill}
  \subfigure[]{ \includegraphics[width=6.0cm]{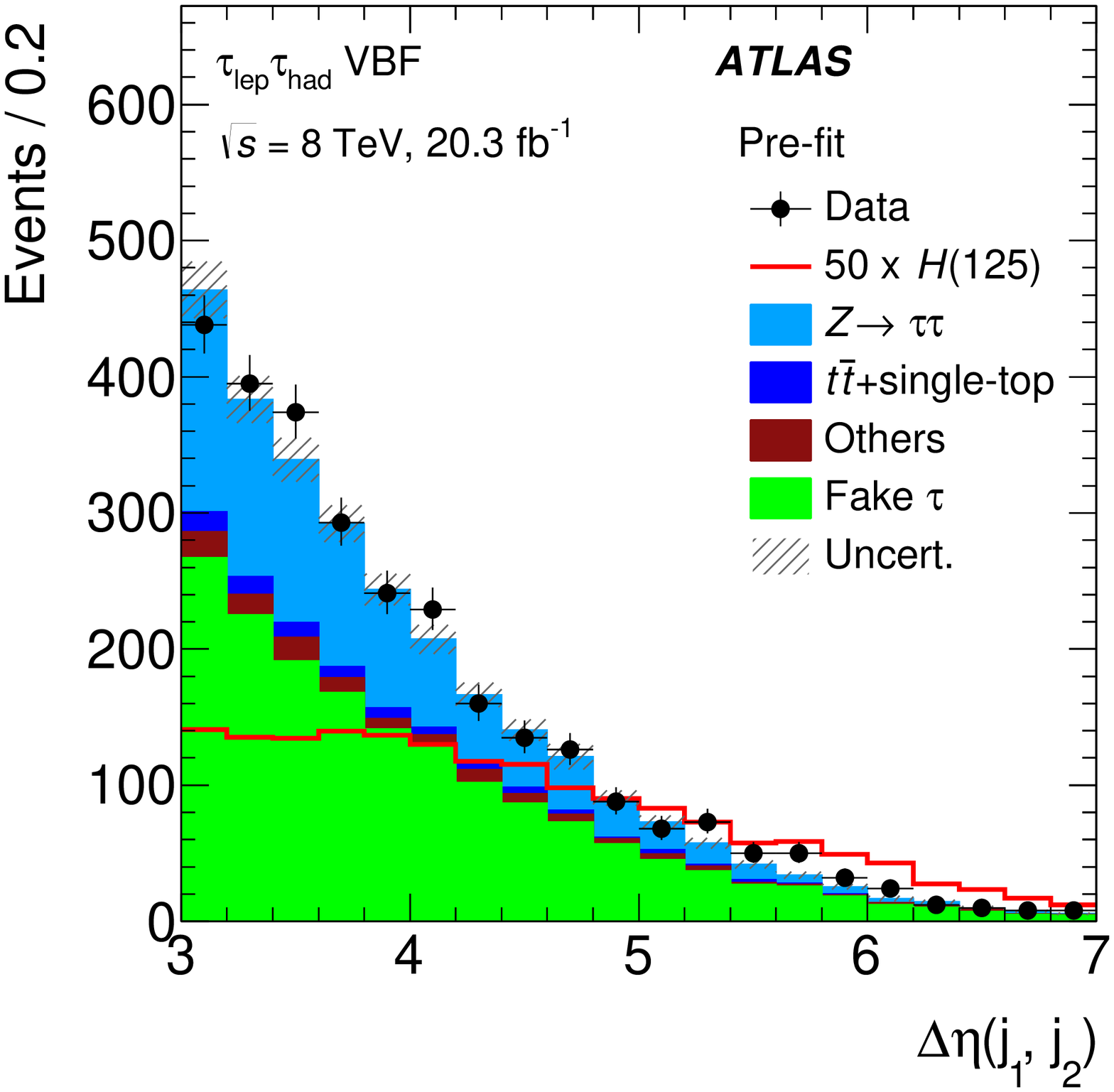}}\hfill
  \subfigure[]{ \includegraphics[width=6.0cm]{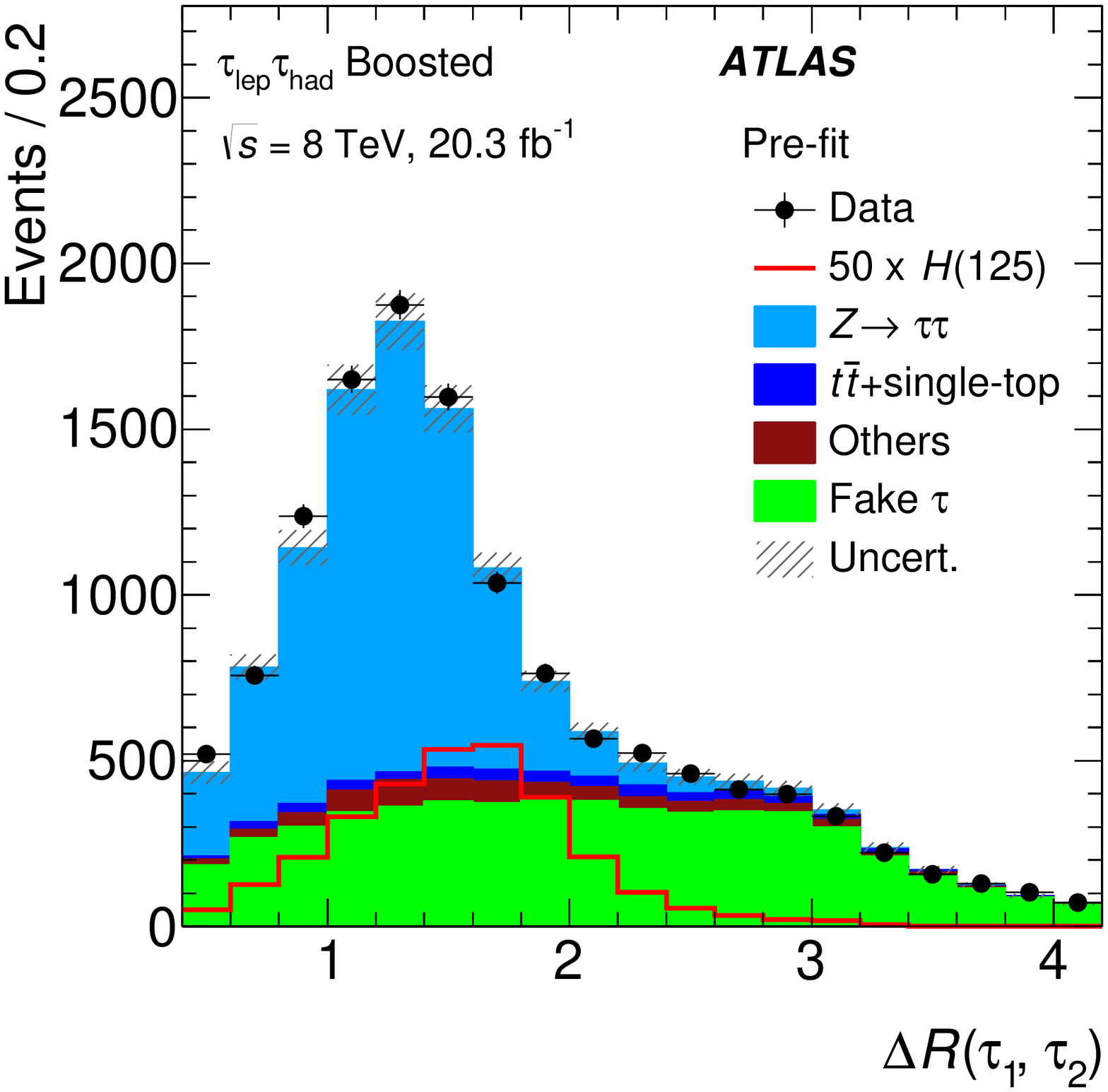}}
  \hspace*{\fill}

  \hspace*{\fill}
  \subfigure[]{ \includegraphics[width=6.0cm]{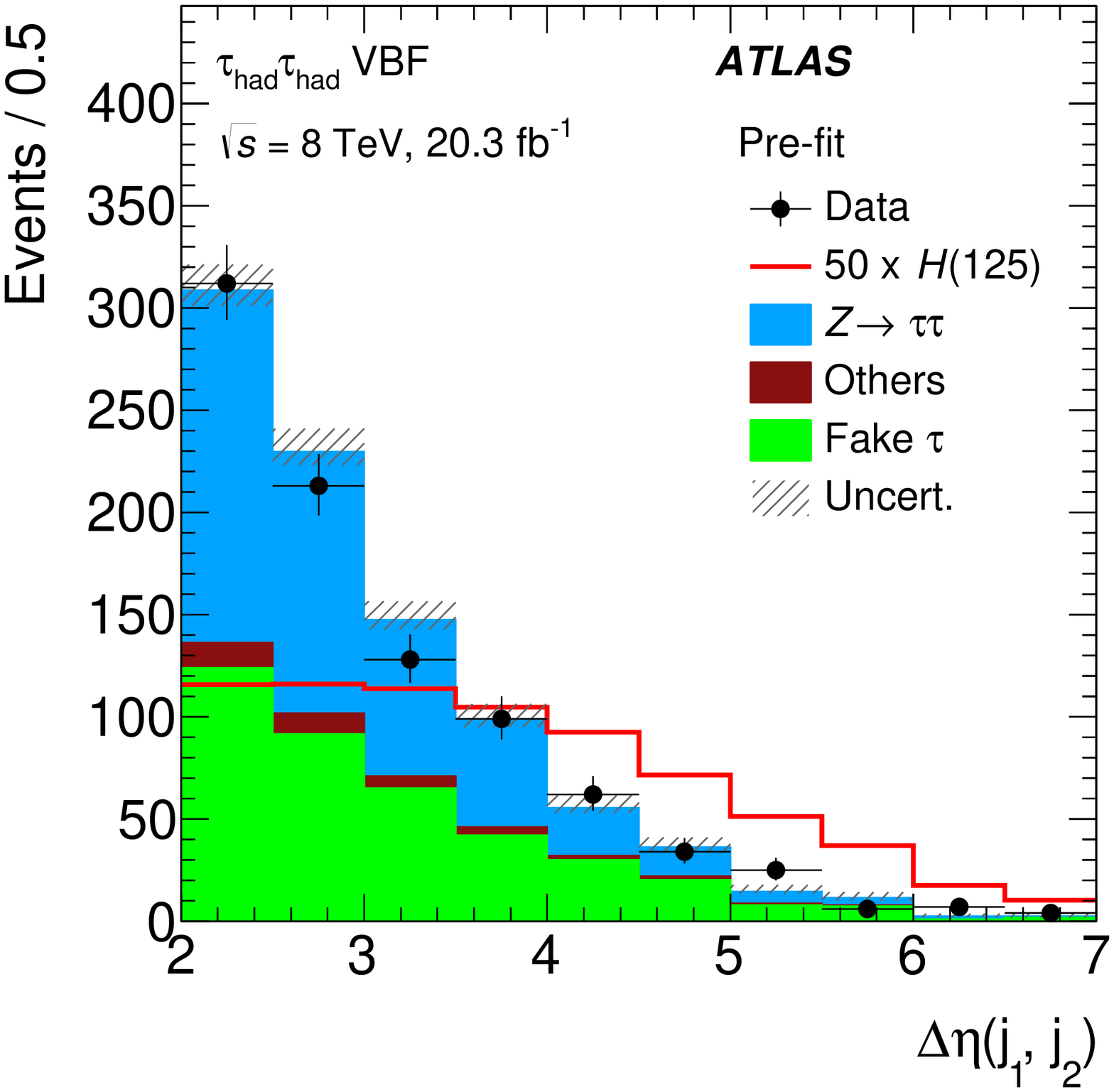}}\hfill
  \subfigure[]{ \includegraphics[width=6.0cm]{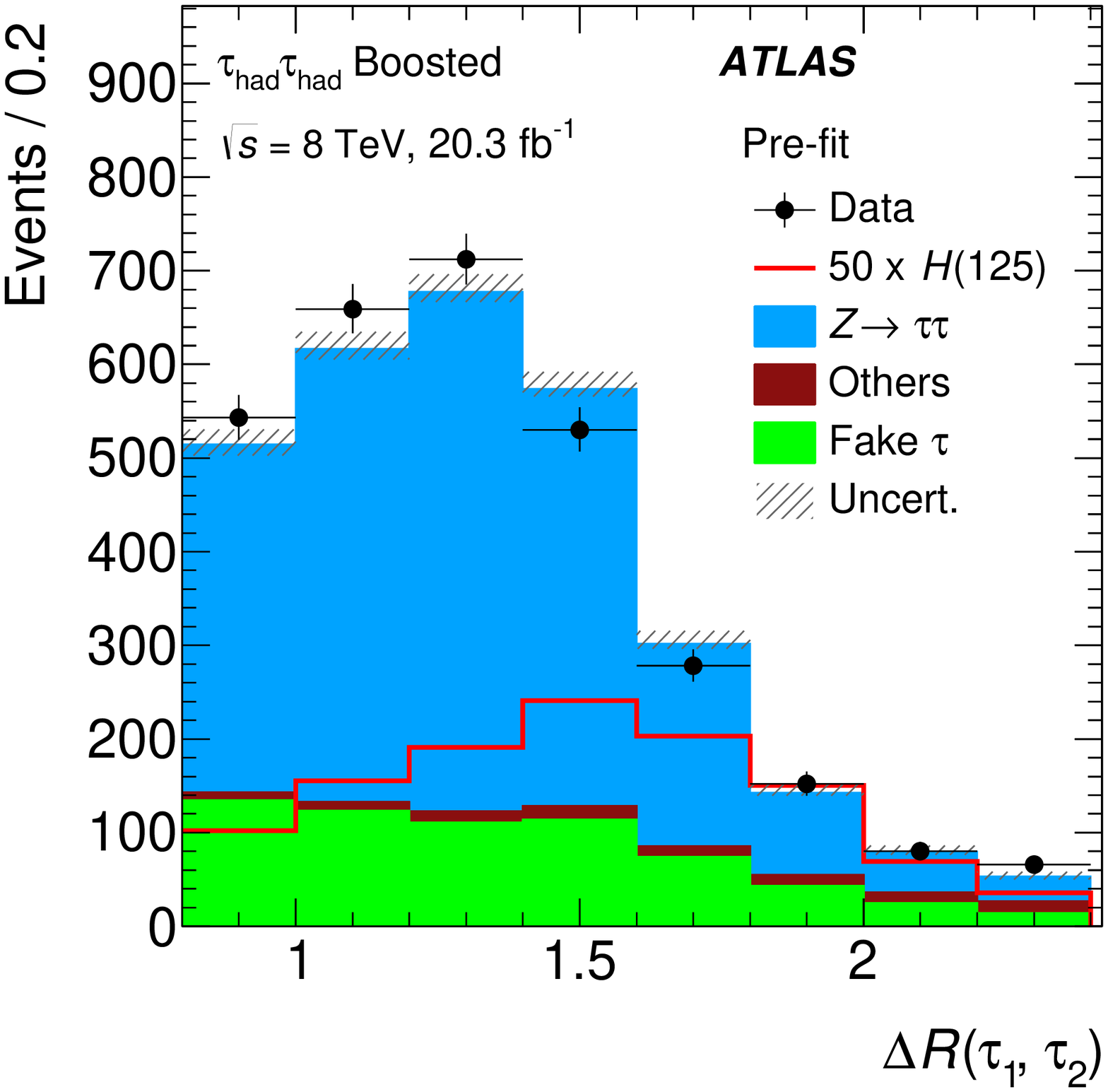}}
  \hspace*{\fill}

  \caption{
Distributions of important BDT input variables for the three channels and the
two categories (VBF, left) and (boosted, right) for data collected
at $\sqrt{s}$~=~8~\TeV. The distributions are shown for (a)
the separation in pseudorapidity of the jets, $\Delta\eta(j_1,j_2)$,  and (b) the
transverse momentum of the leading jet $\pt^{j_1}$ in the~\tll\ channel, for (c)
$\Delta\eta(j_1,j_2)$ and (d) $\Delta{}R(\tau_1,\tau_2)$, the distance
$\Delta{}R$  between the lepton and $\thad$, in the~\tlhad\ channel
and for (e) $\Delta\eta(j_1,j_2)$ and (f)  $\Delta{}R(\tau_1,\tau_2)$, the distance
$\Delta{}R$  between the two $\thad$ candidates, in the
\thadhad\ channel.  The contributions from a Standard Model Higgs boson with
$m_H$~=~125~\GeV\ are superimposed, multiplied by a factor of 50. These figures use background predictions made
without the global fit defined in section~\ref{sec:fit}.  The error band
includes statistical and pre-fit systematic uncertainties.
}
  \label{fig:bdt-inputs}
\end{figure}

\begin{figure}[htbp]
  \centering
  \hspace*{\fill}
  \subfigure[]{ \includegraphics[width=6.0cm]{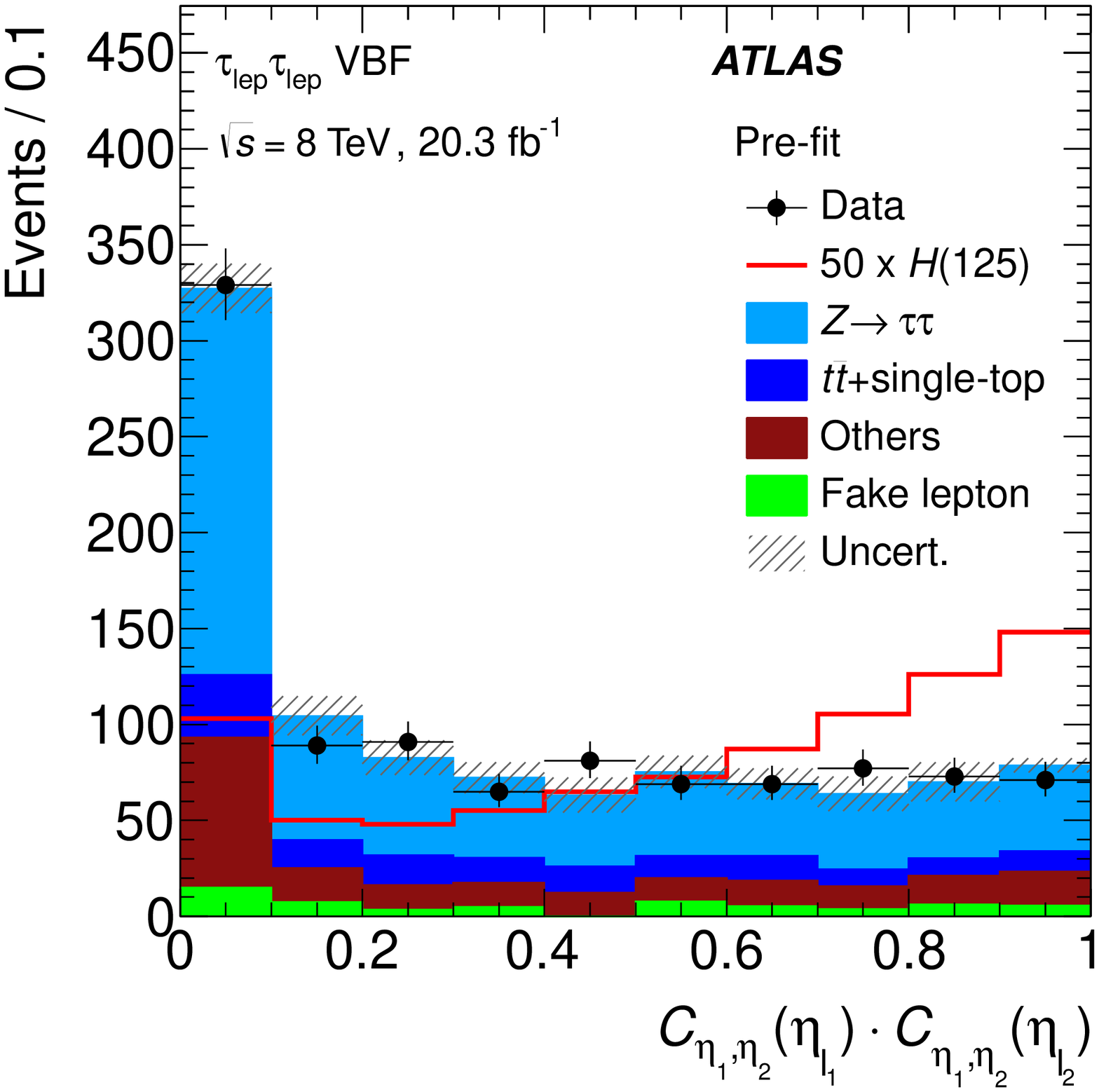}}\hfill
  \subfigure[]{ \includegraphics[width=6.0cm]{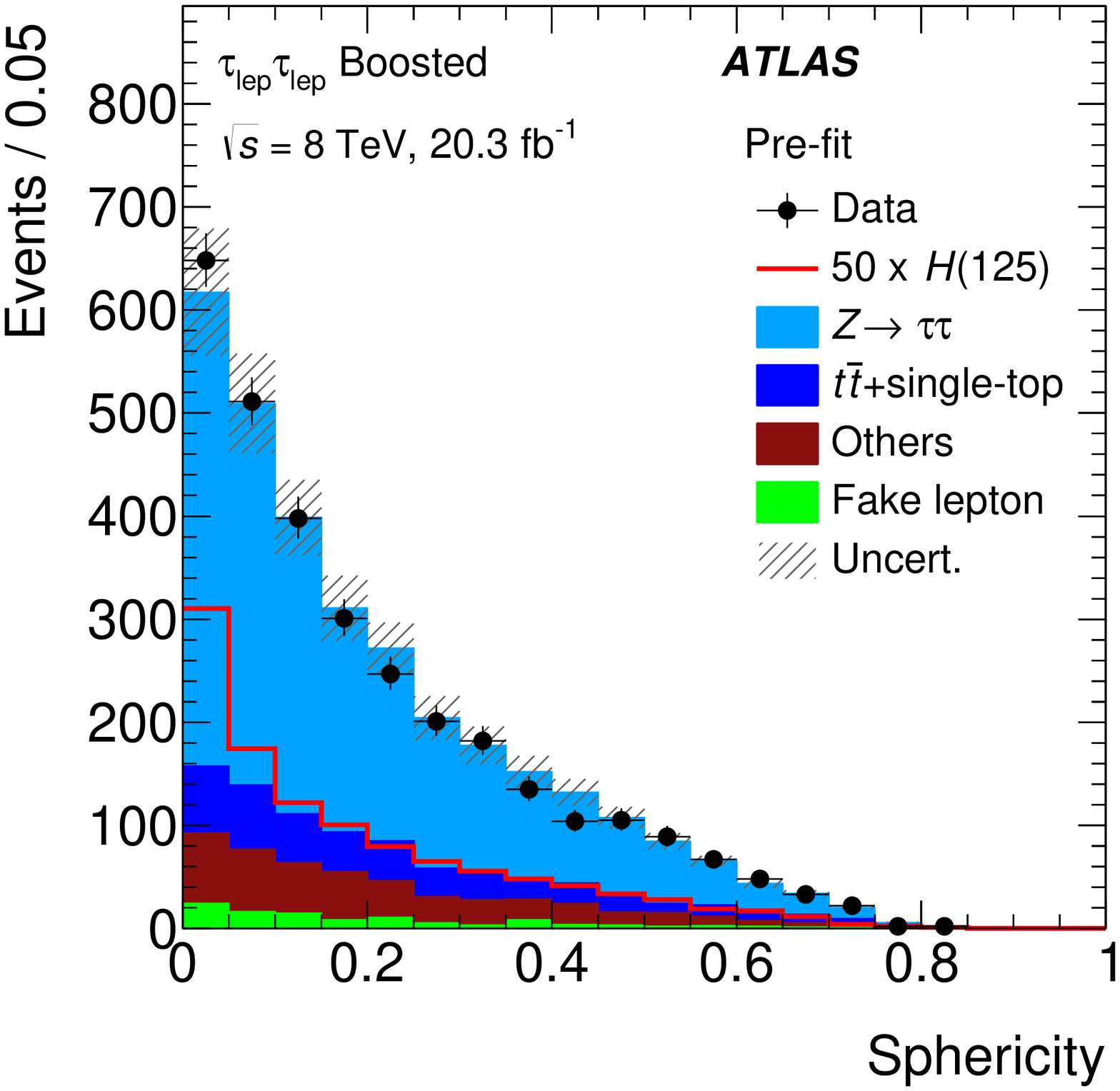}}
  \hspace*{\fill}

  \hspace*{\fill}
  \subfigure[]{ \includegraphics[width=6.0cm]{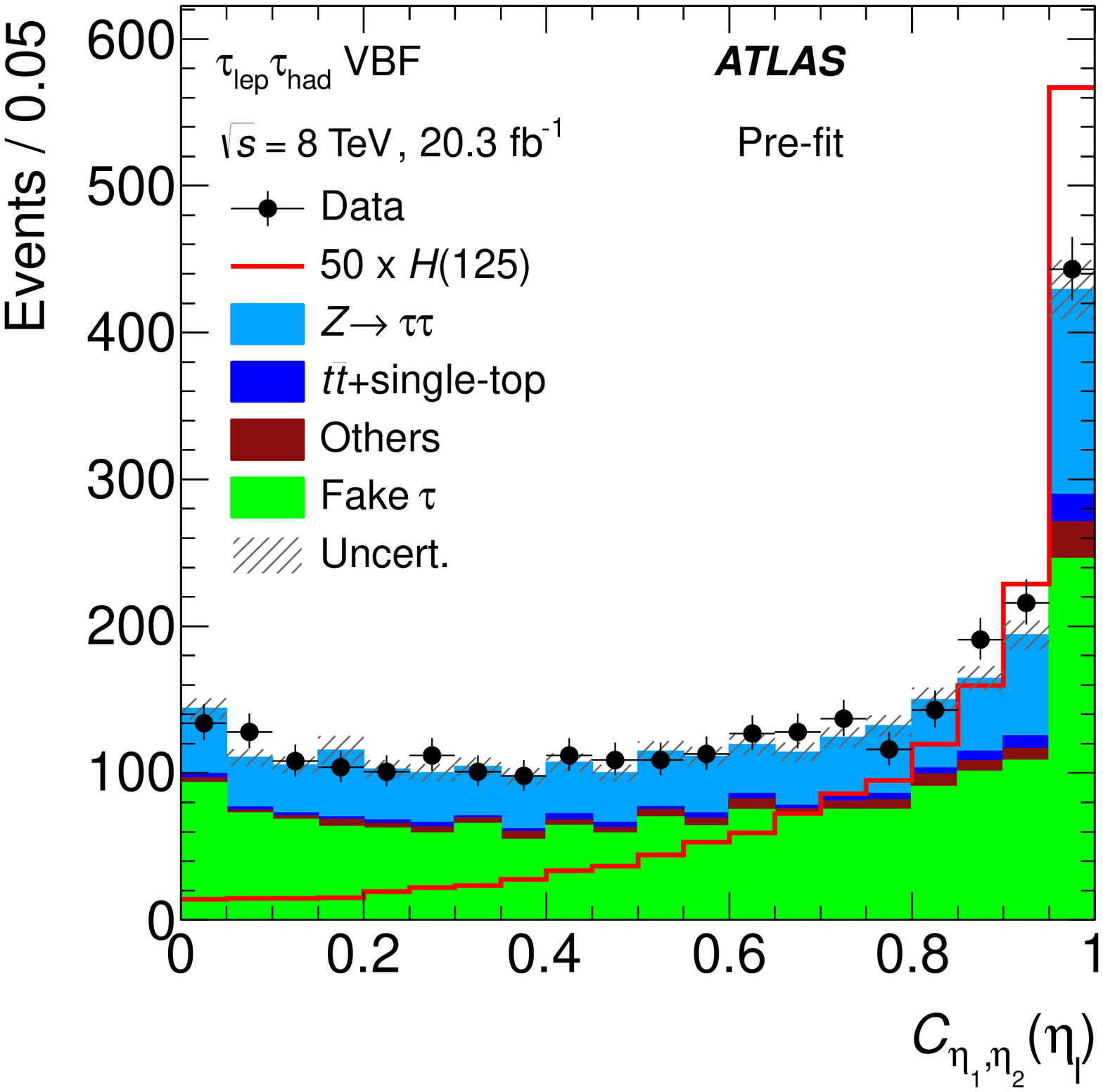}}\hfill
  \subfigure[]{ \includegraphics[width=6.0cm]{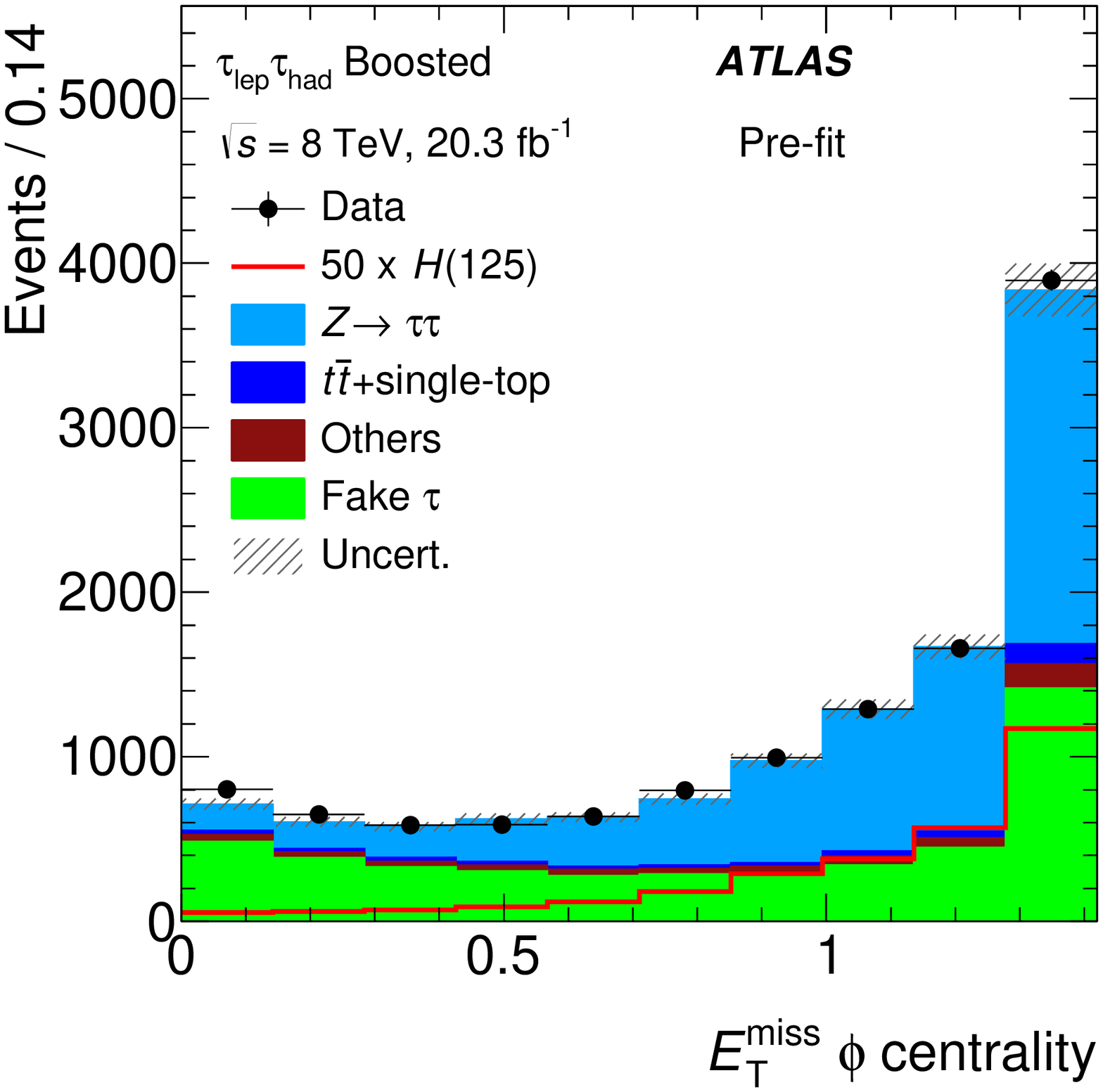}}
  \hspace*{\fill}

  \hspace*{\fill}
  \subfigure[]{ \includegraphics[width=6.0cm]{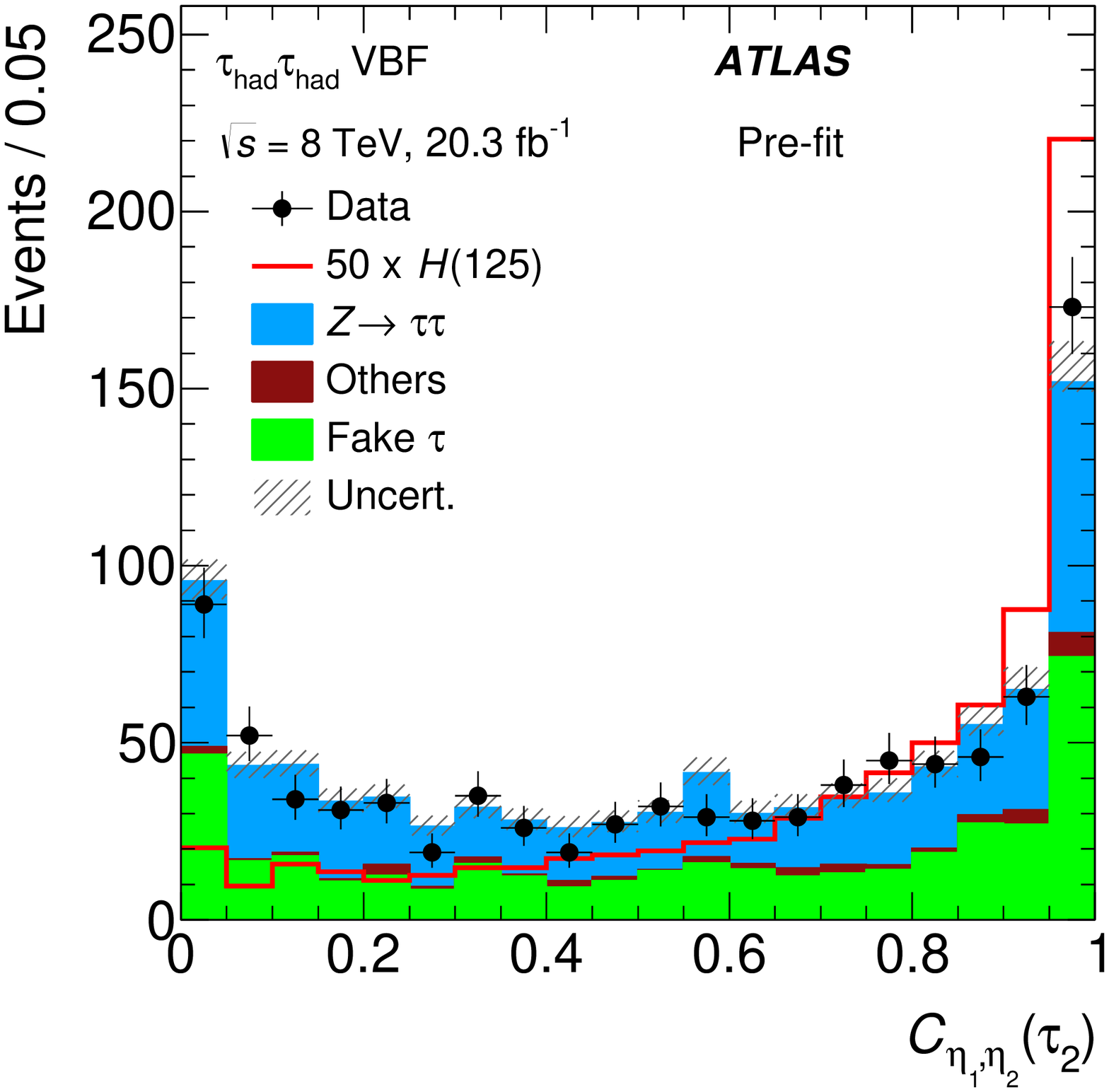}}\hfill
  \subfigure[]{ \includegraphics[width=6.0cm]{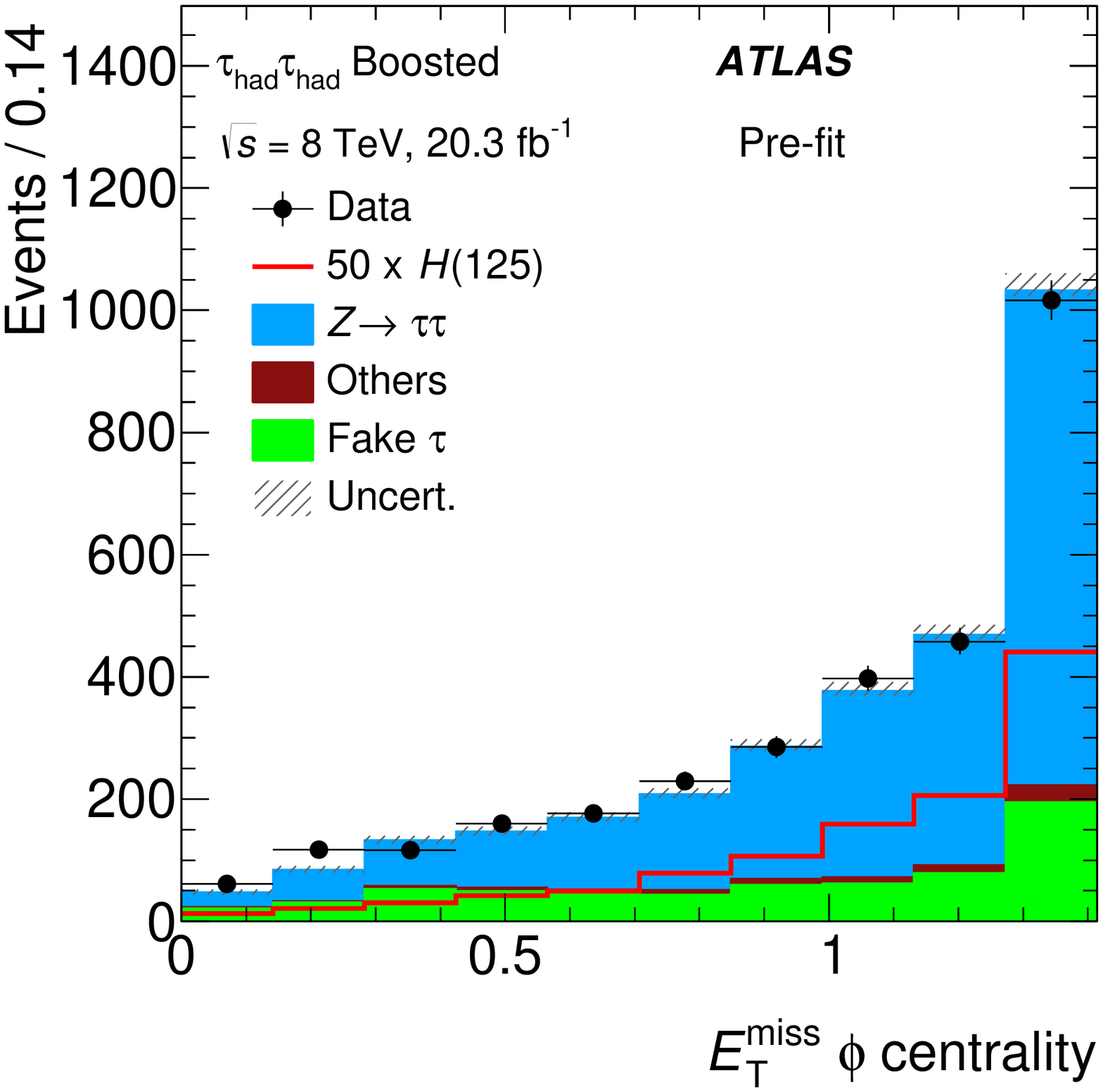}}
  \hspace*{\fill}

  \caption{
Distributions of important BDT input variables for the three channels and the
two categories (VBF, left) and (boosted, right) for data collected
at $\sqrt{s}$~=~8~\TeV. The distributions are shown for (a) the product of
the lepton centralities, $C_{\eta_1, \eta_2}(\eta_{\ell_1})\cdot C_{\eta_1, \eta_2}(\eta_{\ell_2})$,
and (b) the sphericity in the \tll\ channel, for (c) 
the centrality of the lepton, $C_{\eta_1, \eta_2}(\eta_{\ell})$,
 and (d) the $\MET\phi$ centrality  in the \tlhad\ channel,
and for (e) the centrality of the subleading tau, $C_{\eta_1, \eta_2}(\eta_{\tau_2})$,
 and (f) the $\MET\phi$ centrality  in the \thadhad\ channel.
The contributions from a Standard Model Higgs boson with
$m_H$~=~125~\GeV\ are superimposed, multiplied by a factor of 50.
 These figures use background predictions made
without the global fit defined in section~\ref{sec:fit}.  The error band
includes statistical and pre-fit systematic uncertainties.
}
  \label{fig:bdt-inputs-2}
\end{figure}

Among these variables the most discriminating ones include \MMC,
$\Delta{}R(\tau_1,\tau_2)$ and $\Delta\eta(j_1,j_2)$.  Figures
\ref{fig:bdt-inputs} and \ref{fig:bdt-inputs-2} show the 
distributions of selected BDT input variables.
For the VBF category, the distributions of $\Delta\eta(j_1,j_2)$ and centrality are
shown for all three channels.  For the boosted category, the distributions
of  $\Delta{}R(\tau_1,\tau_2)$ and $\MET\phi$ centrality are shown for the $\tlhad$ and
$\thadhad$ channels, and the distribution of the $\pT$ of the leading jet
 and the sphericity are shown for the $\tll$ channel.
 For all distributions, the data are compared to
the predicted SM backgrounds at $\sqrt {s}$~=~8~\TeV. The
corresponding uncertainties are indicated by the shaded bands.  All input
distributions are well described, giving confidence that the background models
(from simulation and data) describe well the relevant input variables of the
BDT. Similarly, good agreement is found for the distributions at $\sqrt
{s}$~=~7~\TeV.

\begin{table}
\begin{center}
\begin{tabular}{|c|ccc|ccc|}
\hline
\multirow{2}{*}{Variable}  & \multicolumn{3}{c|}{VBF}  &\multicolumn{3}{c|} {Boosted}  \\
         & \tll & \tlhad & \thadhad & \tll & \tlhad & \thadhad \\ \hline\hline
\MMC\   &  \tbull & \textbullet & \textbullet & \tbull &\tbull &\tbull   \\\hline
$\Delta{}R(\tau_1,\tau_2)$&  \tbull & \textbullet & \textbullet & &\tbull &\tbull  \\\hline
$\Delta\eta(j_1,j_2)$   &\textbullet & \textbullet  &\textbullet& \multicolumn{3}{c|}{} \\ \hline
$m_{j_1,j_2}$     & \textbullet & \textbullet &\textbullet& \multicolumn{3}{c|}{} \\ \hline
$\eta_{j_1}\times\eta_{j_2}$   & & \textbullet& \textbullet & \multicolumn{3}{c|}{} \\ \hline
$p_\mathrm{T}^{\mathrm{Total}}$  & & \textbullet & \textbullet &  \multicolumn{3}{c|}{} \\ \hline
Sum $p_\mathrm{T}$ & & & & &\tbull &\tbull \\\hline
$p_\mathrm{T}^{\tau_1}/p_\mathrm{T}^{\tau_2}$ & & & & &\tbull &\tbull  \\\hline
$\MET\phi$ centrality &  & \textbullet & \textbullet &\tbull &\tbull &\tbull \\ \hline
$m_{\ell,\ell,j_1}$ & & &  &\textbullet & &  \\ \hline
$m_{\ell_1,\ell_2}$  & & &     & \textbullet & &   \\ \hline
$\Delta \phi(\ell_1,\ell_2)$ & & & & \textbullet &  &  \\ \hline
Sphericity     & & & & \textbullet&  & \\ \hline
$\pt^{\ell_1}$ & & & & \textbullet&  & \\ \hline
$\pt^{j_1}$  & & & & \textbullet &  & \\ \hline
$\MET/\pt^{\ell_2}$ & & & & \textbullet & & \\ \hline
$m_{\mathrm{T}}$        &  & \textbullet & &  & \textbullet&   \\ \hline
min$(\Delta \eta_{\ell_1\ell_2,\mathrm{jets}})$  & \textbullet & &  & \multicolumn{3}{c|}{} \\ \hline
$C_{\eta_1, \eta_2}(\eta_{\ell_1})\cdot C_{\eta_1, \eta_2}(\eta_{\ell_2})$  & \textbullet & &  & \multicolumn{3}{c|}{} \\ \hline
$C_{\eta_1, \eta_2}(\eta_{\ell})$ &  & \textbullet& &\multicolumn{3}{c|}{} \\ \hline
$C_{\eta_1, \eta_2}(\eta_{ j_3})$    & \textbullet & &  & \multicolumn{3}{c|}{} \\ \hline
$C_{\eta_1, \eta_2}(\eta_{\tau_1})$   & &  & \textbullet& &  & \\ \hline
$C_{\eta_1, \eta_2}(\eta_{\tau_2})$   & &  & \textbullet& &  & \\ \hline
 \end{tabular}
\caption{
Discriminating variables used in the training of the BDT for each channel and
category at $\sqrt{s}=8\TeV$. The more complex variables are described in the
text. The filled circles indicate which variables
are used in each case. 
}
\label{tab:bdtvars}
\end{center}
\end{table}

\clearpage

\clearpage

\clearpage
\section{Background estimation}
\label{sec:background}

The different final-state topologies of the three analysis channels have
different background compositions which necessitate different strategies for
the background estimation. In general, the number of expected background events
and the associated kinematic distributions are derived from a mixture of
data-driven methods and simulation. The normalisation of several important
background contributions is performed by comparing the simulated samples of
individual background sources to data in regions which only have a small or
negligible contamination from signal or other background events.
The control regions used in the analysis are summarised in
table~\ref{tab:control-summary}.  

Common to all channels is the dominant $Z \to \tau \tau$ background, for which
the kinematic distributions are taken from data by employing the embedding
technique, as described in section~\ref{sec:samples}.
Background contributions from jets that are misidentified as hadronically
decaying taus (fake backgrounds) are estimated by using either a {\em fake-factor} method
or samples of non-isolated $\thad$ candidates. Likewise, samples of
non-isolated leptons are used to estimate fake-lepton contributions from 
jets or hadronically decaying taus and leptons from other sources, such as
heavy-quark decays.\footnote{For simplicity, leptons from heavy-quark decays are considered
as fake leptons in the following.}

Contributions from various other physics processes with leptons and/or \thad\ candidates in the final state 
are estimated using
the simulation, normalised to the theoretical cross sections, as given in
table~\ref{tab:MCGenerator}.  A more detailed discussion of the estimation of
the various background components in the different channels is given in the
following.

\subsection{Background from $Z\to\tau\tau$ production}
\label{subsec:bkg_Ztt}
\label{subsec:ztt}

\begin{figure}[htbp]
\centering
\subfigure[]{\includegraphics[width=0.49\textwidth]{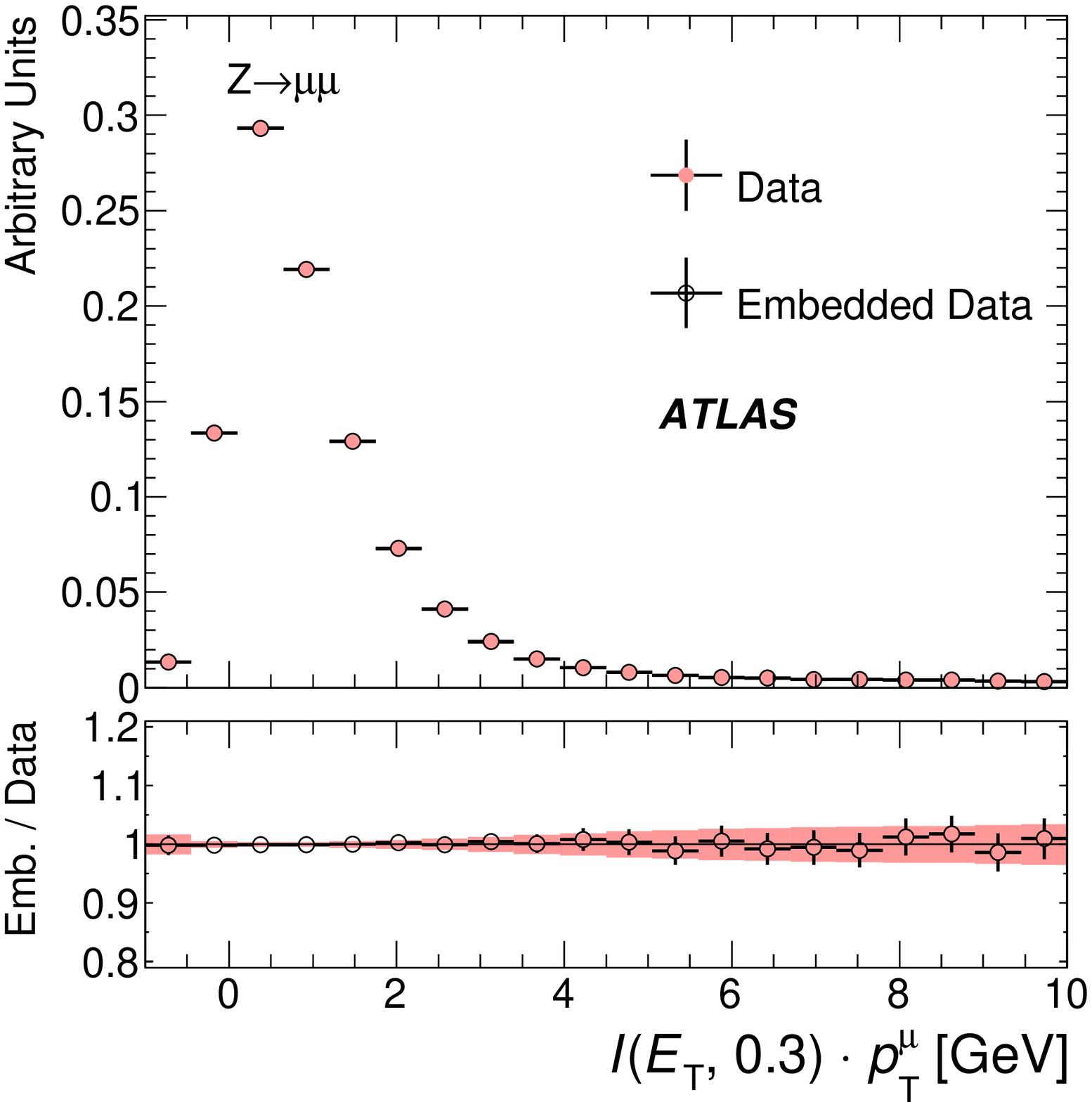}}
\subfigure[]{\includegraphics[width=0.49\textwidth]{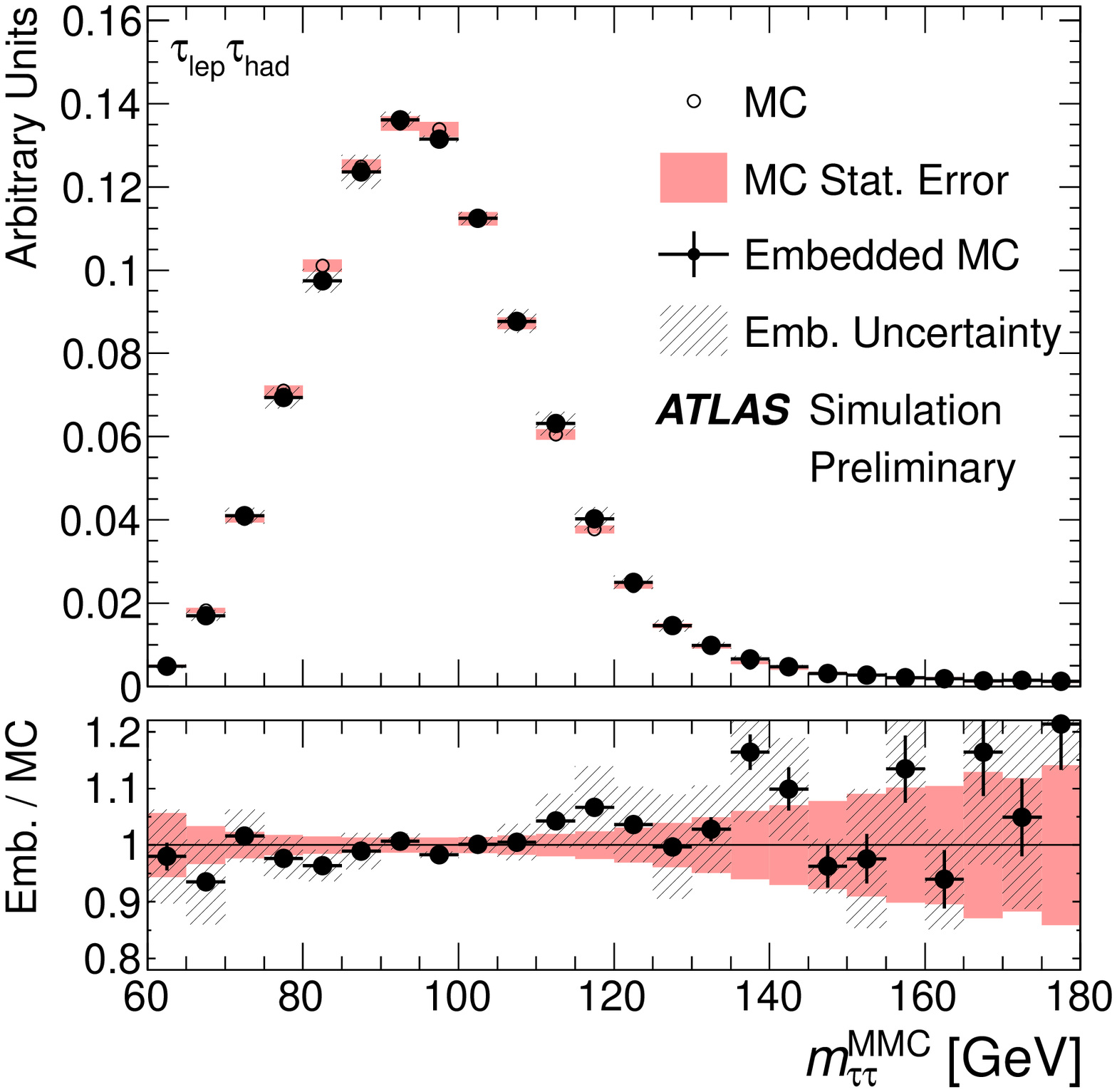}}
\caption{
(a) The distribution of the calorimeter isolation energy $I (\ET, 0.3) \cdot
\pT ^{\mu} $ within a cone of radius $\Delta R$~=~0.3 around the muons in $Z \to
\mu \mu $ events from data, before and after the embedding of simulated muons.  (b)
The distribution of the reconstructed invariant $\tau \tau$ mass, $\MMC$, in the $\tlhad$ final state, for
simulated $Z \to \tau \tau$ events, compared to the one obtained from simulated
$Z \to \mu \mu $ events after tau embedding.  The ratios of the values before
and after the embedding and between the embedded $Z \to \mu \mu $ and  $Z \to \tau
\tau$ events are given in (a) and (b) respectively. The errors in (a) and (b)
on the ratios (points) represent the statistical uncertainties, while the
systematic uncertainties are indicated by the hatched bands in (b). The shaded
bands represent the statistical uncertainties from the $Z \to \mu \mu $ data
events in (a) and from the $Z \to \tau \tau$ simulation in (b).
}
\label{fig:embedding}
\end{figure}

A reliable modelling of the irreducible $Z \to \tau \tau $  background is an
important ingredient of the analysis. It has been shown in other ATLAS analyses
that existing $\Zjets$ Monte Carlo simulation needs to be reweighted to model
data correctly \cite{Aad:2013ysa, Aad:2014xaa, Aad:2014dta}.
 Additionally, it is not possible to select a
sufficiently pure and signal-free $\Ztautau$ control sample from data
to model the background in the signal region. Therefore this 
 background is estimated using  embedded data, as described in section~\ref{sec:samples}.  This
procedure was extensively validated using both data and simulation.  To
validate the subtraction procedure of the muon cell energies and tracks from
data and the subsequent embedding of the corresponding information from
simulation, the muons in $Z \to \mu \mu $ events are replaced by simulated
muons. The calorimeter isolation energy in a cone of $\Delta R$~=~0.3 around
the muons from data before and after embedding is compared in
figure~\ref{fig:embedding}(a).  Good agreement is found, which indicates that
no deterioration (e.g. possible energy biases) in the muon environment is
introduced.  Another important test validates the embedding
of more complex $Z \to \tau \tau$ events, which can only be performed in the
simulation. To achieve a meaningful validation, the same MC generator with
identical settings was used to simulate both the  $Z \to \mu \mu $ and $Z \to \tau
\tau$ events.  The sample of embedded events is corrected for the bias due to
the trigger, reconstruction and acceptance of the original muons. These
corrections are determined from data as a function of $p_{\mathrm{T}}^{\mu}$ and
$\eta(\mu)$, and allow the acceptance of the  original selection to be
corrected. The tau decay products are treated like any other objects obtained
from the simulation, with one important difference due to the absence of
trigger simulation in this sample. Trigger effects are parameterised from the
simulation as a function of the tau decay product $p_{\mathrm{T}}$.
After replacing the muons with simulated taus, kinematic distributions of the embedded sample can be directly
compared to the fully simulated ones. As an example, the reconstructed invariant
mass, $\MMC$, is shown in figure~\ref{fig:embedding}(b), for the $\tlhad$ final state.
Good agreement is found and the observed differences are covered by the systematic uncertainties.
Similarly, good agreement is found for other variables, such as the missing
transverse momentum, the kinematic variables of the hadronically decaying tau
lepton or of the associated jets in the event. A direct comparison of the $Z
\to \tau \tau$ background in data and the modelling using the embedding
technique also shows good agreement. This can be seen in several
kinematic quantity distributions, which are dominated by  $Z \to \tau
\tau$ events, shown in figure~\ref{fig:bdt-inputs}.

The normalisation of this background process is taken from the final fit
described in section~\ref{sec:fit}. The normalisation is 
independent for the $\tll$, $\tlhad$, and $\thadhad$ analysis channels.

\subsection{Background from misidentified leptons or hadronically decaying taus}
\label{subsec:qcdbkg}

For the $\tll$ channel, all background sources resulting from misidentified
leptons are treated together. In this approach, contributions from multijet and
$W$+jets production, as well as the part of the $t\bar{t}$ background resulting
from decays to leptons and hadrons  ($ \ttbar \to \ell \nu b  \  qq b$) are included.  A
control sample is defined in data by inverting the isolation requirements for one
of the two leptons, while applying all other signal region requirements. The
contributions from other background channels (dileptonic $\ttbar$ decays
($\ttbar \to \ell \nu b~\ell \nu b$), $Z \to ee$, $Z \to \mu\mu$, and diboson
production) are obtained from the simulation and are subtracted.  From this
control sample a template is created. The normalisation factor is obtained by
fitting the $\pt$ distribution of the subleading lepton at an early stage of
the preselection.

\begin{figure}[htbp]
\begin{center}
\subfigure[]{\includegraphics[width=0.49\textwidth]{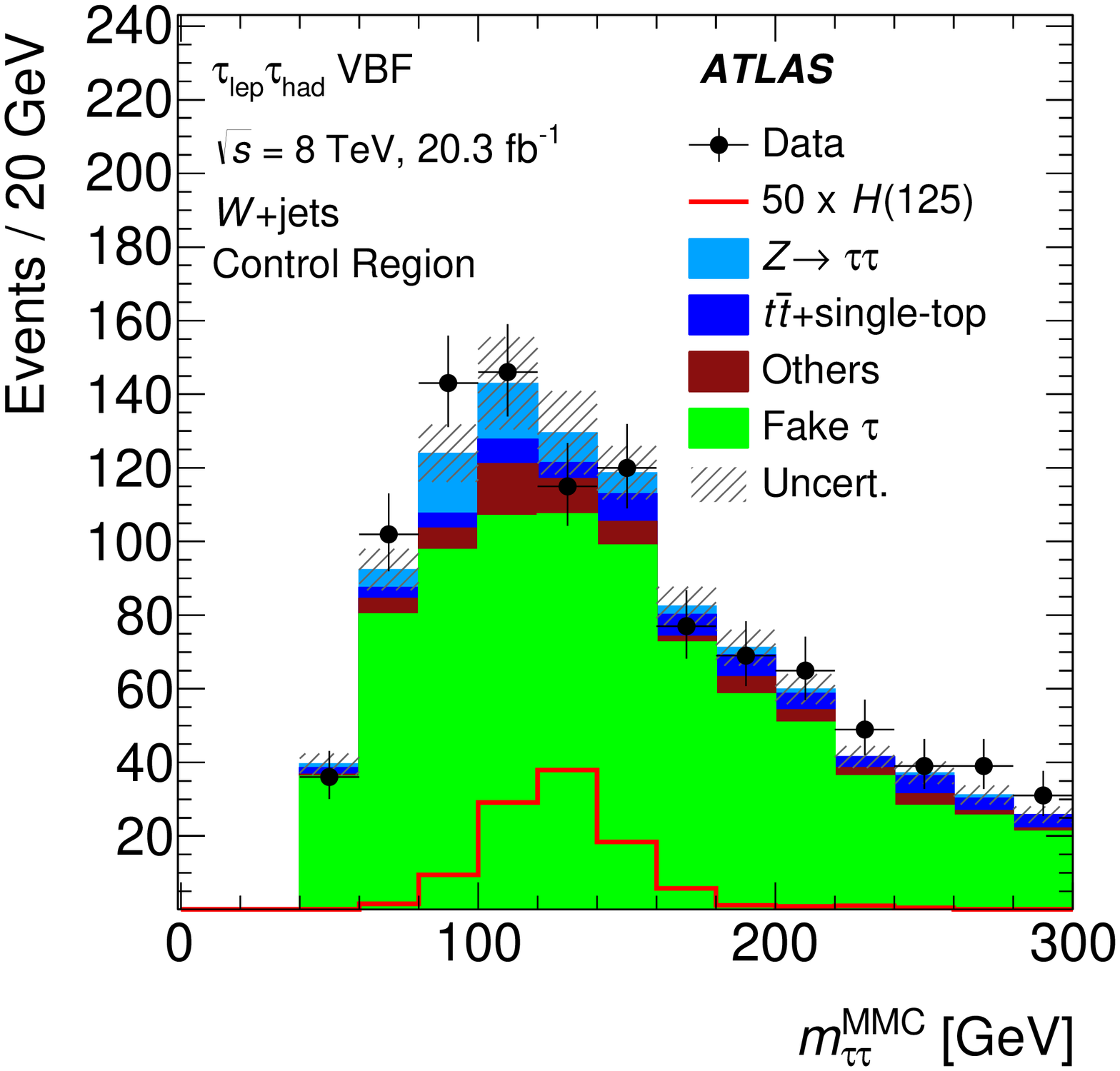}}
\subfigure[]{\includegraphics[width=0.49\textwidth]{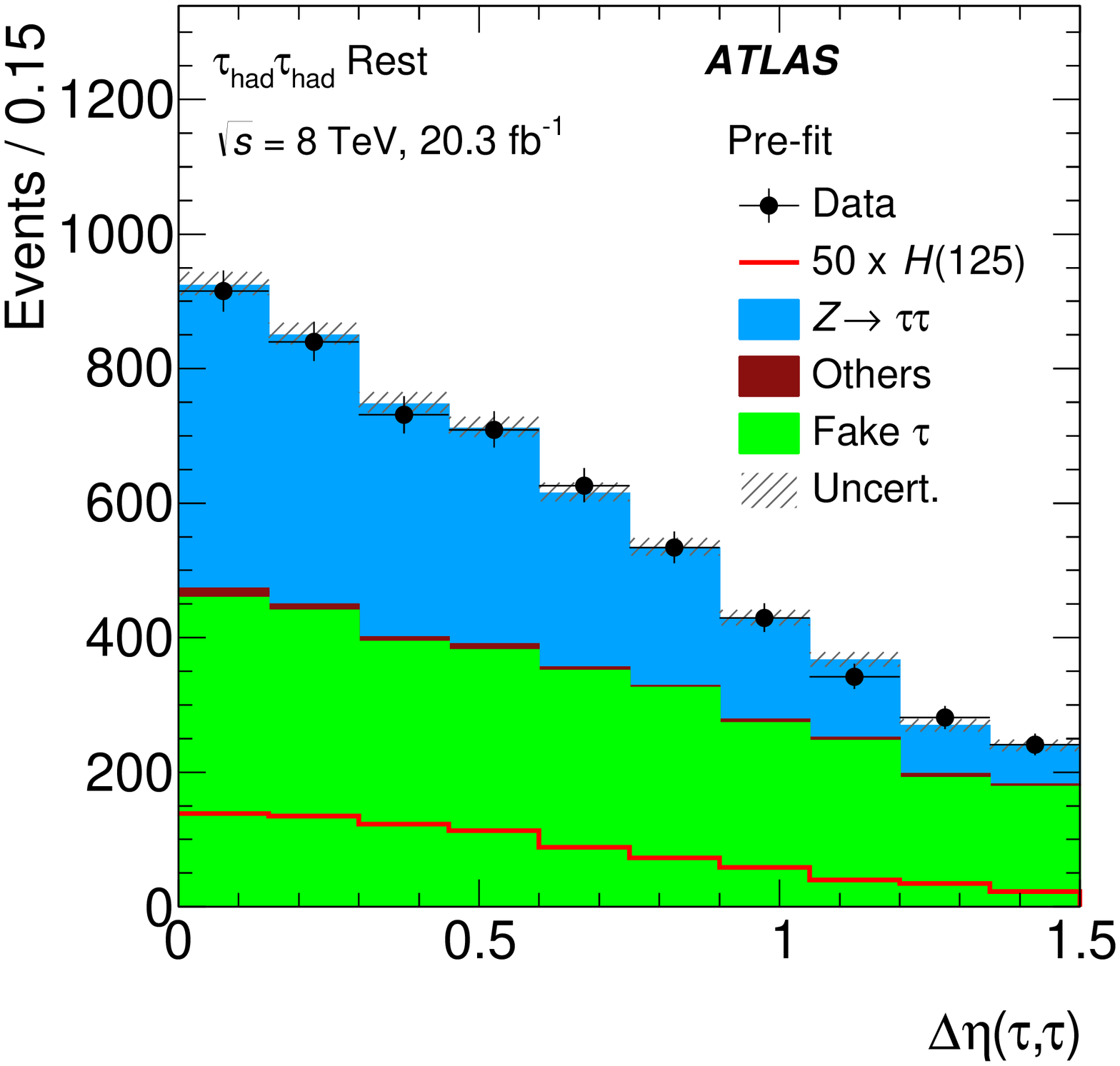}}
\end{center}
\caption{
(a) The distribution of the reconstructed invariant $\tau \tau$ mass, $\MMC$, for events in the
$\Wjets$ control region, for the $\tlhad$ channel. (b) The 
 separation in pseudorapidity of the $\thad$ candidates,
$\Delta{}\eta(\thad,\thad)$, for the $\thadhad$ channel in the rest
control region. The expected SM Higgs boson signal contribution is
superimposed, multiplied by a factor 50.  These figures use background
predictions made without the global fit defined in section~\ref{sec:fit}.  The
error band includes statistical and pre-fit systematic uncertainties.
\label{fig:background-variables}}
\end{figure}

For the $\tlhad$ channel, the fake-factor method
is used to derive estimates for the multijet, $\Wjets$,  $\Zjets$, and
semileptonic $t\bar{t}$ background events that pass the $\tlhad$ selection due
to a misidentified $\thad$ candidate.  The fake factor is defined as the ratio 
of the number of jets identified as medium $\thad$ candidates to the number 
satisfying the loose, but not the medium, criteria.
Since the fake factor depends on the
type of parton initiating the jet and on the $\pT$ of the jet, it is determined as a function of $\pT$
separately for samples enriched in quark- and gluon-initiated jets. In addition, the fake factor
is found to be different for 1-track and 3-track candidates.  Three different,
quark-jet dominated samples are used separately for the $\Wjets$, $t\bar{t}$
and $\Zjets$ background components. They are defined by selecting the
high-$\mT$ region ($\mT > 70\GeV$), by inverting the $b$-jet veto and by
requiring two leptons with an invariant mass consistent with $m_Z$ ($80\GeV < m_{\ell\ell} < 100\GeV$) respectively.  In addition, a multijet sample dominated
by gluon-initiated jets is selected by relaxing the lepton identification and requiring
it to satisfy the loose identification criteria.  The derived fake factors
are found to vary from 0.124 (0.082) for $\pT = 20\GeV$ to 0.088 (0.038) for
\pT~=~150~\GeV\ for 1-track (3-track) candidates in the VBF category. The
corresponding values for the boosted category are 0.146 (0.084)  for
\pT~=~20~\GeV\  and 0.057 (0.033) for \pT~=~150~\GeV. To obtain the fake-background 
estimate for the  VBF and boosted signal regions, these
factors are then applied, weighted by the expected relative $\Wjets$, $\Zjets$,
multijet, and $t\bar{t}$ fractions, to the events in regions defined by
applying the selections of the corresponding signal region, except that the
$\thad$ candidate is required to pass the loose and to fail the medium $\thad$
identification.
As an example, the good agreement between data and background estimates is
shown in figure~\ref{fig:background-variables}(a) for the reconstructed  $\tau
\tau$ mass for events in the high-$\mT$ region, which is dominated by $\Wjets$ production.

For the $\thadhad$ channel, the multijet background is modelled using a
template extracted from data that pass the VBF or boosted category
selection, where, however, the taus fail the isolation and opposite-sign charge
requirements (the number-of-tracks requirement is not enforced).
  The normalisation of the multijet background is first determined
by performing a simultaneous fit of the multijet (modelled by the data sample
just mentioned) and $Z \to \tau \tau $ (modelled by embedding) templates after
the preselection cuts.  The fit is performed for the distribution of the
difference in pseudorapidity between the two hadronic tau candidates,
$\Delta{}\eta(\thad,\thad)$. The signal contribution is expected to be small in
this category.  The agreement between data and the background estimate for this
distribution is shown in figure~\ref{fig:background-variables}(b) for the
 rest category defined in section~\ref{sec:selection}.
The preselection normalisation is used as a reference point and starting value
for the global fit (see below) and is used for validation plots.  The final
normalisations of the two important background components, from multijet and $Z
\to \tau \tau $ events, are extracted from the final global fit, as described
in section~\ref{sec:fit}, in which the $\Delta{}\eta(\thad,\thad)$ distribution
for the rest category is included.

\subsection{$Z\to ee$ and $Z\to \mu\mu$ background}
\label{subsec:zee}
The Drell--Yan $Z/\gamma^{*}\to ee$ and $Z/\gamma^{*} \to \mu\mu$ background
channels are important contributions to the final states with two same-flavour leptons.
They also contribute to the other channels. As described below, a simulation
based on \textsc{Alpgen} is used to estimate these background sources.
Correction factors are applied to account for differences between data and
simulation.

\begin{figure}[htbp]
\begin{center}
\subfigure[]{\includegraphics[width=0.49\textwidth]{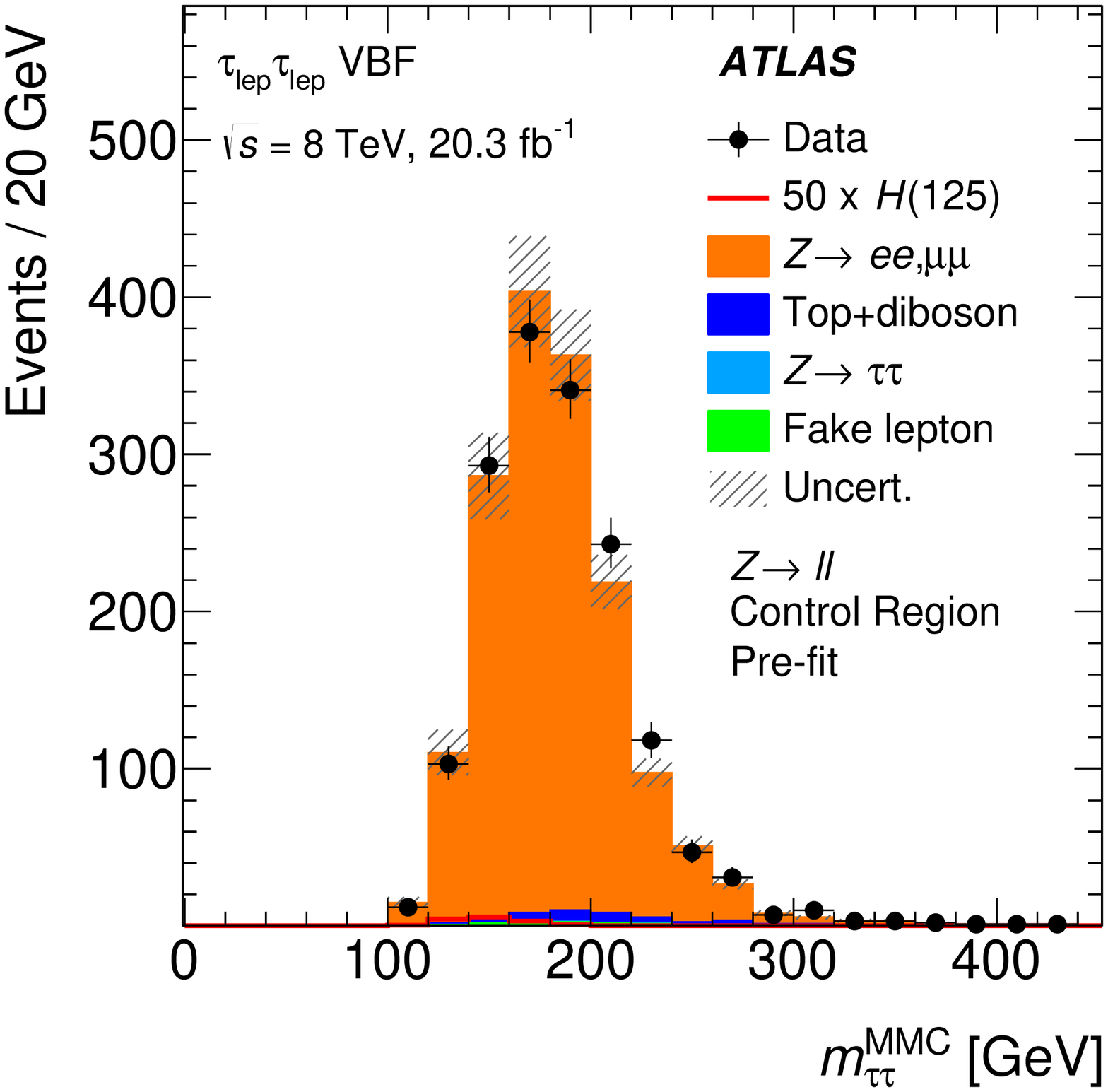}}
\subfigure[]{\includegraphics[width=0.49\textwidth]{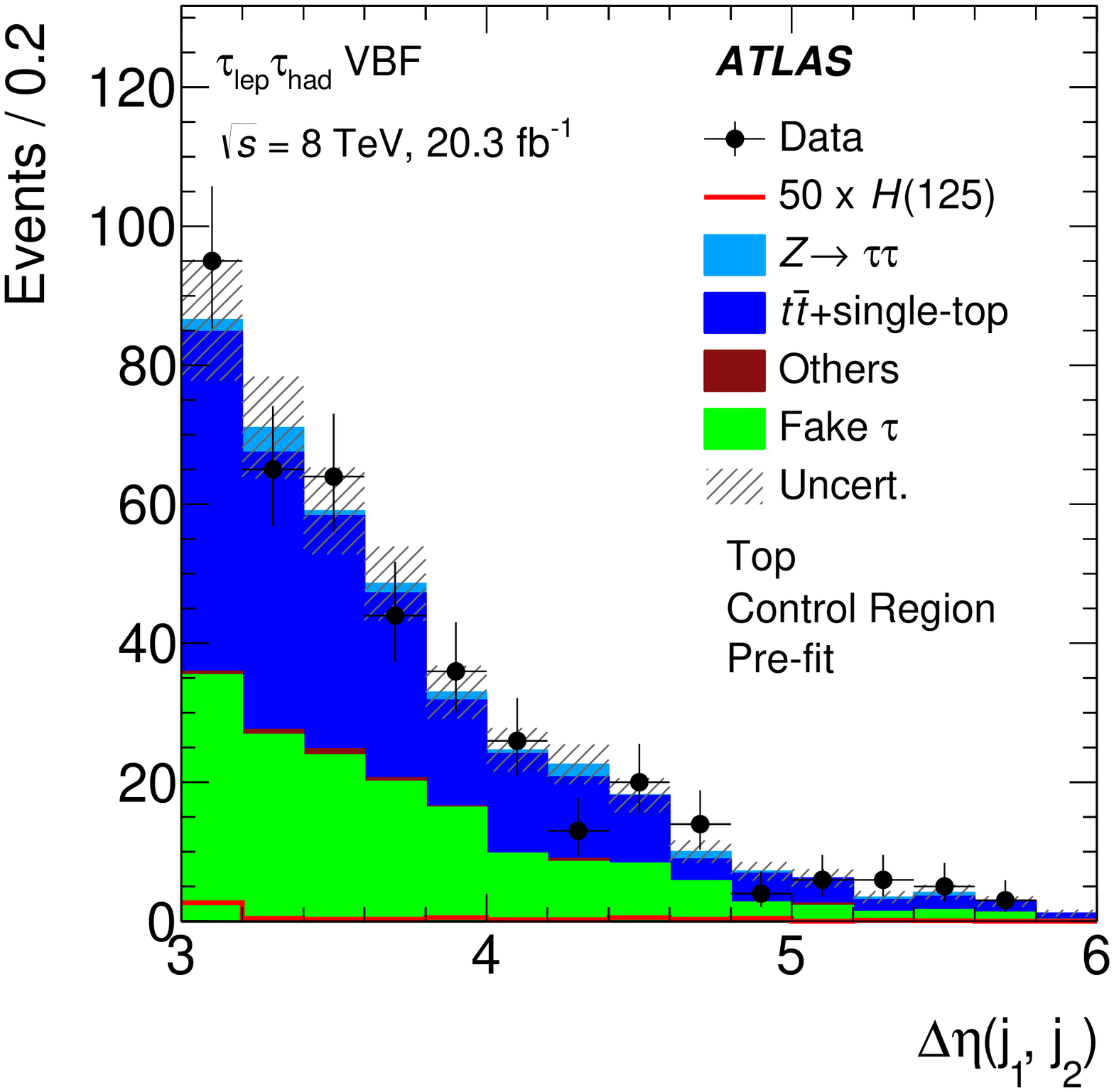}}
\end{center}
\caption{
(a) The distribution of the reconstructed invariant $\tau \tau$ mass, $\MMC$, for events
in the $Z \to \ell \ell$ control region, for the $\tll$ channel. (b) The distribution of the separation in pseudorapidity of the two leading jets,  $\Delta\eta(j_1,j_2)$, for events
in the top control region, for the $\tlhad$ channel. This figure
uses background predictions made without the global fit defined in
section~\ref{sec:fit}.  The error band includes statistical and pre-fit
systematic uncertainties.
\label{fig:background-variables-new}}
\end{figure}

In the $\tll$ channel, the \textsc{Alpgen} simulation is normalised to the data
in the $Z$-mass control region, $80\GeV<m_{\ell\ell}<100\GeV$, for each
category, and separately for $Z\to ee$ and $Z\to \mu\mu$ events.  The
normalisation factors are determined from the final fit described in
section~\ref{sec:fit}.  The distribution of the reconstructed $\tau \tau$ mass
for events in this control region is shown in
figure~\ref{fig:background-variables-new} (a).

In the $\tlhad$ channel, the $Z\to ee$ and $Z\to\mu\mu$ background estimates
are also based on simulation.  The corrections applied for a $\thad$ candidate
depend on whether it originates from a lepton from the $Z$ boson decay or from
a jet.  In the first case, corrections from data, derived from dedicated
tag-and-probe studies, are applied to account for the difference in the rate of
misidentified $\thad$ candidates resulting from leptons \cite{Aad:2012mea, ZtautauXsec}.
This is particularly important for $Z\to ee$ events with a misidentified
$\thad$ candidate originating from a true electron.
In the second case, the fake-factor method described in
section~\ref{subsec:qcdbkg} is applied.

In the $\thadhad$ channel, the contribution of this background is very small
and is taken from simulation.

\subsection{$W+$jets background}

Events with $W$ bosons and jets constitute a background to all channels since
leptonic $W$ decays
can feed into all signatures when the true lepton is accompanied by a jet which
is falsely identified as a $\thad$ or a lepton candidate.  This process can
also contribute via semileptonic heavy quark decays that provide identified
leptons.

As stated in section~\ref{subsec:qcdbkg}, for the $\tll$ and $\tlhad$ channels,
the $\Wjets$ contributions are determined with data-driven methods.
For the $\thadhad$ channel, the $W \to \thad \nu$ background is estimated from
simulation. A correction is applied to account for differences in the $\thad$
misidentification probability between data and simulation.

\subsection{Background from top-quark production}

Background contributions from $t\bar{t}$ and single top-quark production, where
leptons or hadronically decaying taus appear in decays of top quarks, are
estimated from simulation in the $\tll$ and $\tlhad$ channels. The
normalisation is obtained from data control regions defined by requiring a
$b$-jet instead of a $b$-veto. In the $\tlhad$ channel, a large value of the
transverse mass $\mT$ is also required, to enhance the background from
top-quark production and to suppress the signal contribution.  This background is
also found to be small for the  $\thadhad$ channel and it is estimated using
simulation. The distribution of  $\Delta\eta(j_1,j_2)$ for events
in the top control region, for the $\tlhad$ channel, is shown in figure~\ref{fig:background-variables-new} (b).

\subsection{Diboson background}

The production of pairs of vector bosons ($W^+W^-$, $ZZ$ and $W^\pm Z$), with
subsequent decays to leptons or jets, contributes especially to the background
in the $\tll$ channel. For all analysis channels, these contributions are
estimated from simulation, normalised to the NLO cross sections indicated in
table~\ref{tab:MCGenerator}.

\subsection{Contributions from other Higgs boson decays}

In the $\tll$ channel, a non-negligible contribution from $H\to WW \to
\ell\nu\ell\nu$ exists and this process is considered as background. Its
contribution is estimated for $m_{H}=125\GeV$ using simulation.  The
corresponding signal cross section is assumed to be the SM value
and is indicated in table~\ref{tab:MCGenerator}.

\subsection{Validation of background estimates}

As described above, the normalisation for important background sources that are
modelled with simulation are determined by fitting to data in control regions.
These normalisations are compared in table \ref{tab:norms} to predictions
based on the theoretical cross sections for the 8 \TeV\ analysis. 
In most cases, the values obtained are compatible with unity within the statistical uncertainties shown. For the top control region in the
 VBF category of the \tlhad\ channel, the value is also in agreement with unity if the experimental and theoretical systematic uncertainties are included. 
The
control-region normalisations are used for validation plots, and they are used
as starting values in the final global fit described in section \ref{sec:fit}.
The global fit does not change any of these normalisations by more than
2\%.

\begin{figure}[htbp]
  \centering
  \hspace*{\fill}
  \subfigure[]{\includegraphics[width=6.1cm]{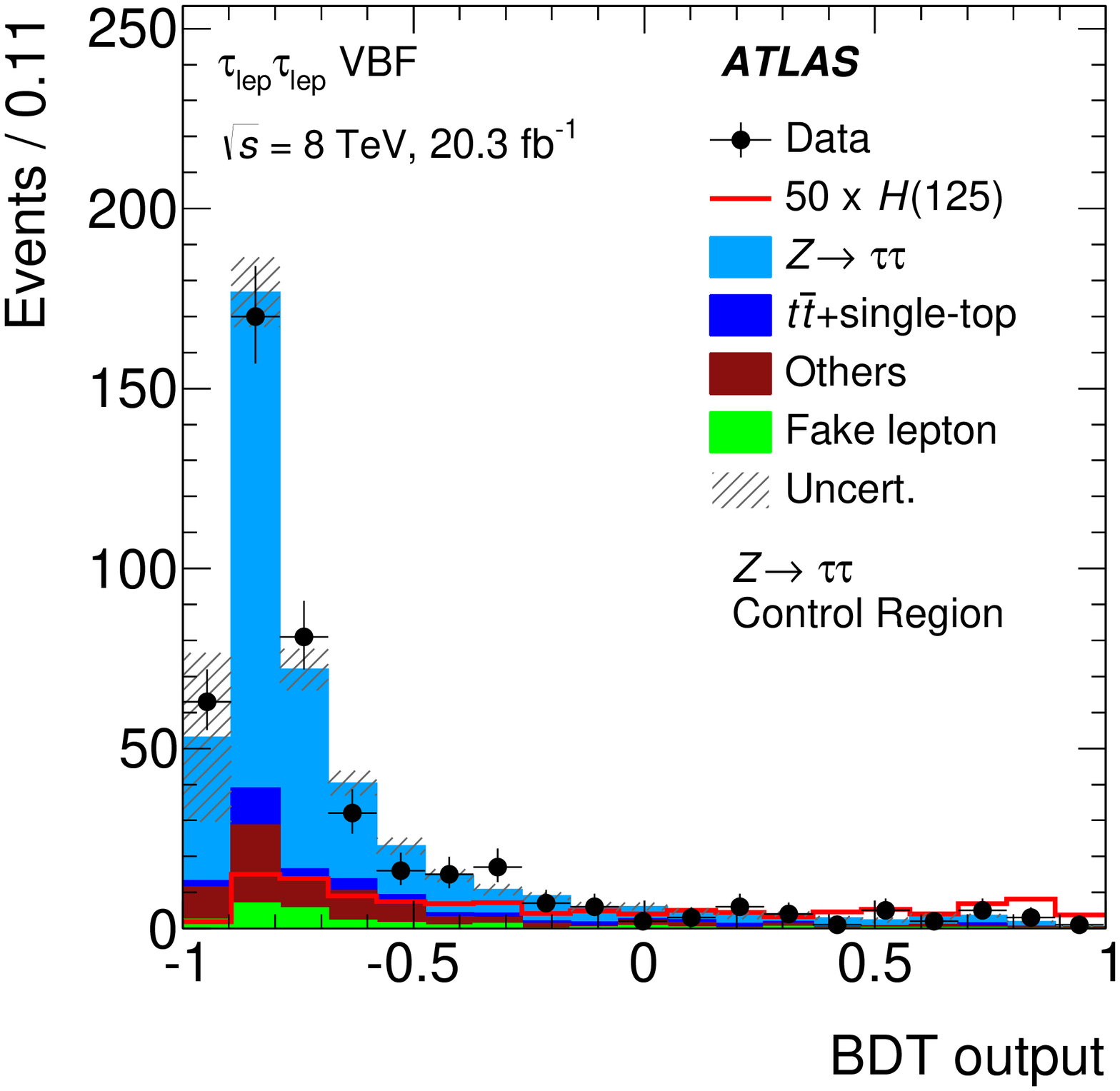}}\hfill
  \subfigure[]{\includegraphics[width=6.1cm]{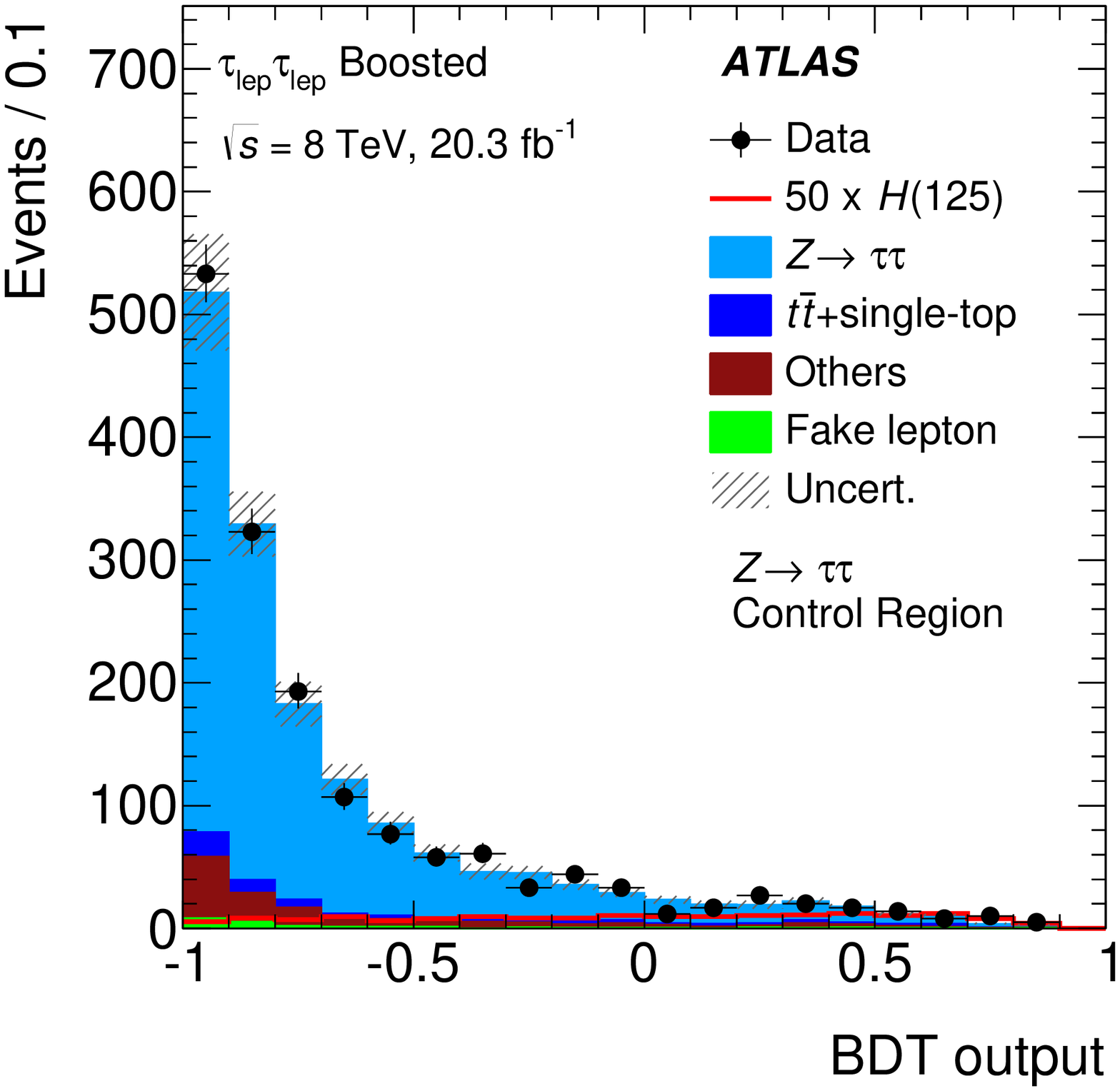}}
  \hspace*{\fill}

  \hspace*{\fill}
  \subfigure[]{\includegraphics[width=6.1cm]{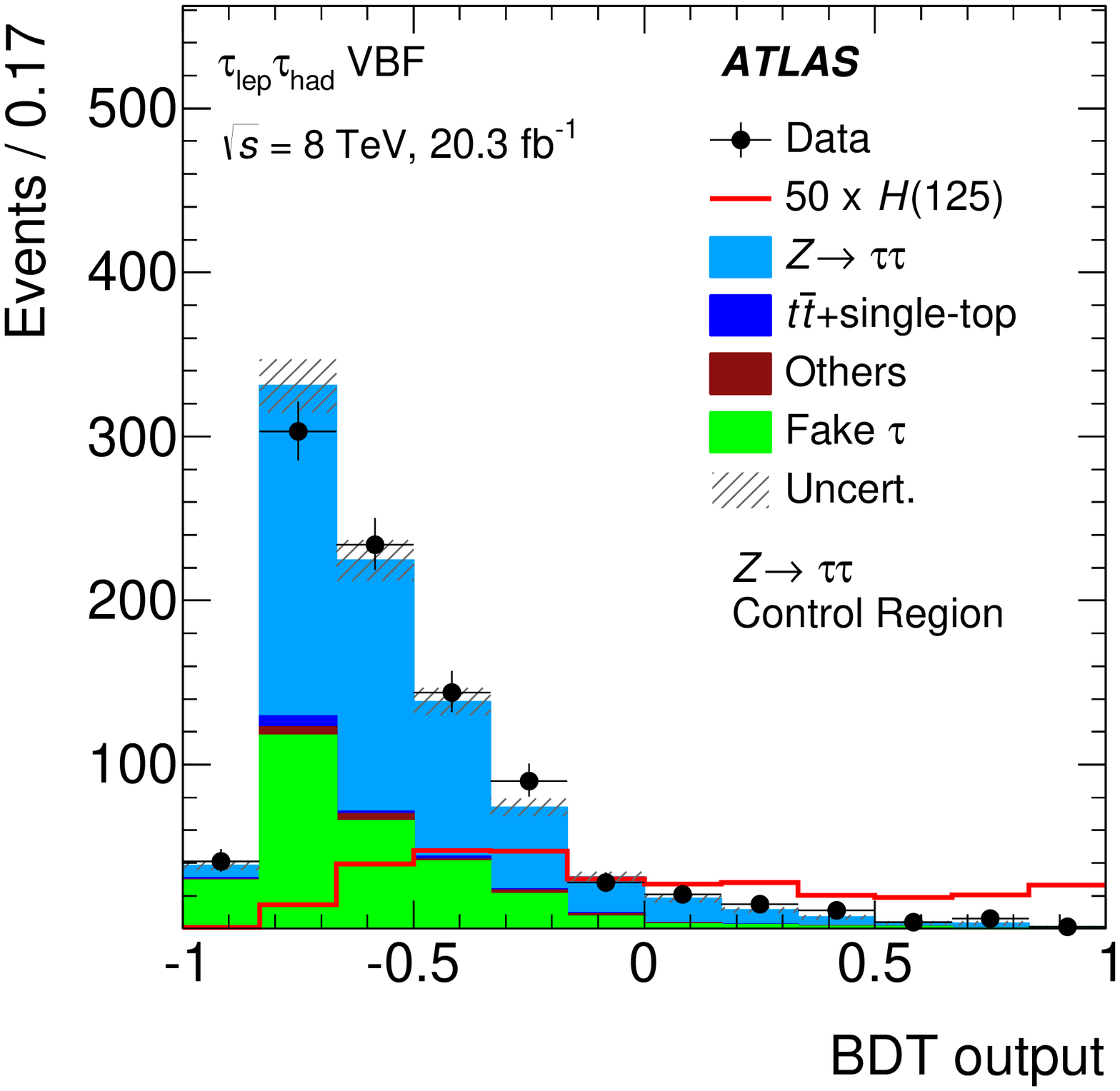}}\hfill
  \subfigure[]{\includegraphics[width=6.1cm]{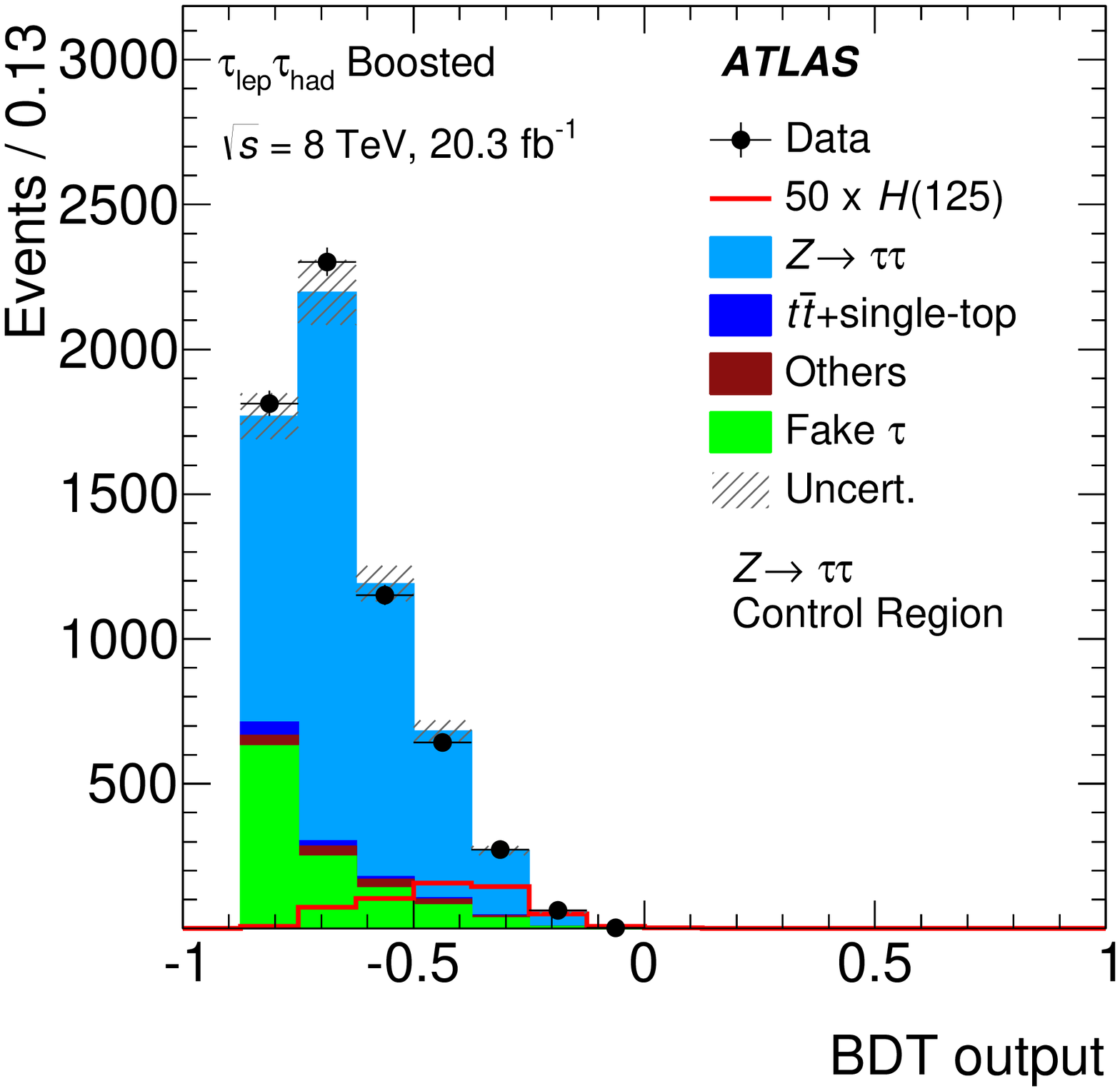}}
  \hspace*{\fill}

  \hspace*{\fill}
  \subfigure[]{\includegraphics[width=6.1cm]{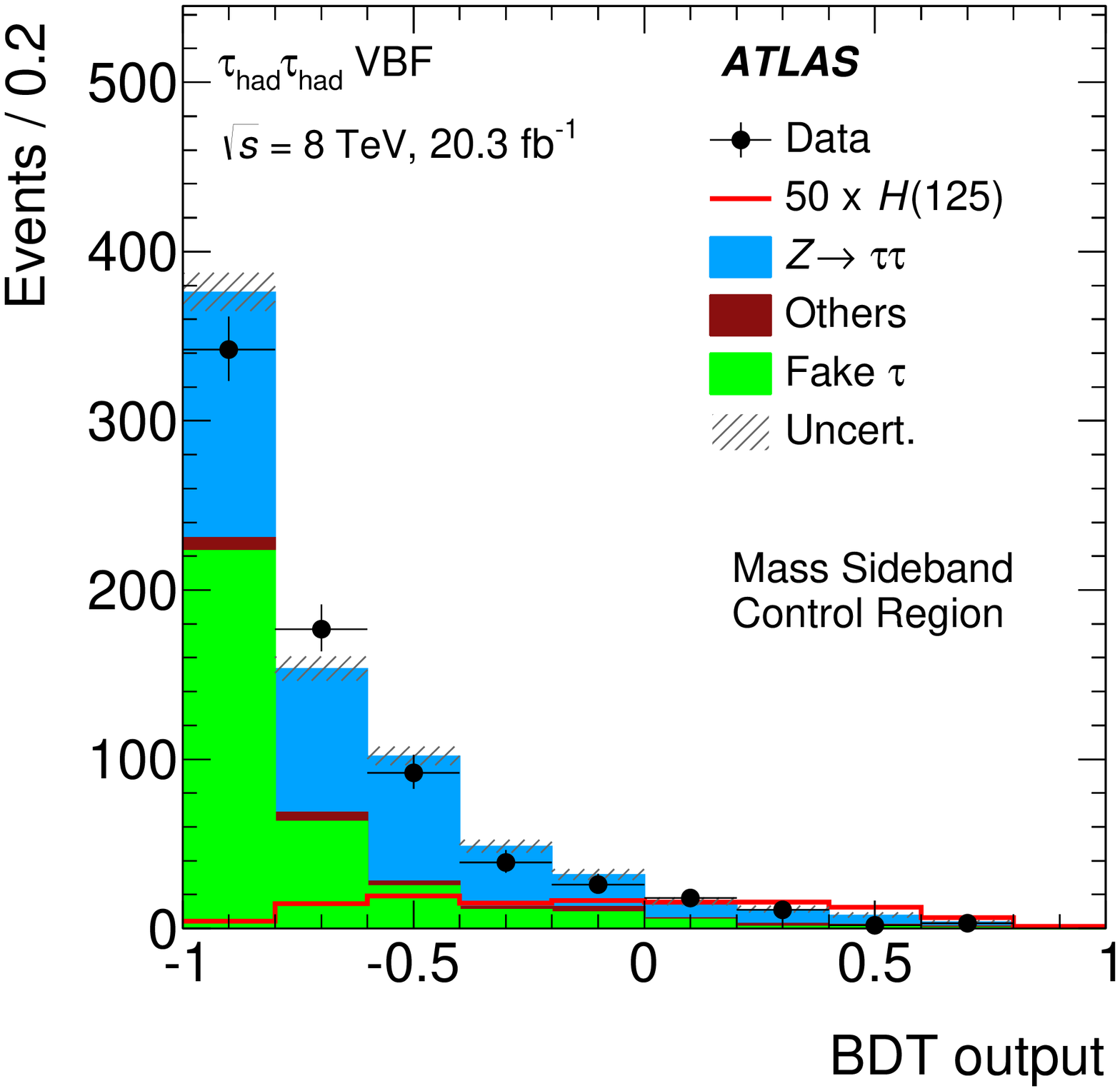}}\hfill
  \subfigure[]{\includegraphics[width=6.1cm]{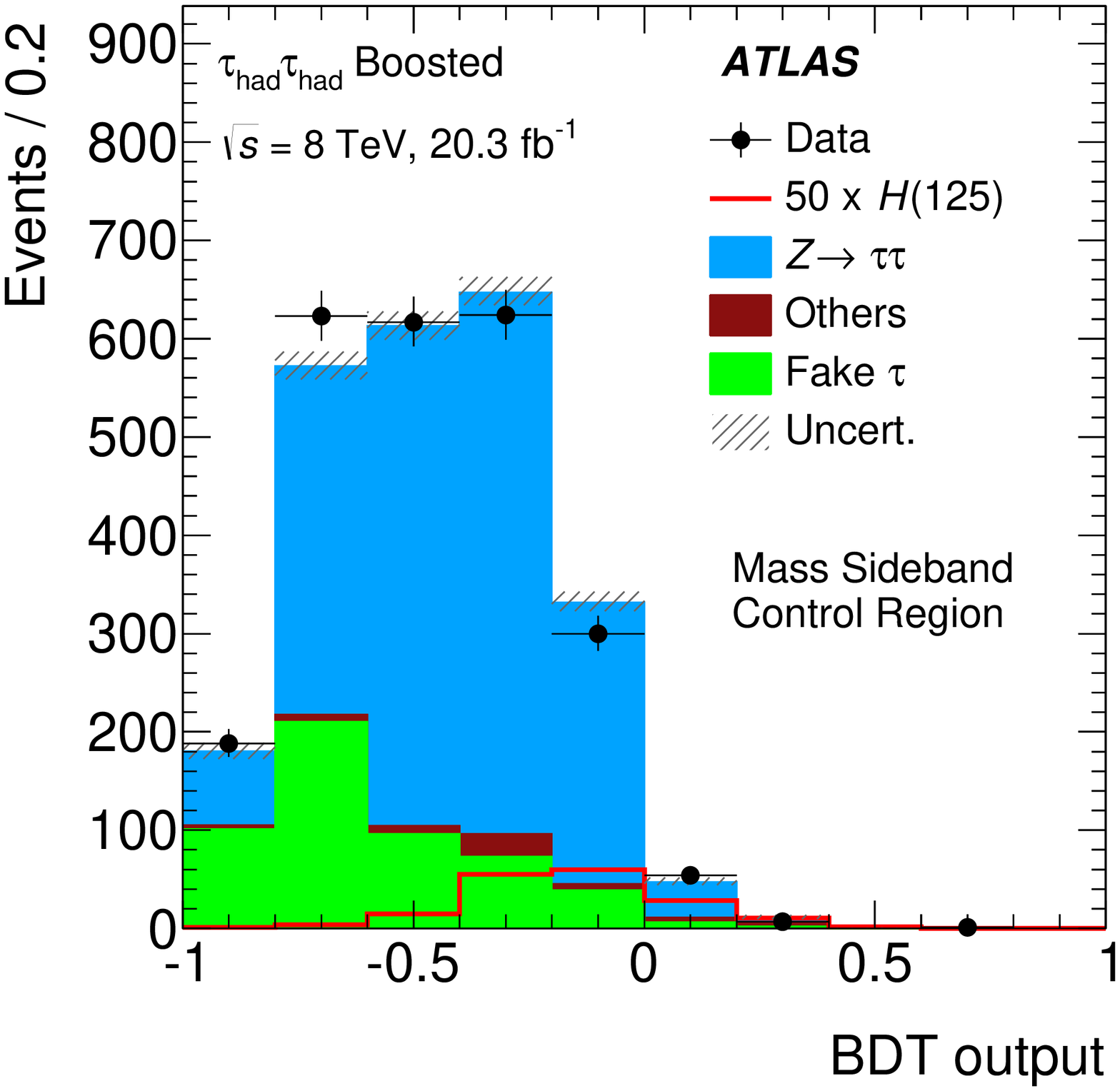}}
  \hspace*{\fill}

  \caption{
Distributions of the BDT output for data collected at $\sqrt {s}=8~\TeV$,
compared to the expected background contributions in important control regions.
The distributions are shown for the VBF (left) and boosted (right)
categories (a,b) for the $Z\to\tau\tau$-enriched control region in the \tll\
channel, (c,d) for the  $Z\to\tau\tau$-enriched control region in the \tlhad\
channel and (e,f) for the $\tau \tau$ invariant mass sideband control region
in the \thadhad\ channel.  The contributions from a Standard Model Higgs boson
with $m_H$~=~125~\GeV\ are superimposed, multiplied by a factor of 50.  These
figures use background predictions made without the global fit defined in
section~\ref{sec:fit}.  The error band includes statistical and pre-fit
systematic uncertainties.
}
  \label{fig:control-bdt}
\end{figure}

It is important to verify that the BDT output distributions in data control
regions are well described after the various background determinations. 
Figure~\ref{fig:control-bdt}
shows distributions from important control regions for the $\sqrt{s}=8\TeV$
dataset, i.e.\ the $Z$-enriched control regions for the  $\tll$  and  $\tlhad$
channels, and the reconstructed $\tau\tau$ invariant mass sideband control
region (defined as $\MMC<100\GeV$ or $\MMC>150\GeV$) for the  $\thadhad$
channel.  The distributions are shown for both the VBF and the 
boosted categories.  All distributions are found to be well described, within
the systematic uncertainties.

\begin{sidewaystable}
\begin{center}
 \begin{tabular}{|c|c|c|c|}
\hline
Process & \tll & \tlhad & \thadhad \\ \hline \hline
$Z \rightarrow \ell\ell$-enriched & $80 < m^\mathrm{vis}_{\tau\tau} < 100~\GeV{}$ & & \\
& (same-flavour) & & \\ \hline
Top control region & Invert $b$-jet veto & Invert $b$-jet veto and $m_\mathrm{T} > 40~\GeV{}$ & \\ \hline
Rest category & & & Pass preselection,  \\
& & &  Fail VBF and  Boosted selections \\ \hline \hline
$Z \rightarrow \tau\tau$-enriched & $m^\mathrm{HPTO}_{\tau\tau} < 100~\GeV{}$  & $m_\mathrm{T} < 40~\GeV{}$ and $\MMC < 110~\GeV{}$ & \\ \hline
Fake-enriched & Same sign $\tau$ decay products & Same sign $\tau$ decay products  & \\ \hline
$W$-enriched & & $m_\mathrm{T} > 70~\GeV{}$ &  \\ \hline
Mass sideband & & & $\MMC < 110~\GeV{}$ or $\MMC > 150~\GeV{}$ \\ \hline
\end{tabular}
\caption{
Summary and definition of the control regions used in the analysis.  The
requirements shown represent modifications to the signal region requirements. 
All other  selections are applied as for the corresponding signal regions.
The variable $m^\mathrm{HPTO}_{\tau\tau}$ is the invariant mass of the
$\tau\tau$-system obtained using the collinear approximation and the same
objects as used in $E_\mathrm{T}^\mathrm{miss,HPTO}$.  Each control region
listed is actually two control regions, corresponding to the VBF or 
boosted categories, with the exception of the rest category.  The first
three control regions listed are used in the global fit defined in
section~\ref{sec:fit}.}
\label{tab:control-summary}
\end{center}
\end{sidewaystable}

\begin{table}[htbp]
\begin{center}
\begin{tabular}{|l |l| cc|}
\hline
Channel & Background &  \multicolumn{2}{c|}{ Scale factors (CR)} \\
&& VBF & Boosted \\
\hline
\tll
& Top & 0.99 $\pm$ 0.07 & 1.01 $\pm$ 0.05 \\
& $Z\rightarrow ee$ & 0.91 $\pm$ 0.16 & 0.98  $\pm$ 0.10\\
& $Z\rightarrow\mu\mu$ & 0.97 $\pm$ 0.13 & 0.96 $\pm$ 0.08 \\
 \hline
 \tlhad & Top  & 0.84 $\pm$ 0.08 &0.96 $\pm$ 0.04 \\
 \hline
\end{tabular}
\caption{
The scale factors calculated in control regions (CR) for background
normalisation. Only the statistical uncertainties are
given. The background contributions shown in validation plots
use the normalisation predicted from simulation multiplied by the
corresponding scale factor. Final normalisations are taken from the global fit,
described in section \ref{sec:fit}.}
\label{tab:norms}
\end{center}
\end{table}

\clearpage

\section{Systematic uncertainties}
\label{sec:sys}

The numbers of expected signal and background events, the input variables to
the BDT, and thereby the BDT output and the final discrimination between signal
and background are affected by systematic uncertainties.  They are discussed
below, grouped into three categories: experimental uncertainties, background
modelling uncertainties, and theoretical uncertainties. 
For all uncertainties, the effects
on both the total signal and background yields and on the shape of the BDT
output distribution are evaluated. Table~\ref{tab:systematics-prefit} gives a
summary of the systematic uncertainties
and their impact on the number of expected events for the signal and the total
background for the analysis of the data taken at $\sqrt{s}$~=~8~\TeV. 
The dominant sources 
that affect the shape of the BDT output distribution are marked in the table. All
uncertainties are treated either as fully correlated or uncorrelated across
channels. The latter are also marked in  table~\ref{tab:systematics-prefit}. The effects of the
systematic uncertainties at $\sqrt{s}$~=~7~\TeV\ are found to be similar and
are not discussed here.  The inclusion of the uncertainties in the profile
likelihood global fit is described in section~\ref{sec:fit} and the effect of
the most significant systematic uncertainties is presented in
table~\ref{tab:systs}.

\subsection{Experimental uncertainties}

The major experimental systematic uncertainties result from uncertainties on
efficiencies for triggering, object reconstruction and identification, as well
as from uncertainties on the energy scale and resolution of jets, hadronically
decaying taus and leptons. In general, the effects resulting from lepton-related
uncertainties are smaller than those from jets and taus. They are not discussed
in detail, however, their impact is included in
table~\ref{tab:systematics-prefit}.  In addition, uncertainties on the
luminosity affect the number of signal and background events from simulation.

\begin{itemize}
\item Luminosity: the uncertainty on the integrated luminosity is $\pm 2.8\%$
    for the 8~\TeV\ dataset and $\pm 1.8\%$ for the 7~\TeV\ dataset.  It is
    determined from a calibration of the luminosity scale derived from
    beam-separation scans performed in 2011 and 2012 using the method described
    in ref.~\cite{LumiPaper}.

\item Efficiencies: the efficiencies for triggering, reconstructing and
    identifying electrons, muons, and $\thad$ candidates are measured in data
    using tag-and-probe techniques.  The uncertainties on the $\thad$
    identification efficiency are $\pm$(2--3)\% for 1-prong and $\pm$(3--5)\%
    for 3-prong tau decays~\cite{ATLAS_Tauperf}.  
    The $b$-jet tagging efficiency has been measured
    from data using \ttbar events, where both top quarks decay to leptons,
    with a total uncertainty of about $\pm$2\% for jets with transverse momenta
    up to 100~GeV \cite{ATLAS-CONF-2014-046, ATLAS-CONF-2014-004}.  The MC
    samples used are corrected for differences in these efficiencies between
    data and simulation and the associated uncertainties are propagated through
    the analysis.

\item Energy scales: the uncertainties on the jet energy scale (JES) arise from
    several sources.  These include, among others, varied response due to the
    jet flavour composition (quark- versus gluon-initiated jets), pile-up,
    $\eta$ intercalibration, and detector response and modelling of in situ
    jet calibration~\cite{Aad:2011he,Aad:2014bia}.  The impact of the JES
    uncertainty in this analysis is reduced because many of the background
    components are estimated using data.
    The tau energy scale is obtained by fitting the reconstructed visible mass
    for $Z\to\tau\tau$ events in data, which can be selected with a
    satisfactory purity. It is measured with a precision of $\pm$(2--4)\%
    \cite{ATLAS-CONF-2013-044}.
    Since systematic uncertainties on the energy scales of all objects affect
    the reconstructed missing transverse momentum, it is recalculated after
    each variation is applied.  The scale uncertainty on $\MET$ due to the energy in calorimeter cells not associated with physics objects is also taken into account.

\item Energy resolutions: systematic uncertainties on the energy resolution of
    taus, electrons, muons, jets, and $\MET$ affect the final discriminant.
    The effects resulting from uncertainties on the tau energy resolution are
    small. The impact of changes in the amount of material (inactive
    material in the detector, e.g. support structures), in the hadronic shower
    model and in the underlying-event tune were studied in the simulation.
    They result in systematic uncertainties below 1$\%$ on the tau energy
    resolution.  The jet energy resolution is determined by in situ
    measurements, as described in ref.~\cite{Aad:2012ag}, and affects signal
    modelling and background components modelled by the simulation.  The
    uncertainty of the  resolution on $\MET$ is estimated by evaluating the
    energy resolution of each of the $\MET$ terms. The largest impact results
    from the soft term (see section~\ref{sec:atlas}), arising both from the MC modelling and the effects of
    pile-up.  It is evaluated using simulated $\Zmumu$ events.

\end{itemize}

\subsection{Background modelling uncertainties}

The most significant systematic uncertainties on the background estimation
techniques, as described in section~\ref{sec:background}, are detailed in the
following for the three decay modes considered.

In the $\tll$ channel, systematic uncertainties on the shape and normalisation
of fake-lepton background sources are estimated by comparing samples of
same-sign lepton events that pass and fail the lepton isolation criteria. These
uncertainties amount to $\pm 33\%$ ($\pm 20\%$) at 8~\TeV\  and  $\pm 10.5\%$
($\pm 13\%$) at 7~\TeV\ for the boosted (VBF) category.  The
extrapolation uncertainty for the $Z \to \ell \ell$ background is obtained by
varying the $m_{\ell\ell}$ window that defines the control region for this
background, and amounts to about $\pm 6\%$. The corresponding
extrapolation uncertainty for top-quark background sources is $\pm$(3--6)$\%$,
obtained from the differences in event yields in the top-quark control regions
when using different MC generators.  Neither of these extrapolation
uncertainties is significant for the final result. The dominant uncertainties on the
normalisation of the \ttbar background, obtained from the global fit, are the systematic 
uncertainties on the $b$-jet tagging efficiency and the jet energy scale.

In the $\tlhad$ channel, an important systematic uncertainty on the background
determination comes from the estimated fake background, for which several sources
of systematic uncertainty are considered.  The statistical uncertainty on the
effective fake factor is $\pm 4.3\%$ ($\pm 2.3\%$) in the 8~\TeV\  VBF
(boosted) category, and about $\pm 22\%$ ($\pm 11\%$) in the 7~\TeV\ VBF (boosted)
 category.  The dominant systematic uncertainty on the
methodology itself arises from the composition of the combined fake background
($W$+jets, $Z$+jets, multijet, and $t\bar{t}$ fractions), which is largely estimated based on simulated event samples as explained in
section~\ref{subsec:qcdbkg}.  The uncertainty is estimated by varying each
fractional contribution by $\pm 50\%$, which affects the effective fake factor
by $\pm 3\%$ ($\pm 6\%$) and by $\pm 10\%$  ($\pm 15\%$)  in the 8~\TeV\ and
7~\TeV\  boosted (VBF) categories respectively.  As a closure test,
the method was also applied in a region of data where the lepton
and \thad\ candidate have the same charge, rich in fake \thad\ candidates. Very good agreement was observed
between data and the method's prediction, so that no additional in situ
uncertainty was deemed necessary.  
In addition, the uncertainties on the
normalisation of the \ttbar background  are
important. As in the case of the  $\tll$ channel,  the dominant contribution obtained from the global fit 
originates from systematic uncertainties on the $b$-jet tagging efficiency and the jet energy scale, 
along with statistical uncertainties on the observed data in the respective control regions.

In the $\thadhad$ channel, the major background from multijet production is
determined using a data-driven template method. The default multijet template,
derived from a sample in data where the \thad\ candidates fail the isolation
and opposite-sign charge requirements, is compared with an alternative template
derived from a sample where the  \thad\ candidates fail just the opposite-sign
charge requirement.  The normalisation of the alternative template is fixed to
that of the default template at preselection; the alternative multijet template is
propagated into the various categories and gives a different set of yields from
the default template. This difference,
along with the difference in shape between the two templates, constitutes the
systematic uncertainty on the background estimate.  This leads to an overall
multijet yield variation of 10\% (3\%) in the VBF (boosted)
category at $\sqrt{s}$\,=\,8\,\TeV\ and of 10\% (30\%) in the VBF
(boosted) category $\sqrt{s}$\,=\,7\,\TeV. However, there is a very
strong shape dependence, such that the uncertainties on the BDT output are much
larger at higher output values.

For the embedding method used in all channels, the major systematic
uncertainties are related to the selection of $Z \to \mu \mu$  events in data
and to the subtraction of the muon energy deposits in the calorimeters. The
selection uncertainties are estimated by varying the muon isolation criteria in
the selection from the nominal value of $I(\pT,0.2) <$ 0.2 (see
section~\ref{sec:selection}) to tighter ($I(\pT,0.4) <$ 0.06 and $I(\ET,0.2) <
$ 0.04) and looser (no isolation requirements) values.  The muon-related cell
energies to be subtracted are varied within $\pm20$\% ($\pm30$\%) for the
8~\TeV\ (7~\TeV) data.
In addition, systematic uncertainties on the corrections for trigger and
reconstruction efficiencies are taken into account. Due to the
combination of single-lepton and dilepton triggers used, the uncertainties are largest for the
$\tll$ channel.  All experimental systematic uncertainties relating to the embedded $\tau$
decay products (such as tau energy scale or identification uncertainties) are
applied normally.  The combined effect of all uncertainties on the signal and
background yields is included in table~\ref{tab:systematics-prefit}.  Because
the $Z \to \tau \tau$ normalisation is determined in the final fit, the impact
on the final result is much smaller.

\subsection{Theoretical uncertainties}

Theoretical uncertainties are estimated for the signal and for all
background contributions modelled with the simulation.  Since the major
background contributions, from $Z\to \tau\tau$ and misidentification of
hadronically decaying $\tau$ leptons, are estimated using data-driven methods, they are
not affected by these uncertainties.  Uncertainties on the signal cross
sections are assigned from missing higher-order corrections, from uncertainties
in the PDFs, and from uncertainties in the modelling of the underlying event.

For VBF and {\em VH} Higgs boson production cross sections, the
uncertainties due to missing higher-order QCD corrections are estimated by
varying the factorisation and renormalisation scales by factors of two around
the nominal scale $m_{W}$, as prescribed by the LHC Higgs Cross Section Working
Group~\cite{LHCHiggsCrossSectionWorkingGroup:2011ti}. The resulting
uncertainties range from $\pm2\%$ to $\pm4\%$, depending on the process and the
category-specific selection considered.  In addition, a $2\%$ uncertainty
related to the inclusion of the NLO EWK corrections (see
section~\ref{sec:samples}) is assigned.

For Higgs boson production via ggF, the uncertainties on the cross sections
 due to missing higher order QCD corrections 
 are estimated by varying the
renormalisation and factorisation scales around the central values
$\mu_\mathrm{R} = \mu_\mathrm{F} = \sqrt{m_{H}^2 + \pt^2}$ in the NLO
cross section calculations of $H+1$-jet and $H+2$-jet production. In the
calculation of the uncertainties, appropriate cuts on the Higgs $\pT$  ($\pT^H >$ 100~\GeV\ )
 and on the jet kinematics ($\Delta \eta, \pT$) are applied at parton level for the boosted and
VBF categories respectively.  The resulting uncertainties on the ggF
contributions are found to be about $\pm24\%$ in the boosted category and
$\pm23\%$ in the VBF category. The ggF contribution is
dominant in the  boosted category, whereas it is only about 
$20\%$ of the signal in the  VBF category.  Since the two categories are
exclusive, their anti-correlation is taken into account
following the prescription in ref.~\cite{StewartAndTackmann}.

In the present analysis, no explicit veto on jets is applied in the VBF
selection, but enough kinematical information is provided as input to the BDT
so that the high BDT-output region corresponds to a more exclusive region,
where the probability of finding a third jet is reduced.  Since the cross
section for gluon-fusion events produced with a third jet is only known at LO,
this could introduce a large uncertainty on the gluon-fusion contamination in
the highest (and most sensitive) BDT-output bins. The uncertainty on the BDT
shape of the ggF contribution is evaluated using the \textsc{Mcfm}
Monte Carlo program~\cite{MCFMVV}, which calculates $H$ + 3 jets at LO.  Scale
variations induce changes of the ggF contribution in the highest BDT bin
of about $\pm 30\%$. They are taken into account in the final fit.

Uncertainties related to the simulation of the underlying event and parton
shower are estimated by comparing the acceptance from \textsc{Powheg+Pythia} to
\textsc{Powheg+Herwig} for both VBF and ggF Higgs boson production modes.
Differences in the signal yields range from $\pm$1\% to $\pm$8\% for the 
VBF and from $\pm$1\% to $\pm$9\% for  ggF production, depending on the
channel and category.  The BDT-score distribution of the \textsc{Powheg+Pythia}
and \textsc{Powheg+Herwig} samples are compatible with each other within
statistical uncertainties.

The PDF uncertainties are estimated by studying the change in the acceptance
when using different PDF sets or varying the \textsc{CT10} PDF set within its
uncertainties.  The standard VBF \textsc{Powheg} sample and a
\textsc{MC@NLO} ~\cite{MCatNLO} ggF sample, both generated with the
\textsc{CT10} PDFs, are reweighted to the MSTW2008NLO ~\cite{top:Martin}, NNPDF
~\cite{Ball:2011mu} and the CT10 eigen-tunes parameterisation.  The largest
variation in acceptance for each category is used as a constant PDF uncertainty; it
varies between approximately $\pm$4.5\% and $\pm$6\% for ggF production
and between about $\pm$0.8\% and $\pm$1.0\% for VBF production.  A shape
uncertainty is also included to cover any difference between the BDT score in
the default sample and the reweighted ones.  The uncertainty on the total
cross section for the VBF, {\em VH} and ggF production modes due to the PDFs
is also considered.

Variations in the acceptance for different Monte Carlo generators are also included,
comparing  \textsc{Powheg+Herwig} samples to  \textsc{MC@NLO+Herwig} samples for
ggF, and to  \textsc{aMC@NLO+Herwig}~\cite{Alwall:2014hca} samples for  VBF. The
generator modelling uncertainty is around $\pm2\%$ for ggF and $\pm4\%$ for
VBF productions modes.

Finally, an uncertainty on the decay branching ratio, BR($\htautau$), of $\pm
5.7\%$~\cite{Dittmaier:12013084} affects the signal rates.

The theoretical systematic uncertainties on the background predictions taken
from the simulation are evaluated by applying the same procedures as used for
the signal samples. Uncertainties resulting from the choice of QCD scales,
 PDF parameterisation and underlying-event model are estimated.  The
results are reported in table~\ref{tab:systematics-prefit}.

\begin{sidewaystable}[h]
\begin{center}
\caption{
Impact of systematic uncertainties on the total signal, $S$, (sum of all
production modes) and on the sum of all background estimates, $B$, for each of
the three channels and the two signal categories for the analysis of the data
taken at $\sqrt{s}$~=~8~\TeV.  Each systematic uncertainty is assumed to be 
correlated across the analysis channels, except those marked with a *.
Uncertainties that affect the shape of the BDT-output distribution in a
non-negligible way are marked with a $\dagger$. All values are given before the
global fit. The notation UE/PS refers to the underlying event and parton shower modelling.
}
\vspace{.2cm}
\small
\begin{tabular}{|p{5cm}|c|c|c|c|c|c|c|c|c|c|c|c|}
\hline
\multirow{4}{*}{Source} & \multicolumn{12}{c|}{Relative signal and background variations [\%]} \\ \cline{2-13}
& \multicolumn{2}{c|}{\tll } & \multicolumn{2}{c|}{\tll} & \multicolumn{2}{c|}{\tlhad} & \multicolumn{2}{c|}{ \tlhad } & \multicolumn{2}{c|}{\thadhad } & \multicolumn{2}{c|}{\thadhad} \\
& \multicolumn{2}{c|}{ VBF} & \multicolumn{2}{c|}{ Boosted} & \multicolumn{2}{c|}{ VBF} & \multicolumn{2}{c|}{ Boosted} & \multicolumn{2}{c|}{ VBF } & \multicolumn{2}{c|}{Boosted} \\

\cline{2-13}
&$S$&$B$&$S$&$B$&$S$&$B$&$S$&$B$&$S$&$B$&$S$&$B$ \\\hline

\textbf{{\small Experimental}} & \multicolumn{12}{c|}{} \\ \cline{2-13}
\hline

Luminosity         &$\pm$2.8 &$\pm$0.1 &$\pm$2.8 &$\pm$0.1 &$\pm$2.8 &$\pm$0.1 &$\pm$2.8 &$\pm$0.1 &$\pm$2.8  &$\pm$0.1 &$\pm$2.8 & $\pm$0.1 \\
Tau trigger*        &   --     &     --   &   --    &   --   & --   &   -- &   --  &  --   &$^{+7.7}_{-8.8}$ & $<0.1$  & $^{+7.8}_{-8.9}$ & $<0.1$ \\
Tau identification  &   --     &     --   &  --      &   --     &$\pm$3.3 &$\pm$1.2 &$\pm$3.3 &$\pm$1.8 &$\pm$6.6  &$\pm$3.8 &$\pm$6.6 & $\pm$5.1 \\
Lepton ident. and trigger* &$^{+1.4}_{-2.1}$ & $^{+1.3}_{-1.7}$ & $^{+1.4}_{-2.1}$ & $^{+1.1}_{-1.5}$
                           &$\pm$1.8 &$\pm$0.5 &$\pm$1.8 &  $\pm$ 0.8  &  --  &   --  &  --  &  -- \\
$b$-tagging          &$\pm$1.3 &$\pm$1.6 &$\pm$1.6 &$\pm$1.6 & $<0.1$ & $\pm$0.2 &$\pm$0.4 & $\pm$0.2 &  -- &   -- &  -- &   --  \\
$\tau$ energy scale$\dagger$  &-- & -- & -- & -- &$\pm$2.4 & $\pm$1.3 & $\pm$2.4 & $\pm$0.9  & $\pm$2.9& $\pm$2.5  &$\pm$2.9 & $\pm$2.5 \\
Jet energy scale and resolution$\dagger$  &$^{+8.5}_{-9.1}$ & $\pm$9.2 &$^{+4.7}_{-4.9}$ &$^{+3.7}_{-3.0}$
                                       &$^{+9.5}_{-8.7}$ & $\pm$1.0       &$\pm$3.9 &$\pm$0.4 &$^{+10.1}_{-8.0}$ & $\pm$0.3      &$^{+5.1}_{-6.2}$ & $\pm$0.2 \\
\MET\ soft scale \& resolution & $^{+0.0}_{-0.2}$ & $^{+0.0}_{-1.2}$ & $ ^{+0.0}_{-0.1}$ & $^{+0.0}_{-1.2}$ & $^{+0.8}_{-0.3}$ & $\pm 0.2$ & $\pm 0.4$ & $ < 0.1$ & $ \pm 0.5$ & $\pm 0.2 $ & $ \pm 0.1$ & $< 0.1$\\
 \hline
\textbf{{\small Background Model}} & \multicolumn{12}{c|}{} \\
\hline
Modelling of fake backgrounds*$\dagger$ & -- &  $\pm$ 1.2 &-- &$\pm$1.2  &-- &$\pm$2.6 &-- &$\pm$2.6 &-- & $\pm$5.2   & --  & $\pm$0.6\\
Embedding$\dagger$  & -- & $^{+3.8}_{-4.3}$ & -- & $^{+6.0}_{-6.5}$  & --  & $\pm$1.5 &  --  & $\pm$1.2 &  --  & $\pm$2.2 & --  & $\pm$3.3 \\
$Z\to\ell\ell$ normalisation* & -- & $\pm$ 2.1 & --  &$\pm$0.7   &-- &-- & -- & -- & -- &-- & -- & --\\
\hline
\textbf{Theoretical} & \multicolumn{12}{c|}{} \\ \cline{2-13}
\hline
Higher-order QCD corrections$\dagger$   &$^{+11.3}_{-9.1}$ & $\pm$0.2  & $^{+19.8}_{-15.3}$ & $\pm$0.2
                       &$^{+9.7}_{-7.6}$  & $\pm$0.2  & $^{+19.3}_{-14.7}$ & $\pm$0.2
                        &$^{+10.7}_{-8.2}$  & $<0.1$ & $^{+20.3}_{-15.4}$ & $<0.1$   \\
UE/PS &  $\pm$ 1.8 & $<0.1$ & $\pm$ 5.9   &   $<0.1$
                                  & $\pm$3.8 & $<0.1$ &  $\pm$2.9  &    $<0.1$
                                  & $\pm$4.6 &$<0.1$  & $\pm$3.8 & $<0.1$  \\
Generator modelling      & $\pm$2.3  & $<0.1$ & $\pm$1.2    &   $<0.1$
            &  $\pm$2.7 & $<0.1$ &  $\pm$1.3  & $<0.1$
            & $\pm$2.4 &$<0.1$  & $\pm$1.2 & $<0.1$  \\
EW corrections  & $\pm$1.1  & $<0.1$ & $\pm$0.4     &   $<0.1$
               &  $\pm$1.3 & $<0.1$ &  $\pm$0.4  & $<0.1$
               & $\pm$1.1  &$<0.1$  & $\pm$0.4  & $<0.1$  \\
PDF$\dagger$              &  $^{+4.5}_{-5.8}$  & $\pm$ 0.3  & $^{+6.2}_{-8.0}$  & $\pm$ 0.2
                 &  $^{+3.9}_{-3.6}$  & $\pm$ 0.2  & $^{+6.6}_{-6.1}$   & $\pm$ 0.2
                 &   $^{+4.3}_{-4.0}$ &$\pm$ 0.2 & $^{+6.3}_{-5.8}$ & $\pm$ 0.1 \\
BR~($H\to\tau\tau$)&  $\pm$ 5.7  &-- &$\pm$5.7  &--
                   & $\pm$5.7& --&$\pm$5.7 &--
                   & $\pm$ 5.7 & --  &$\pm$5.7 & -- \\
\hline
\end{tabular}
\label{tab:systematics-prefit}
\end{center}
\end{sidewaystable}

\clearpage
\section{Signal extraction procedure}
\label{sec:fit}

The BDT output in the six analysis categories provides
the final discrimination between signal and background for both
the 7 and 8~TeV datasets.
A maximum-likelihood fit is performed on all categories simultaneously to
extract the signal strength, $\mu$, defined as the
ratio of the measured signal yield to the Standard Model expectation.
The value $\mu=0$ ($\mu=1$) corresponds to the
absence (presence) of a Higgs boson signal with the SM production cross section.
The statistical analysis of the data employs a binned likelihood
function $\mathcal{L} (\mu,\vec{\theta})$, constructed as the product of
Poisson probability terms, to estimate $\mu$.

The impact of systematic uncertainties on the signal and background
expectations is described by nuisance parameters, $\vec{\theta}$, which
are each parameterised by a Gaussian or log-normal constraint. The expected
numbers of signal and background events in each bin are functions of $\vec{\theta}$.
The test statistic $q_\mu$ is then constructed according to
the profile likelihood ratio: $q_\mu = - 2 \ln [\mathcal{L} (\mu,
\hat{\hat{\vec{\theta}}})/\mathcal{L} (\hat{\mu}, \hat{\vec{\theta}})]$, where
$\hat{\mu}$ and $\hat{\vec{\theta}}$ are the parameters that maximise the
likelihood, and
$\hat{\hat{\vec{\theta}}}$ are the nuisance parameter values that maximise the
likelihood for a given $\mu$.
This test statistic is used to measure
the compatibility of the background-only hypothesis with the observed
data.

The likelihood is maximised on the BDT distributions in the signal regions,
with information from control regions included to constrain background normalisations.
The fit includes the event yields from the $\Zll$ and top control regions in the
 $\tll$ channel, and from the top control region of the $\tlhad$ channel; furthermore the
$\Delta \eta (\thad,\thad)$ distribution in the rest control region of the $\thadhad$ channel is also included.

The $Z\to\tau\tau$ background is constrained
primarily in the signal regions, due to the difference between the BDT distributions
for $Z\to\tau\tau$ events and the signal. For the
$\thadhad$ channel, the $Z\to\tau\tau$ and multijet background rates are constrained by the simultaneous
fit of the two signal regions and the $\Delta \eta (\thadhad)$ distribution in the rest category control region.
 The top and $\Zll$ background components for the
$\tll$ and $\tlhad$ channels are also allowed to float freely, but are primarily constrained
by the inclusion of the respective control regions.

\begin{figure}[tb!]
\begin{center}
\includegraphics[width=0.6\textwidth]{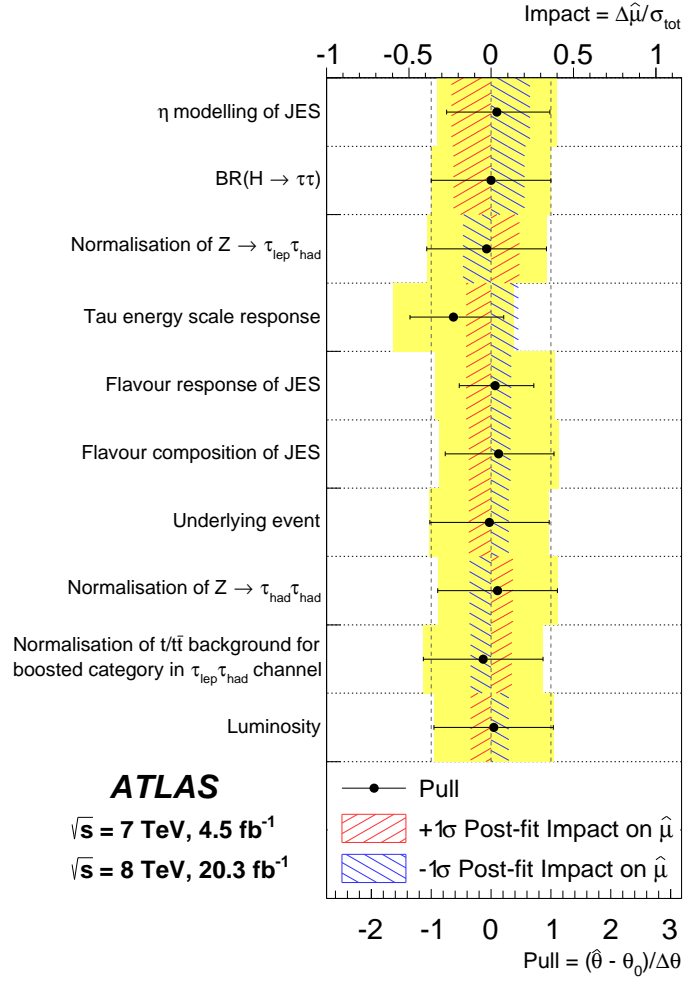}
\end{center}
\caption{
Impact of systematic uncertainties on the fitted signal-strength parameter
$\hat\mu$ for the combined fit for all channels and both centre-of-mass
energies. The systematic uncertainties are listed in decreasing order of their
impact on $\hat\mu$ on the $y$-axis. The hatched blue and red boxes show the
variations of $\hat\mu$ with respect to the total error on $\mu$,
$\sigma_\mathrm{tot}$, referring to the top $x$-axis, when fixing the corresponding
individual nuisance parameter $\theta$ to its post-fit value $\hat\theta$
modified upwards or downwards by its post-fit uncertainty, and repeating the
fit. The filled circles, referring to the bottom $x$-axis, show the pulls of
the fitted nuisance parameters, i.e.\ the deviations of the fitted parameters
$\hat\theta$ from their nominal values $\theta_0$, normalised to their nominal
uncertainties $\Delta\theta$.  The black lines show the post-fit uncertainties
of the nuisance parameters, relative to their nominal uncertainties, which are
indicated by the yellow band. The jet energy scale uncertainties are decomposed
into uncorrelated components.
}
\label{fig:nuisance-parameters}
\end{figure}

As described in section~\ref{sec:sys}, a large number of systematic uncertainties, taken into account via
nuisance parameters, affect the final results. It is important to investigate the behaviour of the global fit
and in particular to investigate how far the nuisance parameters are pulled away from their nominal values and
how well their uncertainties are constrained. Furthermore, it is important to understand which systematic uncertainties
have the most impact on the final result. For this purpose a ranking of nuisance parameters is introduced. For each parameter,
the fit is performed again with the parameter fixed to its fitted value shifted up or down by its fitted
 uncertainty, with all the other parameters allowed to vary.
The ranking obtained for those nuisance parameters contributing most to the uncertainty on the signal strength
is shown in figure~\ref{fig:nuisance-parameters} for the combined fit of the three channels at the
two centre-of-mass energies.
The parameters contributing most
are those related to the jet energy scale, the normalisation uncertainties for $Z\to\tau\tau$  and top-quark events, and the tau energy scale.
The uncertainties on the jet energy scale are decomposed into several uncorrelated
components  (among others: $\eta$ intercalibration of different calorimeter regions, jet energy response, and
response to jets of different flavour).
In addition, theoretical uncertainties on the branching ratio BR~($\htautau$)  are found to have a significant impact.
In general, good agreement is found between the pre-fit and post-fit values for these nuisance parameters and neither
large pulls nor large constraints are observed.

\begin{figure}[htbp]
\centering
\hspace*{\fill}
\subfigure[]{\includegraphics[width=6.1cm]{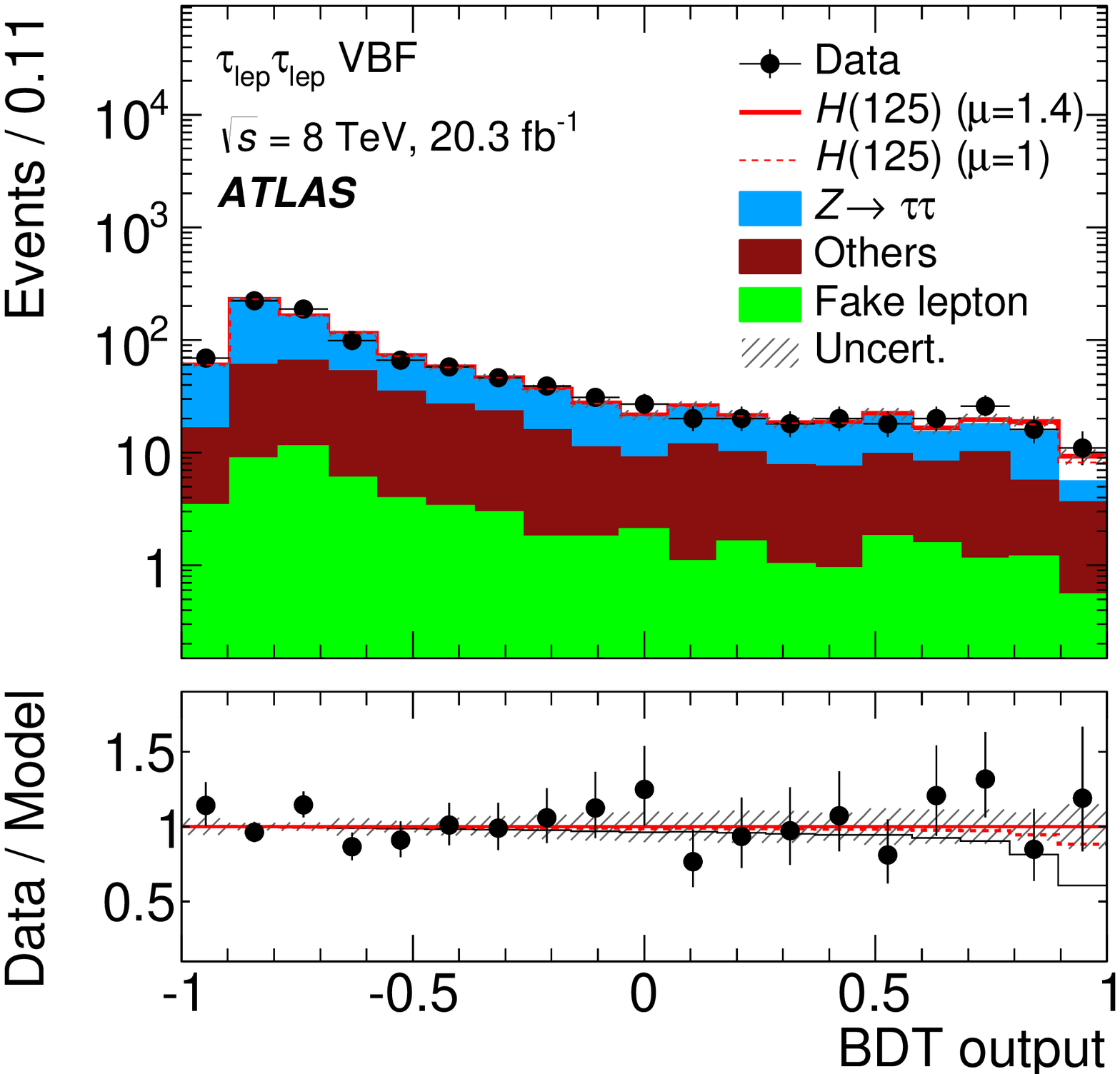}}\hfill
\subfigure[]{\includegraphics[width=6.1cm]{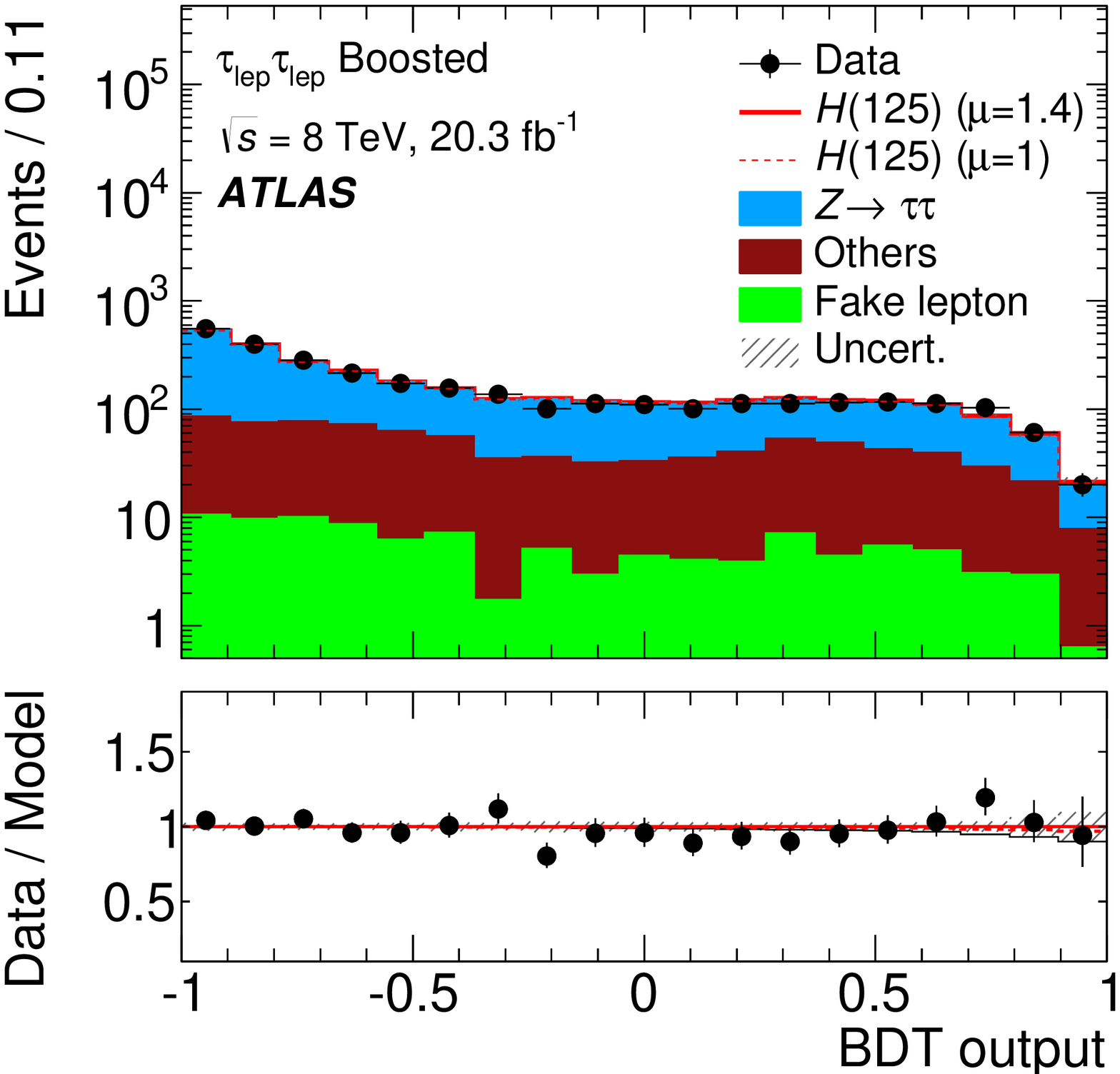}}
\hspace*{\fill}

\hspace*{\fill}
\subfigure[]{\includegraphics[width=6.1cm]{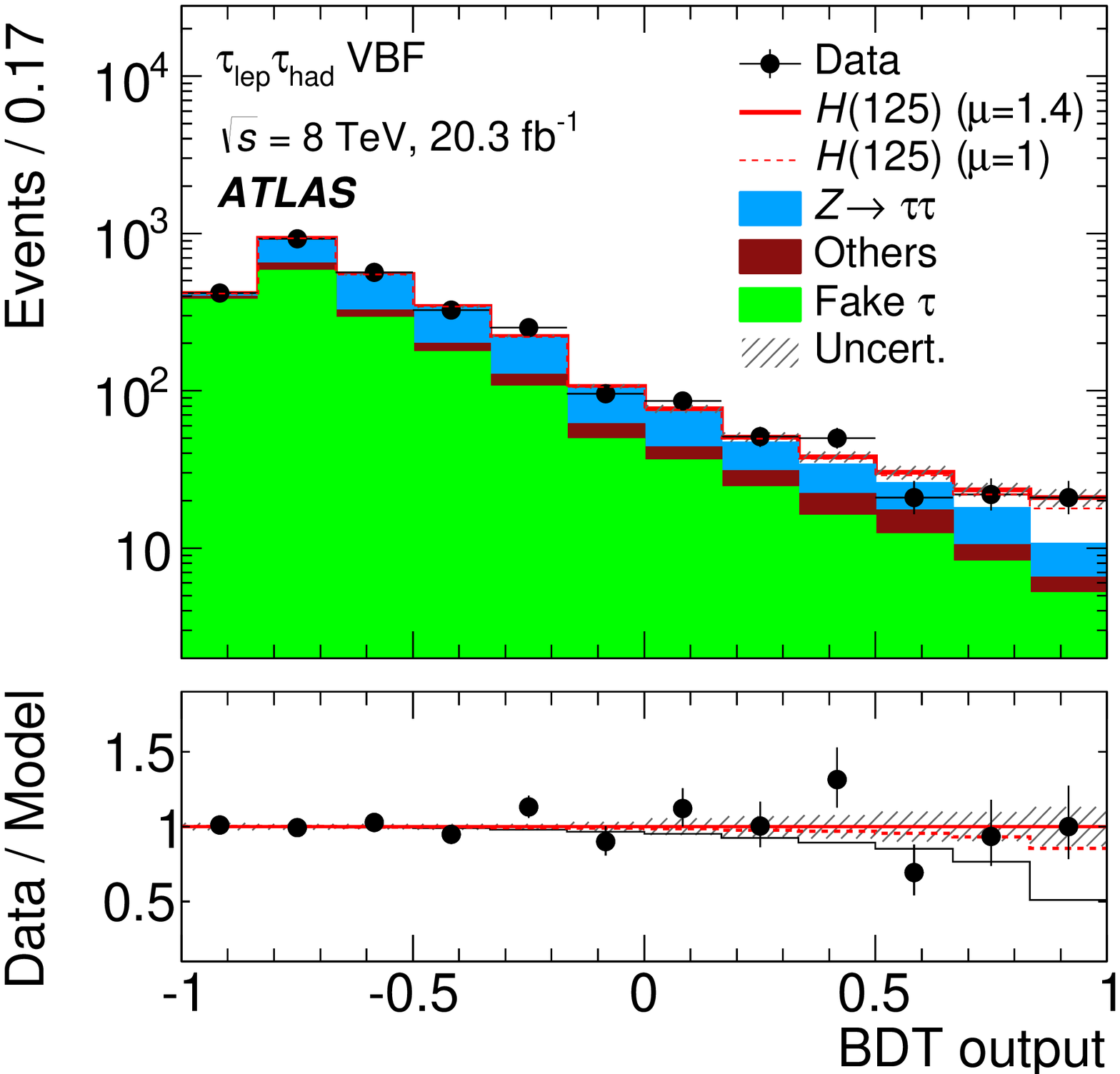}}\hfill
\subfigure[]{\includegraphics[width=6.1cm]{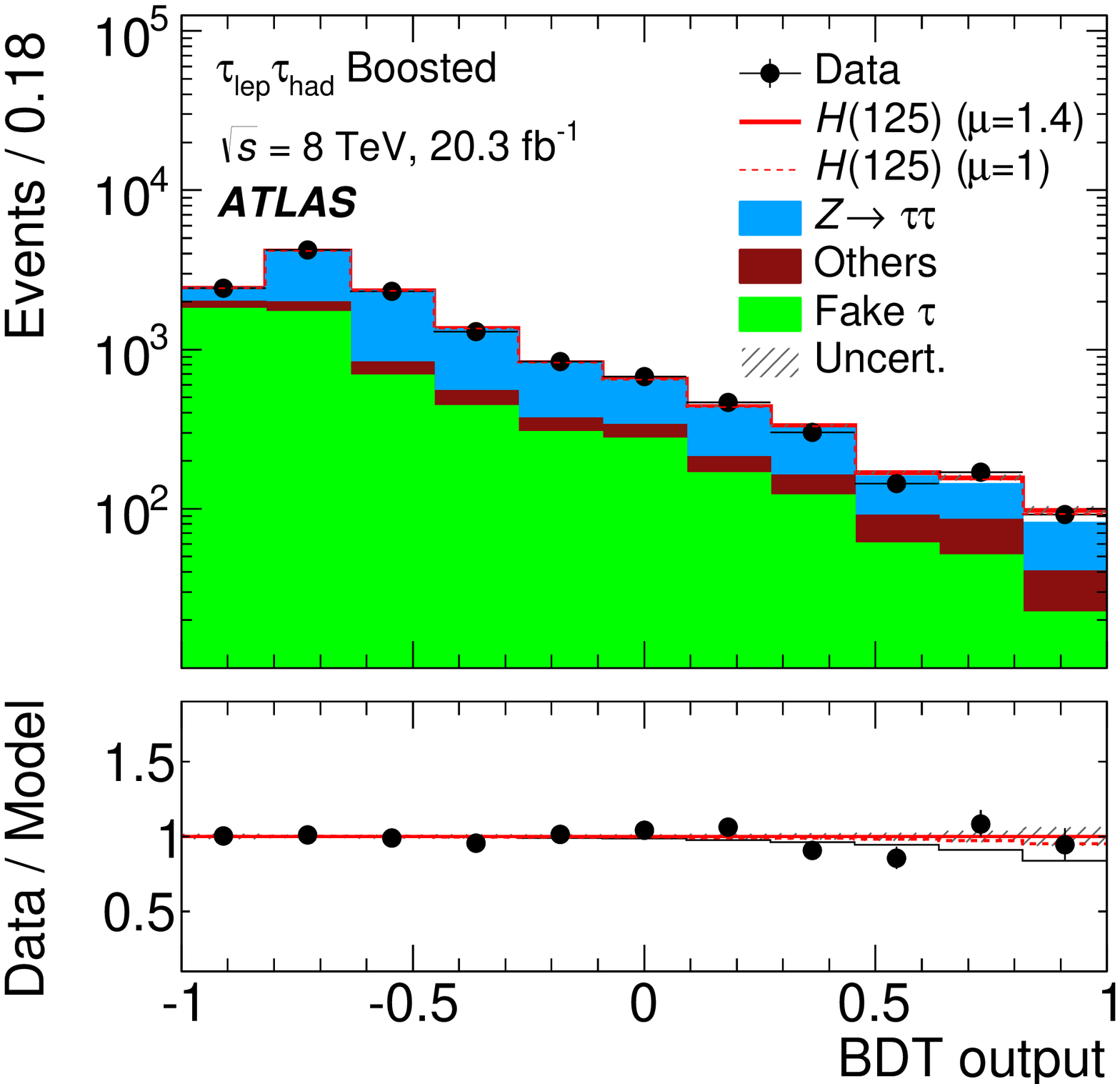}}
\hspace*{\fill}

\hspace*{\fill}
\subfigure[]{\includegraphics[width=6.1cm]{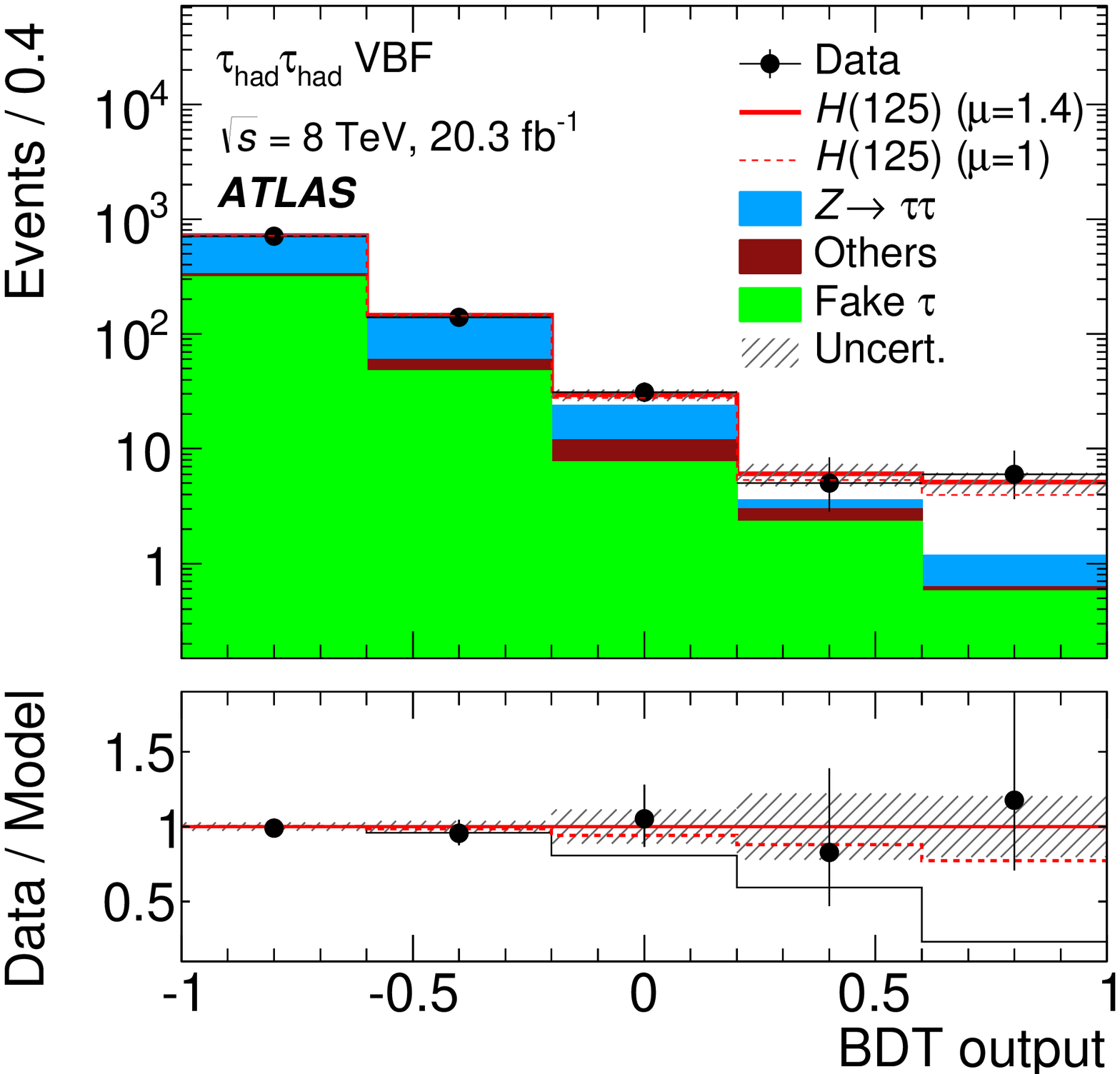}}\hfill
\subfigure[]{\includegraphics[width=6.1cm]{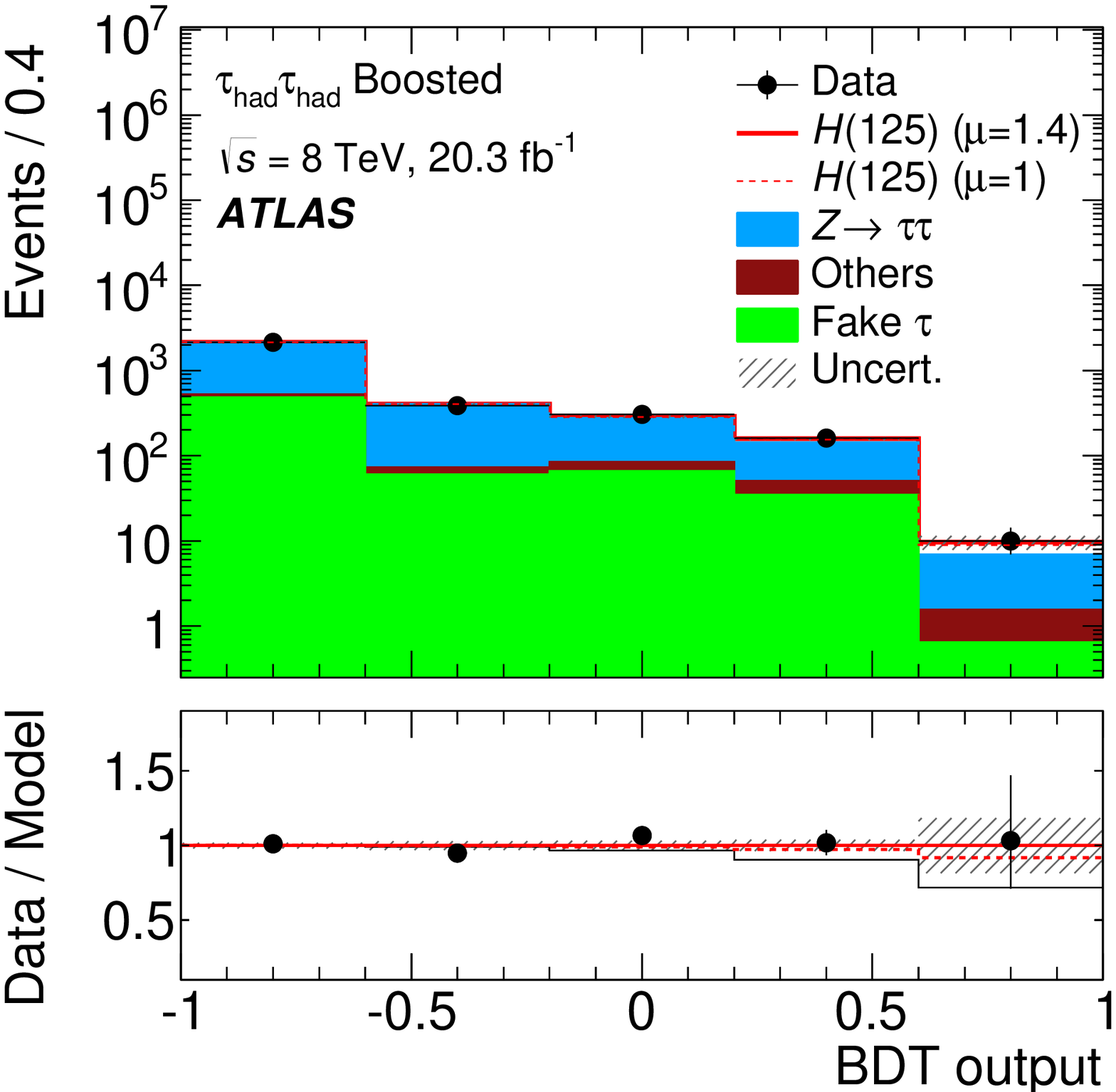}}
\hspace*{\fill}

\caption{
Distributions of the BDT discriminants for the data taken at
$\sqrt{s}$~=~8~\TeV\ in the signal regions of the VBF (left) and 
boosted (right) categories for the $\tlep \tlep$ (top),  $\tlep \thad$
(middle), and  $\thad \thad$ (bottom) channels.  The Higgs boson signal ($m_{H}
= 125$~\GeV) is shown stacked with a signal strength of $\mu=1$ (dashed line)
and $\mu=1.4$ (solid line). The background predictions are determined in the
global fit (that gives $\mu=1.4$).  The size of the statistical and systematic
normalisation uncertainties is indicated by the hashed band.  The ratios
of the data to the model (background plus Higgs boson contributions with
$\mu$~=~1.4) are shown in the lower panels. The dashed red and the solid black
lines represent the changes in the model when $\mu$~=~1.0 or $\mu$~=~0 
are assumed respectively.
}
\label{fig:BDT-SR-postfit}
\end{figure}

The distributions of the BDT discriminants for all channels and categories for the data at 8~\TeV{} are shown
in figure~\ref{fig:BDT-SR-postfit}, with background normalisations, signal normalisation, and nuisance parameters adjusted
by the profile likelihood global fit.

The results for the numbers of fitted signal and background events, split into the various contributions,
are summarised in tables~\ref{tab:lep-lep-event-yields}, \ref{tab:lep-had-event-yields} and \ref{tab:had-had-event-yields} for the three channels separately, for the dataset collected at 8~TeV centre-of-mass energy. In addition to the total number of events, the expected
 number of events in each of the
two highest BDT output bins is given. The number of events observed in the data is also included.
Within the uncertainties, good agreement is observed between the data and the model predictions for the sum of
background components and a Standard Model Higgs boson with $m_H$~=~125~\GeV{}.

\begin{sidewaystable}
\begin{center}
\caption{
The predicted post-fit event yields in the $\tll$ channel for $m_H$~=~125~\GeV\ for the total number of events and for the two highest bins of the BDT distributions for the data taken at $\sqrt{s}$~=~8~\TeV.  The background normalisations, signal normalisation, and their uncertainties represent the post-fit values (see text). The uncertainties on the total background and total signal represent the full statistical and systematic uncertainty, while the uncertainties on the individual background components represent the full systematic uncertainty only.}
\vspace{.5cm}
\begin{tabular}{|l|*{6}{c|}}
\hline
Process/Category & \multicolumn{3}{c|}{VBF} & \multicolumn{3}{c|}{Boosted}\\\hline
BDT output bin & All bins & Second to last bin & Last bin & All bins & Second to last bin & Last bin \\\hline\hline
$Z\to\tau\tau$   & $589\pm24$  & $9.7\pm1.0$  & $1.99\pm0.34$  & $2190\pm80$  & $33.7\pm2.3$  & $11.3\pm1.3$  \\
 Fake background  &$57\pm12$  & $1.2\pm0.6$  & $0.55\pm0.35$  & $100\pm40$  & $2.9\pm1.3$  & $0.6\pm0.4$  \\
 Top   & $131\pm19$  & $0.9\pm0.4$  & $0.89\pm0.33$  & $380\pm50$  & $9.8\pm2.1$  & $4.3\pm1.0$  \\
 Others  & $196\pm17$  & $3.0\pm0.4$  & $1.7\pm0.6$  & $400\pm40$  & $8.3\pm1.6$  & $2.6\pm0.7$  \\
 ggF:$~H\to WW~~(m_H=125~\GeV)$   &  $2.9\pm0.8$  & $0.12\pm0.04$  & $0.11\pm0.04$  & $7.7\pm2.3$  & $0.43\pm0.13$  & $0.24\pm0.08$  \\
 VBF:$~H\to WW$  & $3.4\pm0.4$  & $0.40\pm0.06$  & $0.38\pm0.08$  & $1.65\pm0.18$  & $0.102\pm0.017$  & $<0.1$  \\
 $WH:~H\to WW$   & $<0.1$  & $<0.1$  & $<0.1$  & $0.90\pm0.10$  & $<0.1$  & $<0.1$  \\
 $ZH:~H\to WW$   & $<0.1$  & $<0.1$  & $<0.1$  & $0.59\pm0.07$  & $<0.1$  & $<0.1$  \\
\hline
ggF:$~ H\to\tau\tau $  ($m_H = 125$~GeV) & $9.8\pm3.4$  & $0.73\pm0.26$  & $0.35\pm0.14$  & $21\pm8$  & $2.4\pm0.9$  & $1.3\pm0.5$  \\
VBF:$~ H\to\tau\tau $     & $13.3\pm4.0$  & $2.7\pm0.7$  & $3.3\pm0.9$  & $5.5\pm1.5$  & $0.95\pm0.26$  & $0.49\pm0.13$  \\
$WH:~H\to\tau\tau $     & $0.25\pm0.07$  & $<0.1$  & $<0.1$  & $3.8\pm1.0$  & $0.44\pm0.12$  & $0.22\pm0.06$  \\
$ZH:~H\to\tau\tau$     & $0.14\pm0.04$  & $<0.1$  & $<0.1$  & $2.0\pm0.5$  & $0.21\pm0.06$  & $0.113\pm0.031$  \\
\hline
 Total background  & $980\pm22$  & $15.4\pm1.8$  & $5.6\pm1.4$  & $3080\pm50$  & $55\pm4$  & $19.2\pm2.1$  \\
 Total signal  & $24\pm6$  & $3.5\pm0.9$  & $3.6\pm1.0$  & $33\pm10$  & $4.0\pm1.2$  & $2.1\pm0.6$  \\
\hline
 Data  & 1014  & 16  & 11  & 3095  & 61  & 20  \\
\hline
\end{tabular}
\label{tab:lep-lep-event-yields}
\end{center}
\end{sidewaystable}

\begin{sidewaystable}
\begin{center}
\caption{
The predicted post-fit event yields in the $\tlhad$ channel for $m_H = 125\GeV$
for the total number of events and for the two highest bins of the BDT
distributions for the data taken at $\sqrt{s}$~=~8~\TeV.  The background
normalisations, signal normalisation, and their uncertainties represent the post-fit values (see text). The uncertainties on the total background and total
signal represent the full statistical and systematic uncertainty, while the
uncertainties on the individual background components represent the full
systematic uncertainty only. }
\label{tab:lep-had-event-yields}
\vspace{.5cm}
\begin{tabular}{|l|*{6}{c|}}
\hline
Process/Category & \multicolumn{3}{c|}{VBF} & \multicolumn{3}{c|}{Boosted}\\\hline
BDT output bin & All bins & Second to last bin & Last bin & All bins & Second to last bin & Last bin \\\hline\hline
Fake background  &  $1680\pm50$  & $8.2\pm0.9$  & $5.2\pm0.7$  & $5640\pm160$  & $51.0\pm2.5$  & $22.3\pm1.8$  \\
$Z\rightarrow\tau\tau$  & $877\pm29$  & $7.6\pm0.9$  & $4.2\pm0.7$  & $6210\pm170$  & $57.5\pm2.8$  & $41.1\pm3.2$  \\
Top  & $82\pm15$  & $0.3\pm0.4$  & $0.5\pm0.4$  & $380\pm50$  & $12\pm4$  & $4.8\pm1.5$  \\
$Z\to\ell\ell (\ell\to\thad)$  & $54\pm26$  & $1.0\pm0.7$  & $0.30\pm0.28$  & $200\pm50$  & $13\pm4$  & $8.6\pm3.5$  \\
Diboson  & $63\pm11$  & $1.0\pm0.4$  & $0.48\pm0.20$  & $430\pm40$  & $9.7\pm2.2$  & $4.7\pm1.6$  \\
\hline
 ggF:$~ H\to\tau\tau $  ($m_H = 125$~GeV)  &  $16\pm6$  & $1.0\pm0.4$  & $1.2\pm0.6$  & $60\pm20$  & $9.2\pm3.2$  & $10.1\pm3.4$  \\
VBF:$~ H\to\tau\tau $ & $31\pm8$  & $4.5\pm1.1$  & $9.1\pm2.2$  & $16\pm4$  & $2.5\pm0.6$  & $2.9\pm0.7$  \\
$WH:~H\to\tau\tau$    &$0.6\pm0.4$  & $<0.1$  & $<0.1$  & $9.1\pm2.3$  & $1.3\pm0.4$  & $1.9\pm0.5$  \\
$ZH:~H\to\tau\tau$  &$0.16\pm0.07$  & $<0.1$  & $<0.1$  & $4.6\pm1.2$  & $0.77\pm0.20$  & $0.93\pm0.24$  \\
\hline
 Total background  & $2760\pm40$  & $18.1\pm2.3$  & $10.7\pm2.7$  & $12860\pm110$  & $143\pm6$  & $82\pm6$  \\
 Total signal  & $48\pm12$  & $5.5\pm1.3$  & $10.3\pm2.5$  & $89\pm26$  & $14\pm4$  & $16\pm4$  \\
\hline
 Data  & 2830  & 22  & 21  & 12952  & 170  & 92  \\
\hline
\end{tabular}
\end{center}
\end{sidewaystable}

\begin{sidewaystable}
\begin{center}
\caption{
The predicted post-fit event yields in the $\thadhad$ channel for $m_H =
125\GeV$ for the total number of events and for the two highest bins of the BDT
distributions for the data taken at $\sqrt{s}$~=~8~\TeV.  The background
normalisations, signal normalisation, and their uncertainties represent the post-fit values (see text). The uncertainties on the total background and total
signal represent the full statistical and systematic uncertainty, while the
uncertainties on the individual background components represent the full
systematic uncertainty only. }
\label{tab:had-had-event-yields}
\vspace{.5cm}
\begin{tabular}{|l|*{6}{c|}}
\hline
Process/Category & \multicolumn{3}{c|}{VBF} & \multicolumn{3}{c|}{Boosted}\\\hline
BDT output bin & All bins & Second to last bin & Last bin & All bins & Second to last bin & Last bin \\\hline\hline
Fake background  & $370\pm18$  & $2.3\pm0.9$  & $0.57\pm0.29$  & $645\pm26$  & $35\pm4$  & $0.65\pm0.33$  \\ 
Others  & $37\pm5$  & $0.67\pm0.22$  & $<0.1$  & $89\pm11$  & $15.9\pm2.0$  & $0.92\pm0.22$  \\
$Z\to\tau\tau$   & $475\pm16$  & $0.6\pm0.7$  & $0.6\pm0.4$  & $2230\pm70$  & $93\pm4$  & $5.4\pm1.6$  \\
\hline
 ggF:$~H\to\tau\tau$  ($m_H = 125$~GeV)  &  $8.0\pm2.7$  & $0.67\pm0.23$  & $0.53\pm0.20$  & $21\pm8$  & $9.1\pm3.3$  & $1.6\pm0.6$  \\
VBF:$~H\to\tau\tau$    &  $12.0\pm3.1$  & $1.8\pm0.5$  & $3.4\pm0.9$  & $6.3\pm1.6$  & $2.8\pm0.7$  & $0.52\pm0.13$  \\
$WH:H\to\tau\tau $       & $0.25\pm0.07$  & $<0.1$  & $<0.1$  & $4.0\pm1.1$  & $1.9\pm0.5$  & $0.41\pm0.11$  \\
$ZH:~H\to\tau\tau$   & $0.16\pm0.04$  & $<0.1$  & $<0.1$  & $2.4\pm0.6$  & $1.13\pm0.30$  & $0.23\pm0.06$  \\
\hline
 Total background  & $883\pm18$  & $3.6\pm1.3$  & $1.2\pm1.0$  & $2960\pm50$  & $143\pm6$  & $7.0\pm1.8$  \\
 Total signal  & $20\pm5$  & $2.5\pm0.6$  & $3.9\pm1.0$  & $34\pm10$  & $15\pm4$  & $2.7\pm0.8$  \\
\hline
 Data  & 892  & 5  & 6  & 3020  & 161  & 10  \\
\hline
\end{tabular}
\end{center}
\end{sidewaystable}

\clearpage

\section{Results}
\label{sec:results}

As explained in the previous section, the observed signal strength is
determined from a global maximum likelihood fit to the BDT output distributions
in data, with nuisance parameters that are either free or constrained. The
results are extracted for each channel and for each category individually as
well as for combinations of categories and for the overall combination.

\begin{figure}[htbp]
\begin{center}
\begin{tabular}{c}
\includegraphics[width=0.65\textwidth]{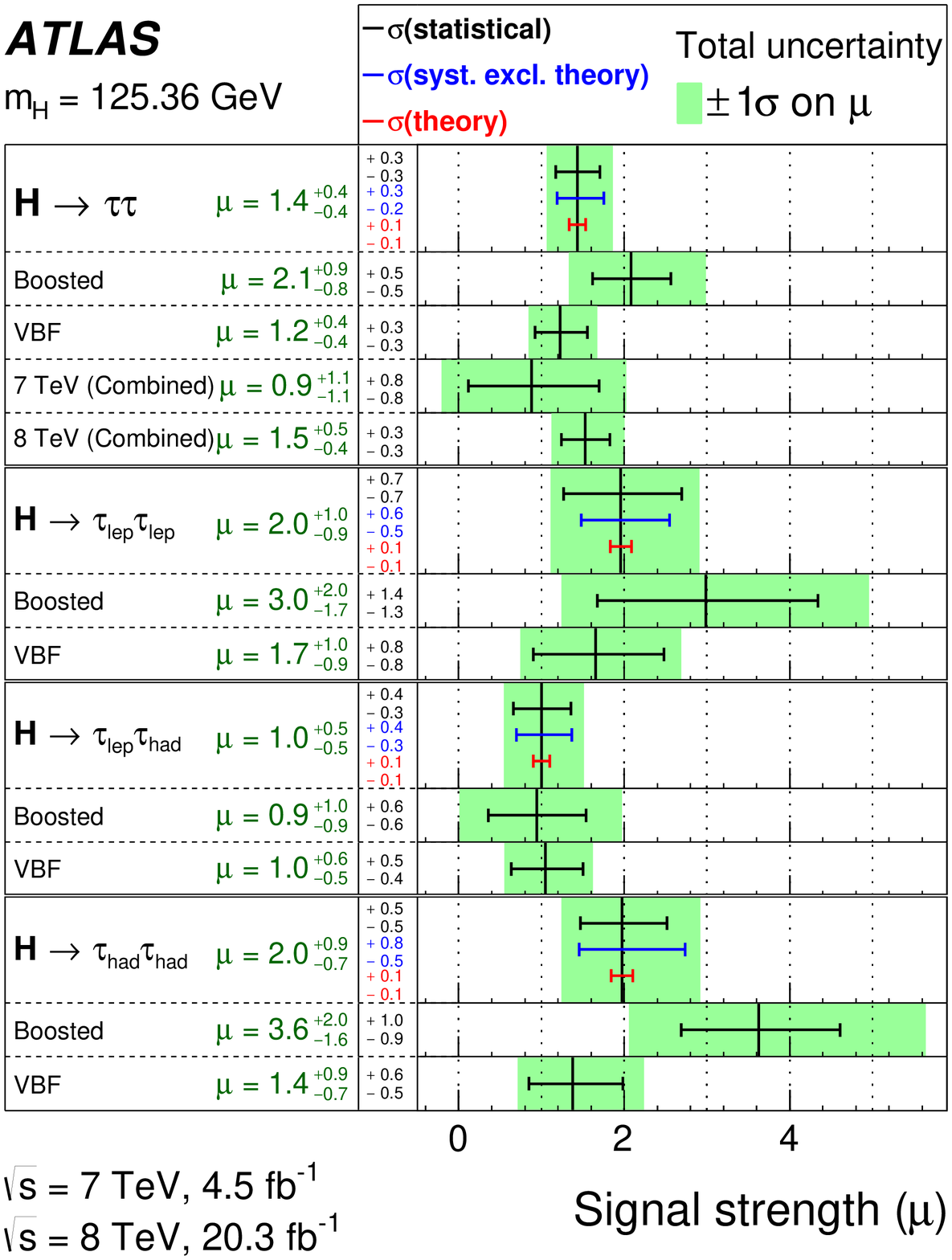}\\
\end{tabular}
\end{center}
\caption{
The best-fit value for the signal strength $\mu$ in the individual channels and
their combination for the full ATLAS datasets at $\sqrt{s}$ =  7~\TeV\ and
$\sqrt{s}$ = 8~\TeV.  The total $\pm{}1\sigma$ uncertainty is indicated by the
shaded green band, with the individual contributions from the statistical
uncertainty (top, black), the experimental systematic
uncertainty (middle, blue), and the theory uncertainty (bottom, red) on the
signal cross section (from QCD scale, PDF, and branching ratios) shown by the
error bars and printed in the central column.
}
\label{fig:signal-strength}
\end{figure}

At the value of the Higgs boson
mass obtained from the combination of the ATLAS \hgg\ and \hzzs\
measurements~\cite{Aad:2014aba}, $m_H$~=~125.36~\GeV, the signal strength obtained from the combined \htautau\ analysis is:
\[
\mu = 1.43 \ ^{+0.27}_{-0.26}\mathrm{(stat.)}\ ^{+0.32}_{-0.25}\mathrm{(syst.)}\ \pm 0.09\mathrm{(theory~syst.)}.
\]
The systematic uncertainties are split into two groups: systematic uncertainties (syst.)  including
 all experimental effects as well as theoretical uncertainties on the signal region acceptance, 
such as those due to the QCD scales, the PDF choice, and the underlying event and parton shower; 
and, separately, theoretical uncertainties on the inclusive Higgs boson production cross section 
and \htautau\ branching ratio (theory syst.). 
The results for each individual
channel and for each category as well as for their combination are shown in
figure~\ref{fig:signal-strength}. They are based on the full dataset, however,
separate combined results are given for the two centre-of-mass energies.

The probability $p_0$ of obtaining a result at least as signal-like as observed
in the data if no signal were present is calculated using the test statistic
$q_{\mu = 0} = -2 \ln (\mathcal{L} (0,\hat{\hat{\vec{\theta}}})/\mathcal{L}
(\hat{\mu},\hat{\vec{\theta}}))$ in the asymptotic
approximation~\cite{Cowan:2010js}.  For $m_H = 125.36\GeV$, the observed $p_0$
value is $2.7 \times 10^{-6}$, which corresponds to a deviation from the
background-only hypothesis of 4.5$\sigma$.  This can be compared to an expected
significance of 3.4$\sigma$. 
This provides evidence at the level of 4.5$\sigma$ for the decay of the  Higgs boson
into  tau  leptons.  Table \ref{tab:Exp_p0} shows the expected and observed
significances for the signal strength measured in each channel separately.

\begin{table}
\begin{center}
\begin{tabular}{|l|c|c|c|}\hline
Channel and Category &Expected Significance ($\sigma$) & Observed  Significance ($\sigma$) \\ \hline
\tll ~VBF & 1.15 & 1.88 \\
\tll ~Boosted & 0.57 & 1.72 \\
\tll ~Total &          1.25 &    2.40  \\
\hline
\tlhad ~VBF & 2.11 & 2.23 \\
\tlhad ~Boosted &  1.11 & 1.01 \\
\tlhad ~Total &       2.33 & 2.33     \\
\hline
\thadhad ~VBF & 1.70 & 2.23 \\
\thadhad ~Boosted & 0.82 & 2.56 \\
\thadhad ~Total &        1.99 &   3.25   \\
\hline
Combined &           3.43 &    4.54    \\
\hline
\end{tabular}
\caption{The expected and observed significances of the signal in each channel and category for the combined 7 and 8~\TeV\ datasets.}
\label{tab:Exp_p0}
\end{center}
\end{table}

\begin{figure}[htbp]
\begin{center}
\includegraphics[width=11cm]{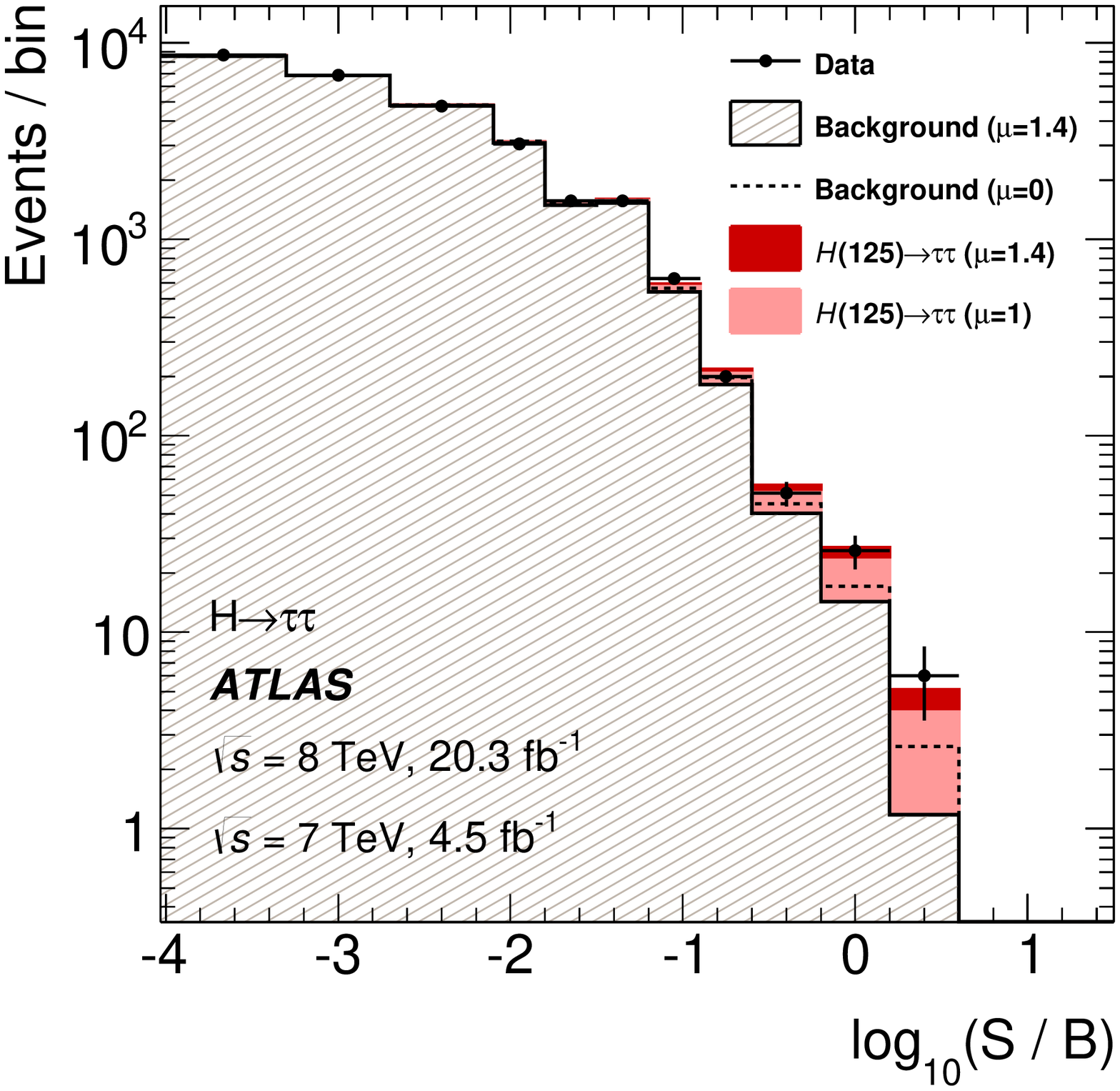}
\end{center}
\caption{
Event yields as a function of $\log_{10} (S/B)$, where $S$ (signal yield) and
$B$ (background yield) are taken from the BDT output bin of each event,
assuming a signal strength $\mu=1.4$.  Events in all categories are included.  The predicted
background is obtained from the global fit (with $\mu=1.4$), and signal yields
are shown for $m_H=125\GeV$ at $\mu=1$ and $\mu=1.4$ (the best-fit value). The
background-only distribution (dashed line) is obtained from the global fit, with
$\mu$ fixed at zero.}
\label{fig:bdt-soverb-combined}
\end{figure}

Figure~\ref{fig:bdt-soverb-combined} shows the expected and observed number of events,
 in bins of $\log_{10}(S/B)$, for all signal region bins.  Here,
$S/B$ is the signal-to-background ratio calculated assuming $\mu=1.4$ for each
BDT bin in the signal regions.  The expected  signal yield for
both $\mu=1$ and the best-fit value $\mu=1.4$ for $m_H=125$~\GeV\ is shown on top of the
background prediction from the best-fit values.  The background
expectation where the signal-strength parameter is fixed to $\mu=0$ is
also shown for comparison.

\begin{figure}[htbp]
\begin{center}
\subfigure[]{\includegraphics[width=7.5cm]{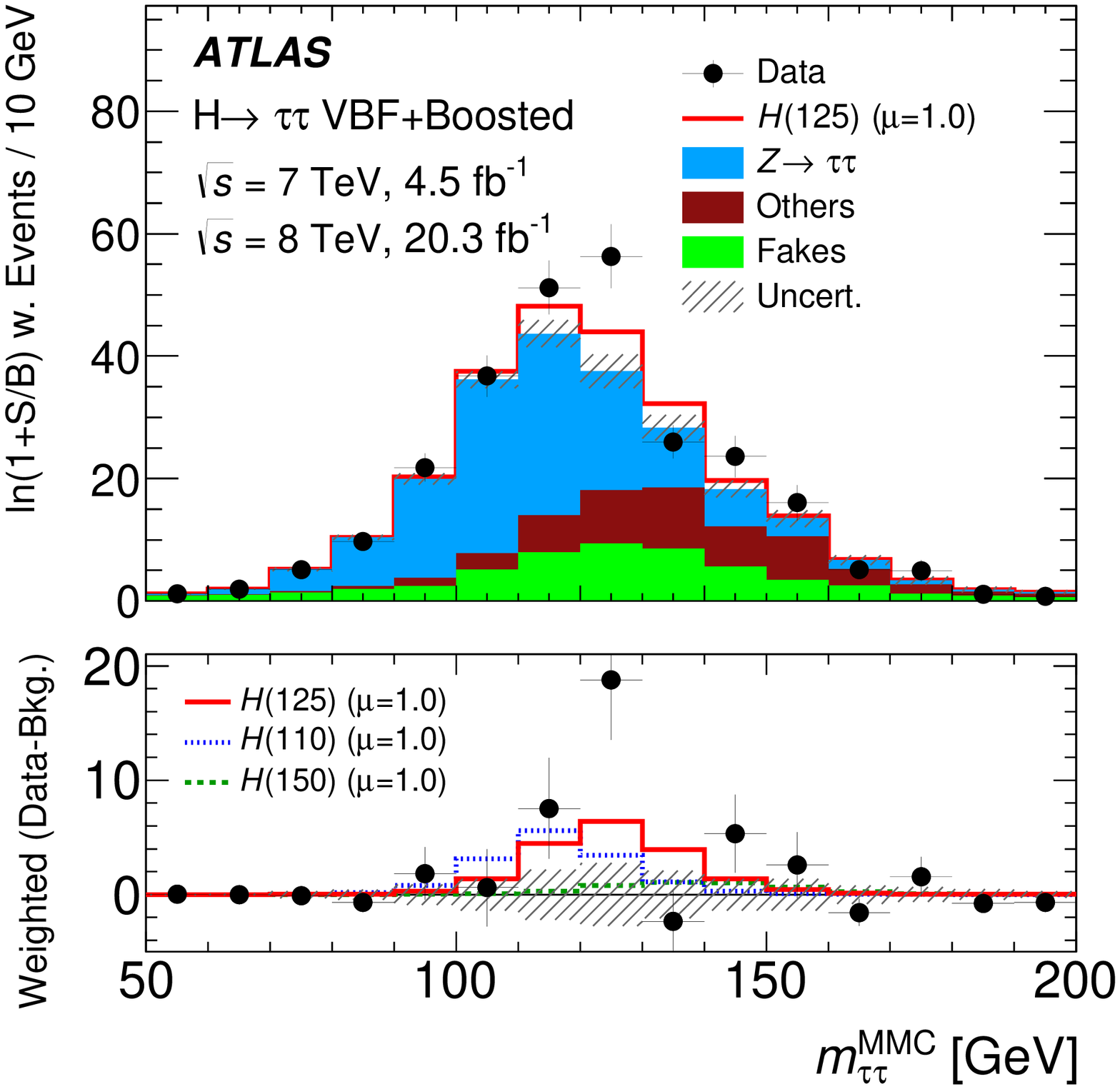}}
\subfigure[]{\includegraphics[width=7.5cm]{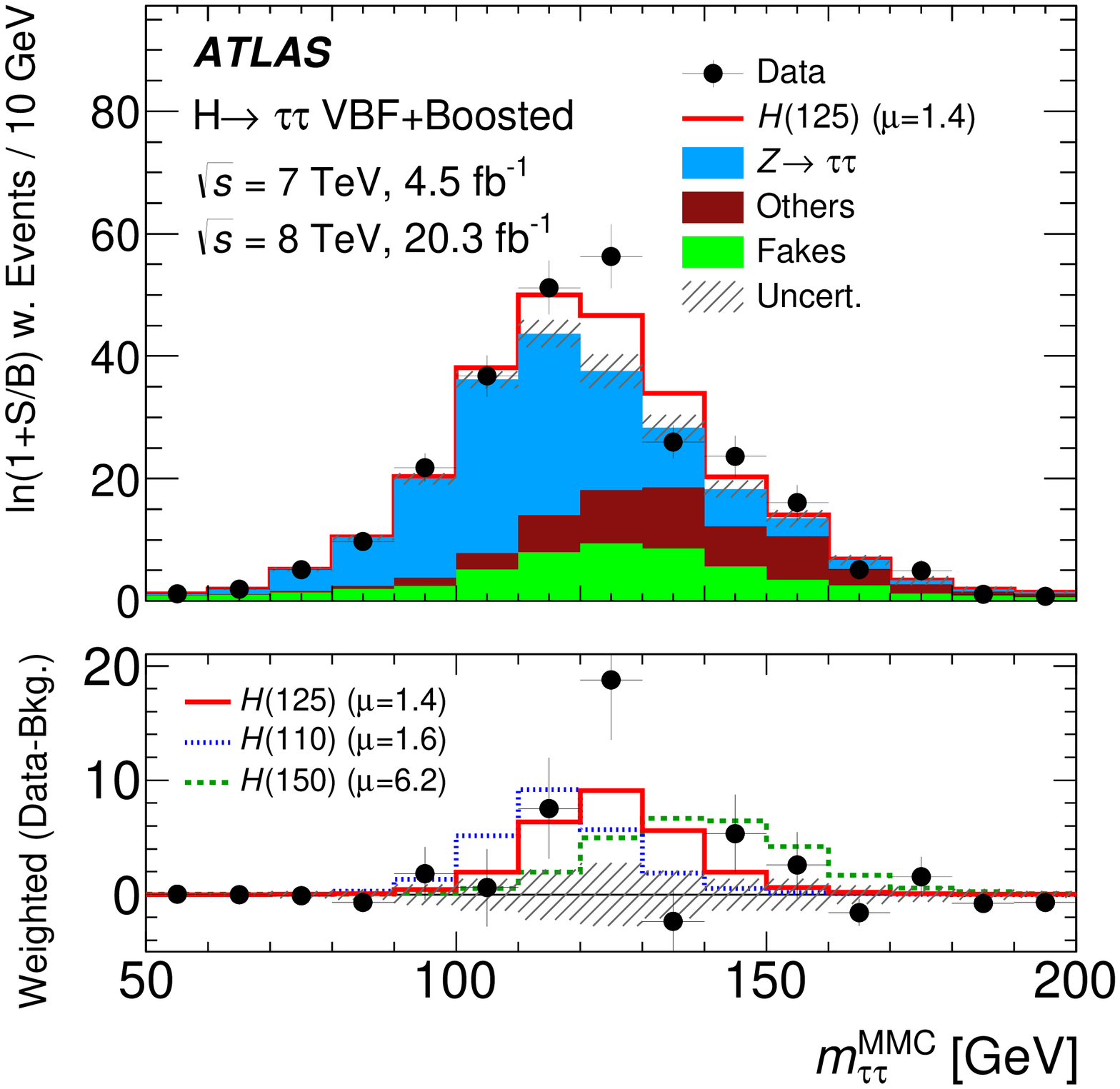}}
\end{center}
\caption{
Distributions of the reconstructed invariant $\tau \tau$ mass, $\MMC$, where events are weighted by $\ln (1+ S/B)$ for all
channels. These weights are determined by the signal ($S$) and background
($B$) predictions for each BDT bin. The bottom panel in each plot shows the
difference between weighted data events and weighted background events (black
points), compared to the weighted signal yields.  The background predictions
are obtained from the global fit with the $m_H$~=~125~\GeV\ signal hypothesis
(signal strength $\mu$~=~1.4).  The $m_H$~=~125~\GeV\ signal is plotted with a solid red line,
and, for comparison, signals for $m_H$~=~110~\GeV\ (blue) and $m_H$~=~150~\GeV\
(green) are also shown. The signal normalisations are taken from fits to data
with the corresponding signal mass hypotheses, and the fitted $\mu$ values are
given in the figure.  The signal strengths are shown for the Standard Model
expectations ($\mu$ = 1) in (a), while in (b) the best-fit values are used.
}
\label{fig:mmc-weighted}
\end{figure}

To visualise the compatibility of this excess of events above
background predictions with the SM Higgs boson at $m_H=125\GeV$, a weighted
distribution of events as a function of $\MMC$ is shown in
figure~\ref{fig:mmc-weighted}.  The events are weighted by a factor of $\ln (1
+ S/B)$, which enhances the events compatible with the signal hypothesis.
The excess of events in this mass distribution is consistent with the
expectation for a Standard Model Higgs boson with $m_H = 125$~GeV. The
distributions for the predicted excess in data over the background are also
shown for alternative SM Higgs boson mass hypotheses of $m_H = 110$~\GeV\ and
$m_H = 150$~\GeV.  The data favour a Higgs boson mass of $m_H=125\GeV$ and are
less consistent with the other masses considered.

As discussed in section~\ref{sec:fit}, the dominant uncertainties on the
measurement of the signal-strength parameters include statistical uncertainties
on the data from the signal regions, uncertainties on the jet and tau energy
scales, uncertainties on the normalisation of the \mbox{$Z \to \tau \tau$} and
$\ttbar$ background components as well as theoretical uncertainties. 
The contributions of each of these significant sources to
the uncertainty of the measured signal strength are summarised in table~\ref{tab:systs}.

\begin{table}
\begin{center}
\def\arraystretch{1.1}
\begin{tabular}{|l|c|}
\hline
Source of Uncertainty & Uncertainty on $\mu$ \\
\hline
\hline
Signal region statistics (data) & $^{+0.27}_{-0.26}$ \\
\hline
Jet energy scale	& $\pm$	0.13	\\
Tau energy scale	& $\pm$	0.07	\\
Tau identification	& $\pm$	0.06	\\
\hline
Background normalisation	& $\pm$	0.12	\\
Background estimate stat.	& $\pm$	0.10	\\
\hline
BR~($H\to\tau\tau$)    	& $\pm$	0.08	\\
Parton shower/Underlying event	& $\pm$	0.04	\\
PDF 	& $\pm$	0.03	\\
\hline\hline
Total sys.        &   $^{+0.33}_{-0.26}$   \\
\hline\hline
Total       &   $^{+0.43}_{-0.37}$   \\
\hline
\end{tabular}
\caption{
Important sources of uncertainty on the measured signal-strength parameter
$\mu$. The contributions are given as absolute uncertainties on the best-fit
value of $\mu$~=~1.43. Various sub-components are combined assuming no
correlations.
}
\label{tab:systs}
\end{center}
\end{table}

The normalisation uncertainties on the $Z\to\tau\tau$ embedded sample are correlated
across the categories in each respective channel.  The global fit also constrains
the normalisation for $Z\to\tau\tau$ more strongly than for the $\Zll$ and
top-quark background components, as the low BDT-score region is dominated by
$Z\to\tau\tau$ events.

\begin{figure}[htbp]
\begin{center}
\includegraphics[width=11cm]{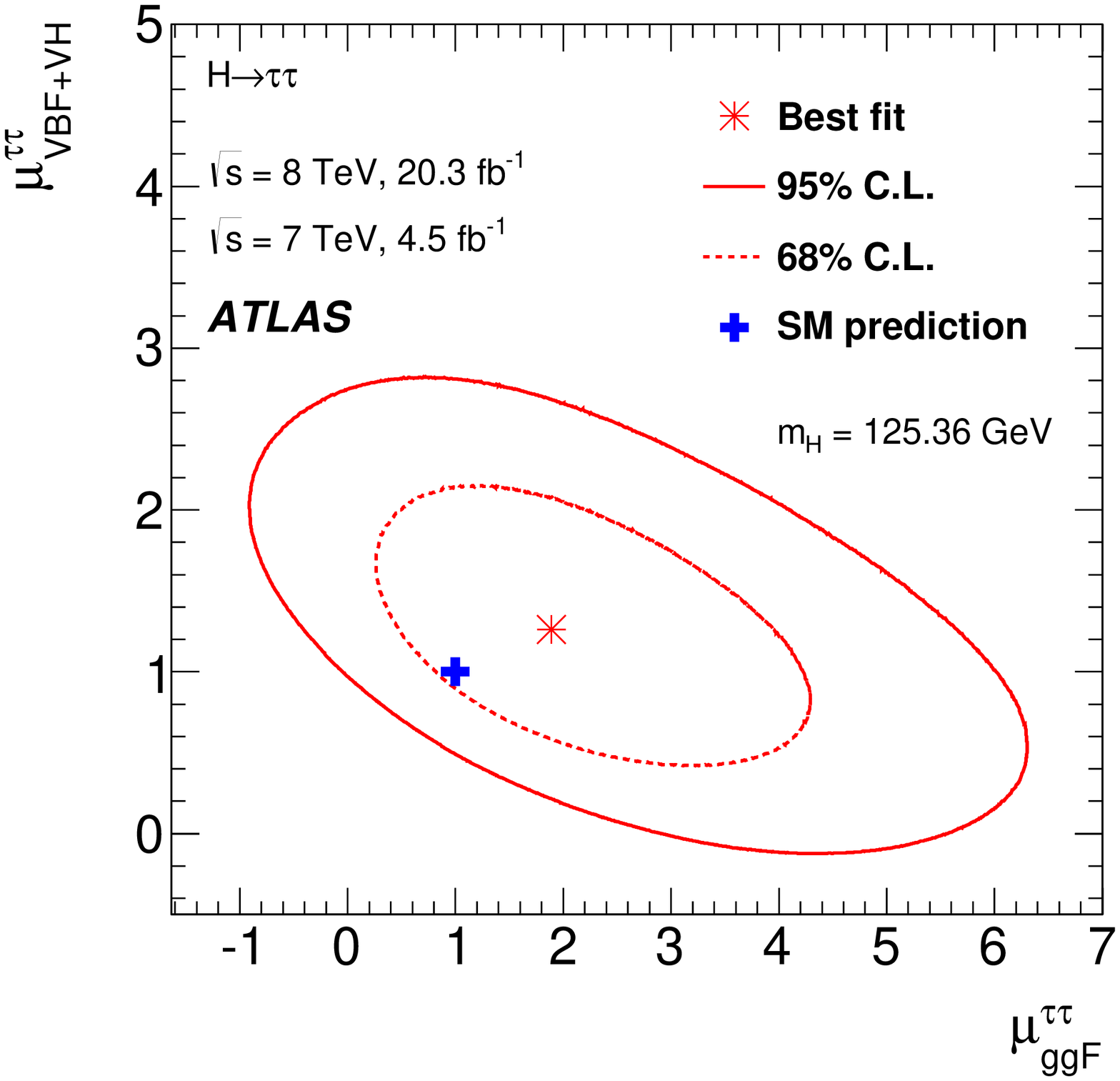}
\end{center}
\caption{
Likelihood contours for the combination of all channels in the
($\mu_{ggF}^{\tau\tau}$, $\mu_\text{\tiny VBF\em+VH}^{\tau\tau}$) plane. The signal strength $\mu$ is the ratio of the measured signal yield to the Standard Model expectation, for each production mode. The
68\% and 95\% CL contours are shown as dashed and solid lines respectively,
for $m_H=125.36\GeV$.  The SM expectation is shown by a filled plus symbol,
and the best fit to the data is shown as a star.
}
\label{fig:2D-contour}
\end{figure}

The measurement of the overall signal strength discussed above does not give
direct information on the relative contributions of the different production
mechanisms.  Therefore,  the signal strengths of different production processes
contributing to the $\htautau$ decay mode are determined, exploiting the
sensitivity offered by the use of the event categories in the analyses of the
three channels.  The data are fitted separating the vector-boson-mediated
 VBF and $VH$ processes from gluon-mediated ggF  processes.  Two
signal strength parameters, $\mu_{ggF}^{\tau\tau}$ and $\mu_\text{\tiny
VBF+\em VH}^{\tau\tau}$, which scale the SM-predicted rates to those observed, are
introduced.  The two-dimensional $68\%$ and $95\%$ confidence level (CL) contours
in the plane of
$\mu_{ggF}^{\tau\tau}$ and $\mu_\text{\tiny VBF\em
+VH}^{\tau\tau}$~\cite{YellowReportIII} are shown in
figure~\ref{fig:2D-contour} for $m_H=125.36$~\GeV.
The best-fit values are
\[
\mu_{ggF}^{\tau\tau} = 2.0 \ \pm 0.8\mathrm{(stat.)}\ ^{+1.2}_{-0.8}\mathrm{(syst.)}\ \pm 0.3\mathrm{(theory~syst.)}
\]
and
\[
\mu_\text{\tiny VBF\em+VH}^{\tau\tau}=1.24\ ^{+0.49}_{-0.45}\mathrm{(stat.)}\ ^{+0.31}_{-0.29}\mathrm{(syst.)}\ \pm 0.08\mathrm{(theory~syst.)} ,
\]
in agreement with the predictions from the Standard Model.  The two results are strongly anti-correlated (correlation coefficient of $-48$\%). The observed (expected) significances of the $\mu_{ggF}^{\tau\tau}$ and
 $\mu_\text{\tiny VBF\em+VH}^{\tau\tau}$  signal strengths are 1.74$\sigma$ (0.95$\sigma$) and 2.25$\sigma$ (1.72$\sigma$) respectively.

A total cross section times branching ratio for $H\to\tau\tau$ with
$m_H$~=~125~\GeV\ can also be measured. The central
value is  obtained from the product of the measured $\mu$ and the predicted
cross section used to define it. The uncertainties are similarly obtained by
scaling the uncertainties on $\mu$ by the predicted cross section, noting that
theoretical uncertainties on the inclusive cross section cancel between $\mu$
and the predicted cross section and thus are not included for the production processes under consideration. 
These include the uncertainties on the inclusive cross section due to the QCD scale and the
PDF choice as  well as the uncertainty on the branching ratio $H\to\tau\tau$;
however, theoretical uncertainties on the acceptance of the signal regions from the QCD
scale and PDF choice are retained, along with the uncertainties due to 
underlying event and parton shower, and the electroweak correction on VBF production.
Table~\ref{tab:xsec} gives the measured values for the total cross section at 7
and at 8~\TeV, as well as the measured values  at 8~\TeV\ for gluon fusion production and for VBF and $VH$
production separately.

\begin{table}
\begin{center}
\def\arraystretch{1.4}
\begin{tabular}{|l|c|c|}
\hline
 & Measured $\sigma \times$BR [pb]  &Predicted $\sigma \times$BR [pb] \\
\hline
\hline
7~\TeV &  $1.0\ ^{+0.9}_{-0.8}\mathrm{(stat.)}^{+0.9}_{-0.8}\mathrm{(syst.)}$ & 1.09 $\pm$ 0.11 \\
\hline
8~\TeV &  $2.1 \pm 0.4\mathrm{(stat.)}^{+0.5}_{-0.4}\mathrm{(syst.)}$ & 1.39 $\pm$ 0.14 \\
\hline
Gluon fusion, 8~\TeV &  $1.7 \pm 1.1\mathrm{(stat.)}^{+1.5}_{-1.1}\mathrm{(syst.)}$ & 1.22 $\pm$ 0.14 \\
\hline
VBF+$VH$, 8~\TeV &  $0.26 \pm 0.09\mathrm{(stat.)}^{+0.06}_{-0.05}\mathrm{(syst.)} $& 0.17 $\pm$ 0.01  \\
\hline
\end{tabular}
\caption{
Measured and predicted total cross section times branching ratio for
$H\to\tau\tau$ with $m_H$~=~125~\GeV, at 7~\TeV\ and at 8~\TeV\ for all production
modes, as well as for gluon fusion and for VBF and $VH$ production
separately, at 8~\TeV\ only. The theoretical predictions are obtained as
described in section~\ref{sec:samples}.
}
\label{tab:xsec}
\end{center}
\end{table}

\clearpage
\section{Cross-check with cut-based analysis}
\label{sec:cutbased}

The search for the SM Higgs boson presented above is cross-checked for the dataset collected at
$\sqrt{s}$~=~8~\TeV\ in an  analysis where cuts on kinematic variables are applied.
This search uses improved definitions of event categories and
an improved fit model with respect to results previously published for the  $\sqrt{s}$~=~7~\TeV\ dataset
\cite{Aad:2012mea}.
To allow a straightforward comparison of results, the multivariate and cut-based analyses have common
components. The two analyses are performed for  the same three channels,
$\tlep \tlep$,  $\tlep \thad$ and $\thad \thad$, they use the same preselection and share the same strategy
for the estimation of background contributions and systematic uncertainties. As in the multivariate analysis,
the irreducible $Z \to \tau  \tau$ background is estimated using the embedding procedure and the reducible ones are
estimated using similar data-driven methods, as described in section~\ref{sec:background}.
Finally, the same statistical methods are used to extract the results, although these are applied to
different discriminating variables. While the multivariate analysis performs a fit to the BDT output distribution,
the cut-based analysis relies on a fit to the $\tau \tau$ invariant
mass distribution. The $\tau \tau$ invariant mass is calculated using the missing mass calculator, as described in
section~\ref{sec:mmc}. The analysis is not designed to be sensitive to a specific value of the Higgs
 boson mass $m_H$. The use of the  mass as the discriminating variable is motivated not only by its power
to separate the irreducible $Z \to \tau  \tau$ background from signal, but also by its sensitivity to
the mass of the signal itself.

In the cut-based analysis, a categorisation is performed similar to that in the multivariate analysis,
i.e.\ VBF and boosted categories are defined. To increase the separation power,
subcategories are introduced for the $\tlhad$ and $\thadhad$ channels. These
subcategories target events produced via the same production mode, but select different
phase-space regions with different signal-to-background ratios. With this strategy the most sensitive
subcategories have a small number of events, but a high signal-to-background ratio. Although the combined
sensitivity is dominated by the few highly sensitive subcategories, the others are important not just to
increase the sensitivity but also to constrain the various background components.

\begin{table}[h]
\begin{center}
\small
\def\arraystretch{1.1}
\begin{tabular}{|c|p{3.8cm}|p{4.3cm}|p{4.0cm}|}
\hline
\rule[-3mm]{0mm}{8mm} Channel & \multicolumn{3}{|c|}{VBF category selection criteria} \\
\hline\hline 
\multirow{5}{*}{\tll}
& \multicolumn{3}{|l|}{At least two jets with $p_{\rm T}^{j_1} > 40 \GeV$ and $p_{\rm T}^{j_2}> 30 \GeV$}\\
& \multicolumn{3}{|l|}{$|\Delta \eta_{j_1, j_2}| >3.0$}\\
& \multicolumn{3}{|l|}{$m_{j_1,j_2} >400 \GeV$}\\
& \multicolumn{3}{|l|}{$b$-jet veto for jets with $\pt > 25 \GeV$}\\
& \multicolumn{3}{|l|}{Jet veto: no additional jet with $\pt > 25 \GeV$ within $|\eta| < 2.4$}\\
\hline
\multirow{11}{*}{\tlhad}
    & \multicolumn{3}{|l|}{At least two jets with $p_{\rm T}^{j_1} > 40 \GeV$ and $p_{\rm T}^{j_2}> 30 \GeV$} \\
    & \multicolumn{3}{|l|}{$\MET >$ ~20~\GeV }\\
    & \multicolumn{3}{|l|}{ $|\Delta \eta_{j_1, j_2}| >3.0$ and $\eta(j_1) \cdot \eta(j_2) <0$, \qquad $ m_{j_1,j_2} > $~300~\GeV }\\
    & \multicolumn{3}{|l|}{$p_{\rm T}^{\rm Total} = |\vec{p}_{\rm T}^{\ell} + \vec{p}_{\rm T}^{\tau_{\rm had}} + \vec{p}_{\rm T}^{j_1} + \vec{p}_{\rm T}^{j_2} + \vec{E}_{\rm T}^{\rm miss}| < 30 \GeV $ }\\
& \multicolumn{3}{|l|}{$b$-jet veto for jets with $\pt > 30 \GeV$}\\
& \multicolumn{3}{|l|}{${\rm min} (\eta_{(j_1)}, \eta_{(j_2)}) < \eta_{(\ell)} ,  \eta_{(\tau_{\rm had})} < \rm{max} (\eta_{(j_1)},\eta_{(j_2)})$}\\
\cline{2-4}
\rule{0pt}{2ex}  
& \textbf{VBF tight} & \textbf{VBF loose}&\\
& $  m_{j_1,j_2}  >500 \GeV$& {Non tight VBF} &\\
& $p_{\rm T}^{H}>100 \GeV$  & &\\
& $p_{\rm T}^{\tau_{\rm had}} > 30 \GeV$  & &\\
& $m_{\rm vis}> 40 \GeV$ & &\\
\hline
\multirow{8}{*}{\thadhad}&\multicolumn{3}{|l|}{At least two jets with $p_{\rm T}^ {j_1}>50 \GeV$~and $p_{\rm T}^{j_2}>30 \GeV$ }\\
&\multicolumn{3}{|l|}{ $|\Delta \eta (\tau_1, \tau_2)|<1.5$ }\\
&\multicolumn{3}{|l|}{ $|\Delta \eta_{j_1, j_2} | >2.6$ and $m_{j_1,j_2} >250 \GeV$ }\\
&\multicolumn{3}{|l|}{ ${\rm min} (\eta_{(j_1)}, \eta_{(j_2)})< \eta_{(\tau_1)} ,  \eta_{(\tau_2)} < {\rm max} (\eta_{(j_1)},\eta_{(j_2)}) $}\\
\cline{2-4}
\rule{0pt}{2ex}  
&\textbf{VBF high $p_{\rm T}^{H}$} & \textbf{VBF low $p_{\rm T}^{H}$, tight } & \textbf{VBF low $p_{\rm T}^{H}$, loose}\\
& $\Delta R (\tau_1, \tau_2) < 1.5 $ and & $\Delta R (\tau_1, \tau_2) > 1.5 $ or & $\Delta R (\tau_1, \tau_2) > 1.5 $ or\\
& $p_{\rm T}^{H}>140 \GeV$ & $p_{\rm T}^{H} < 140 \GeV$ & $p_{\rm T}^{H} < 140 \GeV$\\
& & $m_{j_1,j_2} [\rm{GeV}] > (-250 \cdot |\Delta\eta_{j_1,j_2}| +1550)$& $m_{j_1,j_2} [\rm{GeV}] < (-250 \cdot |\Delta\eta_{j_1,j_2}| +1550)$\\
\hline\hline
\rule[-3mm]{0mm}{8mm} Channel & \multicolumn{3}{|c|}{Boosted category selection criteria} \\ \hline\hline
\multirow{3}{*}{\tll}
&\multicolumn{3}{|l|}{Exclude events passing the VBF selection}\\
&\multicolumn{3}{|l|}{$\pt^{H} > 100 \GeV$}\\
&\multicolumn{3}{|l|}{$b$-jet veto for jets with $\pt > 25 \GeV$}\\
\hline
\multirow{5}{*}{\tlhad}
    &\multicolumn{3}{|l|}{Exclude events passing the VBF selection}\\
    &\multicolumn{3}{|l|}{$\MET > 20 \GeV$}\\
&\multicolumn{3}{|l|}{$\pt^H > 100 \GeV$}\\
&\multicolumn{3}{|l|}{$\pt ({\tau_{\rm had}}) > 30 \GeV$}\\
&\multicolumn{3}{|l|}{$b$-jet veto for jets with $\pt > 30 \GeV$}\\
\hline
\multirow{6}{*}{\thadhad}
    &\multicolumn{3}{|l|}{Exclude events passing the VBF selection}\\
&\multicolumn{3}{|l|}{$\Delta \eta(\tau_1,\tau_2)<1.5$}\\
&\multicolumn{3}{|l|}{ $\pt^{H} > 100 \GeV$}\\
\cline{2-4}
\rule{0pt}{2ex}  
&\textbf{Boosted high $\pt^{H}$ }&\multicolumn{1}{|l}{\textbf{Boosted low $\pt^{H}$}}&\\
& $\Delta R(\tau_1,\tau_2)<1.5$ and &\multicolumn{1}{|l}{ $\Delta R(\tau_1,\tau_2)>1.5$ or}&\\
& $\pt^H > 140 \GeV$ &\multicolumn{1}{|l}{ $\pt^H < 140 \GeV$}&\\
\hline\hline
\end{tabular}
\caption{Summary of the selection criteria used to define the VBF and boosted subcategories in the cut-based analysis for
the three analysis channels. The labels (1) and (2) refer to the leading (highest $\pT$) and subleading 
final-state objects (leptons, \thad, jets). The variables are defined in the text.
}
\label{tab:cb_categories}
\end{center}
\end{table}

An overview of the defined categories in the three channels is given in table~\ref{tab:cb_categories}.
In all channels, the event categorisation is designed by splitting events first according to the
production mode, either VBF-like or boosted ggF-like, and second, for the $\tlep \thad$ and
$\thad \thad$ channels, by signal-to-background ratio.
The events accepted in the VBF categories pass a common selection that
requires the presence of the two forward jets distinctive of VBF production.
In the $\tlep \thad$ channel, tight and loose {\em VBF } subcategories are defined, via cuts on the
mass of the dijet system, $m_{jj}$, and $p_\mathrm{T}^H$, the transverse momentum of the Higgs boson candidate.
In the $\thad \thad$ channel, the variables used to select the most sensitive categories for both
production modes are $p_\mathrm{T}^H$ and the
separation $ \Delta R ( \tau_1 , \tau_2 )$ between the two $\thad$ candidates. In the VBF-like events,
correlations between the invariant mass of the selected jets $m_{jj}$ and $\Delta \eta_{jj}$
of the jets characteristic of VBF production are also used. The subcategory with the highest purity
is the {\em VBF high-$p_\mathrm{T}^H$} subcategory, where tight cuts on $p_\mathrm{T}^H$ and 
$ \Delta R (\tau_1 , \tau_2)$ reject almost all non-resonant background sources.
The other two VBF-like subcategories are distinguished by a different signal-to-background ratio due to a
tighter selection applied to the forward jets.
For the $\thad \thad$ channel,  boosted subcategories are also defined.
The division is based on the same cuts on  $p_\mathrm{T}^H$ and $ \Delta R ( \tau_1 , \tau_2 )$ 
as used in the  VBF high-$p_\mathrm{T}^H$  category. Events with low transverse momentum
are not used in any category because in such events the signal cannot be effectively distinguished
from background channels. The proportion of the signal yield produced via VBF in the VBF-like subcategories is found to be 
80\% in the $\tll$ channel, between 67\% and 85\% in the $\tlep \thad$ channel and between 58\% and 78\% in the $\thad \thad$ channel.

\begin{table}
\begin{center}
\vspace{.5cm}

\begin{tabular}{|l|*{5}{c|}}
\hline
\tll & \multicolumn{3}{c|}{VBF}  &\multicolumn{2}{c|}{Boosted} \\ \hline
Total signal  &     \multicolumn{3}{c|}{$11 \pm 4$  }  &\multicolumn{2}{c|}{$38 \pm 13$  } \\
Total background  &     \multicolumn{3}{c|}{$130 \pm 7$ }  &\multicolumn{2}{c|}{$3400 \pm 64$} \\
Data       &     \multicolumn{3}{c|}{$152$       }  &\multicolumn{2}{c|}{$3428$       } \\
\hline
\hline
\tlhad &  Tight VBF   & \multicolumn{2}{c|}{  Loose VBF}  &\multicolumn{2}{c|}{Boosted} \\ \hline
Signal  & $8.8 \pm 3$   & \multicolumn{2}{c|}{$17 \pm 6$   }  &\multicolumn{2}{c|}{$52 \pm 17$  } \\
Background  & $52 \pm 4$    & \multicolumn{2}{c|}{$398 \pm 17$ }  &\multicolumn{2}{c|}{$4399 \pm 73$} \\
Data       & $62$          & \multicolumn{2}{c|}{$407$        }  &\multicolumn{2}{c|}{$4435$       } \\
\hline
\hline
\multirow{2}{*}{\thadhad} & VBF high $p_{\rm T}^{H}$ & \multicolumn{2}{c|}{VBF low $p_{\rm T}^{H}$}      & \multicolumn{2}{c|}{Boosted}  \\
                          &                          &       \multicolumn{1}{c}{tight}             &       \multicolumn{1}{c|}{loose}           &  \multicolumn{1}{c}{high $\pt^{H}$}  &  \multicolumn{1}{c|}{low $\pt^{H}$ } \\ \hline
Signal                                      &  $5.7 \pm 1.9$     &  $5.2 \pm 1.9$          &  $3.7 \pm 1.3$           &  $17 \pm 6$       &  $20 \pm 7$     \\
Background                                      &  $59 \pm 4$        &  $86 \pm 5$             &  $156 \pm 7$             &  $1155 \pm 28$    &  $2130 \pm 41$  \\
Data                                           &  $65$              &  $94$                   &  $157$                   &  $1204$           &  $2121$         \\
\hline
\end{tabular}
\caption{The measured signal and background yields of the cut-based analysis at $\sqrt{s}$~=~8~\TeV\ 
in the $\tll$, $\tlhad$ and $\thadhad$ channels for $m_H$~=~125~\GeV. The normalisations and uncertainties are 
taken from the global fit. The uncertainties on the predicted yields reflect the full statistical and systematic
uncertainties.
}
\label{tab:cba-event-yields}
\end{center}
\end{table}

\begin{figure}[htbp]
\begin{center}
\begin{tabular}{c}
\includegraphics[width=0.60\textwidth]{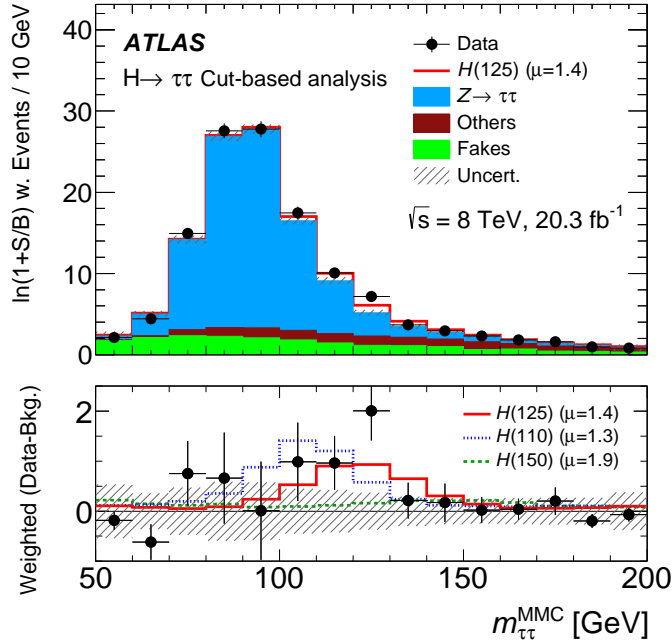}
\end{tabular}
\end{center}
\caption{Distribution of the reconstructed invariant $\tau \tau$ mass, $\MMC$, for the sum of all channels in the cut-based analysis for the
  data taken at $\sqrt{s}$~=~8~\TeV. The events are weighted by a factor of $\ln (1 + S/B)$
  based on the signal ($S$) and background ($B$) yields in each category. 
  The bottom panel shows the difference between weighted data events and weighted
  background events (black points), compared to the weighted signal
  yields.  The background predictions are obtained from the global fit
  with the $m_H$~=~125~\GeV\ signal hypothesis (signal strength $\mu$~=~1.4).
  The $m_H$~=~125~\GeV\ signal is plotted as a solid red
  line, and, for comparison, signals for $m_H$~=~110~\GeV\ (blue) and
  $m_H$~=~150~\GeV\ (green) are also shown. The signal normalisations are taken from fits to 
  data with the corresponding signal mass hypotheses and the fitted $\mu$ values are given in the figure.  
}
\label{fig:CBA_m-tautau}
\end{figure}

The final results are derived from the combined fit of the $m_{\tau \tau}$ distributions observed in
the various subcategories. 
The number of fitted signal and background events in each channel and category  is given in 
table~\ref{tab:cba-event-yields}.
The combined mass distribution for the three channels is shown in
figure~\ref{fig:CBA_m-tautau}, where events are weighted by $\ln (1 + S/B)$, based on the signal and 
background content of their channel and category.
An excess of events above the expected SM background is visible in the mass region around 125~\GeV.

The signal strengths extracted in the three analysis channels and their 
combination are given in table~\ref{tab:cb_mu-values}.
This cut-based analysis also provides evidence for \htautau\ decays, giving a combined signal strength of
\[
\mu_{\mathrm{Cut-based}} = 1.43 \ ^{+0.36}_{-0.35}\mathrm{(stat.)}\ ^{+0.41}_{-0.33}\mathrm{(syst.)}\ \pm 0.10\mathrm{(theory~syst.)}
\]
for $m_H$~=~125.36~\GeV. For comparison, the results obtained in the multivariate analysis
 for the dataset at $\sqrt{s}$~=~8~\TeV\ are also included in table~\ref{tab:cb_mu-values}. 
Good agreement between the results of the two analyses is found for the individual channels as well as for their combination.
To further quantify the level of agreement, the correlation $\rho$ and the uncertainties 
on the differences between the $\mu$ values obtained, i.e.\ $\Delta \mu \pm \delta ( \Delta \mu )$, 
were evaluated using the so-called jackknife technique \cite{jackknife1, jackknife2}. Using this method, 
the correlation between the $\mu$ values obtained in the two analyses is found to be  between 
0.55 and 0.75 for each of the three analysis channels. The results of the analyses are found to be fully 
compatible, with deviations $\Delta \mu /  \delta ( \Delta \mu )$ below 1 for all analysis channels 
as well as for the combined result. 

\begin{table}
\begin{center}
\begin{tabular}{|c|c|c|c|}
\hline
\rule[-3mm]{0mm}{8mm} &\multicolumn{3}{|c|}{Fitted $\mu$ values} \\
\hline
& $\sqrt{s}$ & Multivariate  & Cut-based \\
&            & analysis      & analysis  \\
\hline\hline
 & & &  \\
{\tll} & 8~\TeV      & $1.9^{+1.0}_{-0.9}$   & $3.2^{+1.4}_{-1.3}$ \\
& & &  \\
\hline
& & &  \\
{\tlhad} & 8~\TeV    & $1.1^{+0.6}_{-0.5}$ & $0.7^{+0.7}_{-0.6}$ \\
& & &  \\
\hline
& & &  \\
{\thadhad} & 8~\TeV  & $1.8^{+0.9}_{-0.7}$ & $1.6^{+0.9}_{-0.7}$ \\
& & &  \\
\hline\hline
& & &  \\
{All channels} & 8~\TeV  & $1.53^{+0.47}_{-0.41}$ & $1.43^{+0.55}_{-0.49}$ \\
& & &  \\
\hline
\end{tabular}
\caption{Fitted values of the signal strength for the different channels at
$\sqrt{s}=8 \TeV$ for the multivariate and cut-based analyses,
measured at $m_H$=125.36~\GeV. The results for the combinations of all channels
are also given. The total uncertainties (statistical and systematic) are quoted.
}
\label{tab:cb_mu-values}
\end{center}
\end{table}

\begin{figure}[htbp]
\begin{center}
\begin{tabular}{c}
 \includegraphics[width=0.70\textwidth]{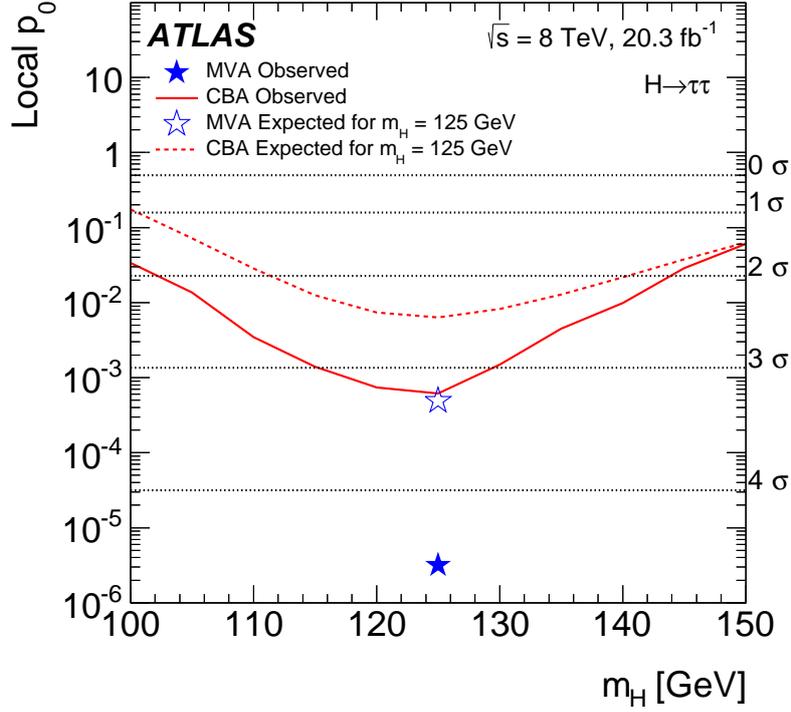}
\end{tabular}
\end{center}
\caption{Observed (solid red) and expected (dashed red) $p_0$ values as a function of $m_H$ for the combination of all channels in the
cut-based analysis (CBA) for the data taken at $\sqrt{s}$~=~8~\TeV. The expected  $p_0$ values are given for the
background-only hypothesis. The corresponding observed and expected $p_0$ values for the multivariate analysis (MVA)
are indicated for $m_H$~=~125~\GeV\ by a full and open star respectively. The axis labels on the right hand side and the 
dotted lines display the significance in units of Gaussian standard deviations. 
}
\label{fig:CBA_p0-mass}
\end{figure}

The probability $p_0$ of obtaining a result at least as signal-like as observed if no signal were present
is shown as a function of the mass in figure~\ref{fig:CBA_p0-mass} for the cut-based analysis for the combined 
dataset at $\sqrt{s}$~=~8~\TeV. The observed  $p_0$ values show a shallow minimum around 125~\GeV, corresponding 
to a significance
of 3.2$\sigma$. The expected significance for the cut-based analysis is superimposed on the figure and reaches a
significance of 2.5$\sigma$ at $m_H$~=~125.36~\GeV. The corresponding significance values
for the multivariate analysis for the dataset at $\sqrt{s}$~=~8~\TeV\
are found to be 4.5$\sigma$ (observed) and 3.3$\sigma$ (expected).
 They are also indicated in the figure.

\begin{figure}[htbp]
  \centering
  \subfigure{\includegraphics[width=0.6\textwidth]{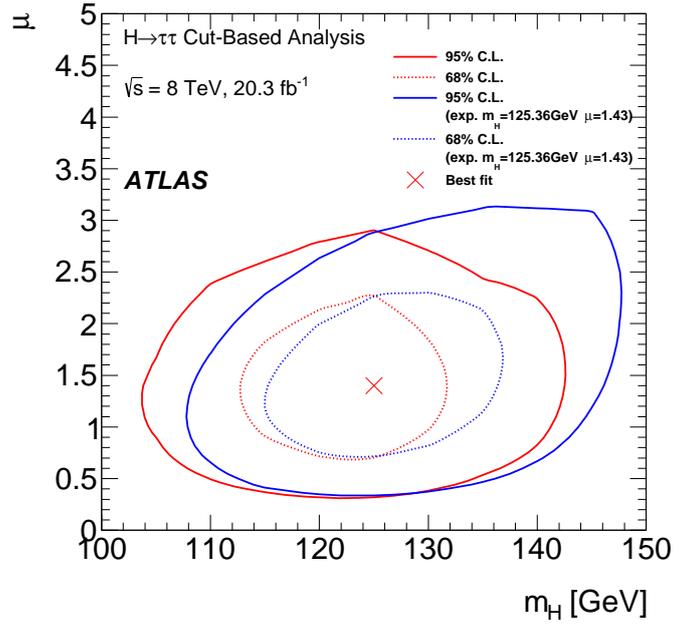}}
  \caption{The results of the two-dimensional likelihood fit in the ($m_H$, $\mu$) plane for the cut-based 
analysis for the data taken at $\sqrt{s}$~=~8~\TeV. The signal strength $\mu$ is the ratio of the measured signal yield to the Standard Model expectation. 
The 68\% and 95\% CL contours are shown as dashed and solid red lines respectively. The best-fit value is indicated as a red cross. The dashed and solid blue lines correspond to the expected 68\% and 95\% CL contours for $m_H$~=~125.36~\GeV\ and $\mu=1.43$.  
}
  \label{fig:cut-based-mass_mu-mH}
\end{figure}

Given the mass sensitivity of the cut-based analysis, a two-dimensional likelihood fit for the signal 
strength $\mu$ and the mass $m_H$ is performed. The mass points are tested in steps of 5~\GeV\ in the 
range between 100~\GeV\ and 150~\GeV. The best fit value is found at $\mu = 1.4$ and $m_H$~=~125~\GeV. 
The result is shown in the ($m_H$, $\mu$) plane in figure~\ref{fig:cut-based-mass_mu-mH} 
together with the 68\% and 95\% CL contours. This result indicates that the observation is compatible with the decay of 
a Standard Model Higgs boson with a mass of 125~\GeV.  

\clearpage
\section{Conclusions}
\label{sec:conclusions}

Evidence for decays of the recently discovered Higgs boson into pairs of tau
leptons is presented.  The analysis is based on the full set of proton--proton
collision data recorded by the ATLAS experiment at the LHC during Run 1. The
data correspond to integrated luminosities of 4.5~\fbs\ and 20.3~\fbs\ at
centre-of-mass energies of $\sqrt{s}=7\TeV$  and $\sqrt{s}=8\TeV$
respectively.  All combinations of leptonic and hadronic tau decay channels are
included and event categories selecting both the vector boson fusion and  highly
boosted $\tau \tau$ signatures are considered in a multivariate analysis.  An
excess of events over the expected background from other Standard Model
processes is found with an observed (expected) significance of 4.5 (3.4)
standard deviations. This excess is consistent with resulting from $\htautau$
decays with $m_H=125.36\GeV$. The measured signal strength, normalised to the
Standard Model expectation, is
\[
\mu = 1.43 \ ^{+0.27}_{-0.26}\mathrm{(stat.)}\ ^{+0.32}_{-0.25}\mathrm{(syst.)}\ \pm 0.09\mathrm{(theory~syst.)}.
\]
This value is consistent with the predicted Yukawa coupling strength of the
Higgs boson in the Standard Model.

The results of the multivariate analysis are cross-checked for the data
collected at $\sqrt{s}=8\TeV$ using a cut-based analysis. The results confirm
the findings of the multivariate analysis, and an excess with a compatible
signal strength is found.  A two-dimensional fit of the signal strength $\mu$
and $m_H$ in the cut-based analysis indicates that the observed excess is
compatible with the $\tau\tau$ decay of the Higgs boson with a mass of 125~\GeV.

\section*{Acknowledgements}

We thank CERN for the very successful operation of the LHC, as well as the
support staff from our institutions without whom ATLAS could not be
operated efficiently.

We acknowledge the support of ANPCyT, Argentina; YerPhI, Armenia; ARC,
Australia; BMWFW and FWF, Austria; ANAS, Azerbaijan; SSTC, Belarus; CNPq and FAPESP,
Brazil; NSERC, NRC and CFI, Canada; CERN; CONICYT, Chile; CAS, MOST and NSFC,
China; COLCIENCIAS, Colombia; MSMT CR, MPO CR and VSC CR, Czech Republic;
DNRF, DNSRC and Lundbeck Foundation, Denmark; EPLANET, ERC and NSRF, European Union;
IN2P3-CNRS, CEA-DSM/IRFU, France; GNSF, Georgia; BMBF, DFG, HGF, MPG and AvH
Foundation, Germany; GSRT and NSRF, Greece; RGC, Hong Kong SAR, China; ISF, MINERVA, GIF, I-CORE and Benoziyo Center, Israel; INFN, Italy; MEXT and JSPS, Japan; CNRST, Morocco; FOM and NWO, Netherlands; BRF and RCN, Norway; MNiSW and NCN, Poland; GRICES and FCT, Portugal; MNE/IFA, Romania; MES of Russia and ROSATOM, Russian Federation; JINR; MSTD,
Serbia; MSSR, Slovakia; ARRS and MIZ\v{S}, Slovenia; DST/NRF, South Africa;
MINECO, Spain; SRC and Wallenberg Foundation, Sweden; SER, SNSF and Cantons of
Bern and Geneva, Switzerland; NSC, Taiwan; TAEK, Turkey; STFC, the Royal
Society and Leverhulme Trust, United Kingdom; DOE and NSF, United States of
America.

The crucial computing support from all WLCG partners is acknowledged
gratefully, in particular from CERN and the ATLAS Tier-1 facilities at
TRIUMF (Canada), NDGF (Denmark, Norway, Sweden), CC-IN2P3 (France),
KIT/GridKA (Germany), INFN-CNAF (Italy), NL-T1 (Netherlands), PIC (Spain),
ASGC (Taiwan), RAL (UK) and BNL (USA) and in the Tier-2 facilities
worldwide.

\bibliographystyle{JHEP}
\bibliography{bibliography}

\onecolumn
\clearpage
\begin{flushleft}
{\Large The ATLAS Collaboration}

\bigskip

G.~Aad$^{\rm 85}$,
B.~Abbott$^{\rm 113}$,
J.~Abdallah$^{\rm 152}$,
S.~Abdel~Khalek$^{\rm 117}$,
O.~Abdinov$^{\rm 11}$,
R.~Aben$^{\rm 107}$,
B.~Abi$^{\rm 114}$,
M.~Abolins$^{\rm 90}$,
O.S.~AbouZeid$^{\rm 159}$,
H.~Abramowicz$^{\rm 154}$
% ...
\bigskip
\\
% ...
$^{11}$ Institute of Physics, Azerbaijan Academy of Sciences, Baku, Azerbaijan\\
% ...
$^{85}$ CPPM, Aix-Marseille Universit{\'e} and CNRS/IN2P3, Marseille, France\\
% ...
$^{90}$ Department of Physics and Astronomy, Michigan State University, East Lansing MI, United States of America\\
% ...
$^{107}$ Nikhef National Institute for Subatomic Physics and University of Amsterdam, Amsterdam, Netherlands\\
% ...
$^{113}$ Homer L. Dodge Department of Physics and Astronomy, University of Oklahoma, Norman OK, United States of America\\
$^{114}$ Department of Physics, Oklahoma State University, Stillwater OK, United States of America\\
% ...
$^{117}$ LAL, Universit{\'e} Paris-Sud and CNRS/IN2P3, Orsay, France\\
% ...
$^{152}$ Institute of Physics, Academia Sinica, Taipei, Taiwan\\
% ...
$^{154}$ Raymond and Beverly Sackler School of Physics and Astronomy, Tel Aviv University, Tel Aviv, Israel\\
% ...
$^{159}$ Department of Physics, University of Toronto, Toronto ON, Canada\\
% ...
\end{flushleft}

\end{document}